\documentclass[longauth]{aa}
\usepackage{graphicx}
\usepackage[varg]{txfonts}

\usepackage{silence}
\newcommand{\Nai}{\ion{Na}{i}\,}

\newcommand{\Feii} {\ion{Fe}{ii}\,}

\newcommand{\Caii} {\ion{Ca}{ii}\,}

\newcommand{\Cii} {\ion{C}{ii}\,}
\newcommand{\Ciii} {\ion{C}{iii}\,}

\newcommand{\Hei} {\ion{He}{i}\,}
\newcommand{\Heii} {\ion{He}{ii}\,}

\newcommand{\Nii} {\ion{N}{ii}\,}
\newcommand{\Niii} {\ion{N}{iii}\,}
\newcommand{\Oi} {\ion{O}{i}\,}
\newcommand{\Oii} {\ion{O}{ii}\,}

\newcommand{\Sii} {\ion{S}{ii}\,}

\newcommand{\SiII} {\ion{Si}{ii}\,}

\newcommand{\Mgi} {\ion{Mg}{i}\,}
\newcommand{\Mgii} {\ion{Mg}{ii}\,}

\newcommand{\msun}{\mbox{M$_{\odot}$\,}}

\newcommand{\kms}{\mbox{$\rm{km}\,s^{-1}$}}

\begin{document}

   \title{Massive stars exploding in a He-rich circumstellar medium $-$ X.}

   \subtitle{Flash spectral features in the Type Ibn SN 2019cj and observations of SN 2018jmt}

   \authorrunning{Z.-Y. Wang et al.} 
   \titlerunning{Two Ibn SNe 2018jmt and 2019cj}
   
   \author{
Z.-Y. Wang\orcid{0000-0002-0025-0179}\inst{\ref{inst1},\ref{inst2}} \and
A. Pastorello\orcid{0000-0002-7259-4624}\inst{\ref{inst3}} \and
K. Maeda\orcid{0000-0003-2611-7269}\inst{\ref{inst4}} \and
A. Reguitti\orcid{0000-0003-4254-2724}\inst{\ref{inst5},\ref{inst3}} \and
Y.-Z. Cai\orcid{0000-0002-7714-493X}\inst{\ref{inst6},\ref{inst7},\ref{inst8}}\thanks{Corresponding authors: yzcai789@163.com (CYZ)} \and
D. Andrew Howell\inst{\ref{inst9},\ref{inst10}} \and \\
S. Benetti\orcid{0000-0002-3256-0016}\inst{\ref{inst3}} \and
D. Buckley\orcid{0000-0002-7004-9956}\inst{\ref{inst11}} \and
E. Cappellaro\inst{\ref{inst3}} \and
R. Carini\orcid{0000-0003-1604-2064}\inst{\ref{inst12}} \and
R. Cartier\inst{\ref{inst13}} \and
T.-W. Chen\orcid{0000-0002-1066-6098}\inst{\ref{inst14}} \and
N. Elias-Rosa\orcid{0000-0002-1381-9125}\inst{\ref{inst3},\ref{inst15}} \and \\
Q.-L. Fang\inst{\ref{inst4}} \and
A. Gal-Yam\orcid{0000-0002-3653-5598}\inst{\ref{inst16}} \and
A. Gangopadhyay \inst{\ref{inst17}} \and
M. Gromadzki\orcid{0000-0002-1650-1518}\inst{\ref{inst18}} \and
{W.-P. Gan}\inst{\ref{inst19add}} \and
D. Hiramatsu\orcid{0000-0002-1125-9187}\inst{\ref{inst19},\ref{inst20}} \and
{M.-K. Hu}\inst{\ref{inst36}}\and \\
C. Inserra\orcid{0000-0002-3968-4409}\inst{\ref{inst21}} \and
C. McCully\inst{\ref{inst9}} \and
M. Nicholl\orcid{0000-0002-2555-3192}\inst{\ref{inst22}} \and
F. Olivares E.\orcid{0000-0002-5115-6377}\inst{\ref{inst23}} \and
G. Pignata\inst{\ref{inst24}} \and
J. Pineda-Garc\'ia\orcid{0000-0003-0737-8463}\inst{\ref{inst25}} \and
M. Pursiainen\orcid{0000-0003-4663-4300}\inst{\ref{inst26}} \and \\
F. Ragosta\orcid{0000-0003-2132-3610}\inst{\ref{inst12},\ref{inst27}} \and
A. Rau\inst{\ref{inst28}} \and
R. Roy\inst{\ref{inst29}} \and
J. Sollerman\orcid{0000-0003-1546-6615}\inst{\ref{inst17}} \and
L. Tartaglia\orcid{0000-0003-3433-1492}\inst{\ref{inst30}} \and
G. Terreran\orcid{0000-0003-0794-5982}\inst{\ref{inst9},\ref{inst10}} \and
G. Valerin\orcid{0000-0002-3334-4585}\inst{\ref{inst3}} \and
Q. Wang\orcid{0000-0001-5233-6989}\inst{\ref{inst31}} \and 
{S.-Q. Wang}\orcid{0000-0001-7867-9912}\inst{\ref{inst19add}} \and
D. R. Young\orcid{0000-0002-1229-2499}\inst{\ref{inst22}} \and
A. Aryan\inst{\ref{inst14}} \and
M. Bronikowski\inst{\ref{inst32}} \and 
E. Concepcion\inst{\ref{inst32}} \and
L. Galbany\orcid{0000-0002-1296-6887}\inst{\ref{inst15},\ref{inst33}} \and
H. Lin\inst{\ref{inst6},\ref{inst7},\ref{inst8}} \and \\
A. Melandri\inst{\ref{inst12}} \and 
T. Petrushevska\inst{\ref{inst32}} \and 
M. Ramirez\orcid{0009-0001-3293-7741}\inst{\ref{inst25},\ref{inst34}} \and
D.-D Shi\inst{\ref{inst39},\ref{inst35}} \and
B. Warwick\inst{\ref{inst26}} \and
J.-J. Zhang\orcid{0000-0002-8296-2590}\inst{\ref{inst6},\ref{inst7},\ref{inst8}} \and
B. Wang\inst{\ref{inst6},\ref{inst7},\ref{inst8}} \and
X.-F.~Wang\orcid{0000-0002-7334-2357}\inst{\ref{inst36}}~\thanks{wang\_xf@mail.tsinghua.edu.cn (WXF)} \and
X.-J. Zhu\orcid{0000-0001-7049-6468}\inst{\ref{inst2},\ref{inst37},\ref{inst38}}\thanks{zhuxj@bnu.edu.cn (ZXJ)}
}
   \institute{
\label{inst1}School of Physics and Astronomy, Beijing Normal University, Beijing 100875, P.R. China \and
\label{inst2}Department of Physics, Faculty of Arts and Sciences, Beijing Normal University, Zhuhai 519087, P.R. China \and
\label{inst3}INAF - Osservatorio Astronomico di Padova, Vicolo dell'Osservatorio 5, 35122 Padova, Italy \and
\label{inst4}Department of Astronomy, Kyoto University, Kitashirakawa-Oiwake-cho, Sakyo-ku, Kyoto 606-8502, Japan \and
\label{inst5}INAF - Osservatorio Astronomico di Brera, Via E. Bianchi 46, 23807 Merate (LC), Italy \and
\label{inst6}Yunnan Observatories, Chinese Academy of Sciences, Kunming 650216, P.R. China \and
\label{inst7}International Centre of Supernovae, Yunnan Key Laboratory, Kunming 650216, P.R. China 
\and
\label{inst8}Key Laboratory for the Structure and Evolution of Celestial Objects, Chinese Academy of Sciences, Kunming 650216, P.R. China \and
\label{inst9}Las Cumbres Observatory, 6740 Cortona Drive, Suite 102, Goleta, CA 93117-5575, USA \and
\label{inst10}Department of Physics, University of California, Santa Barbara, CA 93106-9530, USA \and
\label{inst11}{South African Astronomical Observatory, PO Box 9, Observatory 7935, Cape Town, South Africa} \and
\label{inst12}INAF - Osservatorio Astronomico di Roma, via Frascati 33, I-00078, Monte Porzio Catone, Italy \and
\label{inst13}Instituto de Estudios Astrof\'isicos, Facultad de Ingenier\'ia y Ciencias, Universidad Diego Portales, Av. Ej\'ercito Libertador 441, Santiago, Chile \and
\label{inst14}Graduate Institute of Astronomy, National Central University, 300 Jhongda Road, 32001 Jhongli, Taiwan \and
\label{inst15}Institute of Space Sciences (ICE, CSIC), Campus UAB, Carrer de Can Magrans, s/n, E-08193 Barcelona, Spain \and
\label{inst16}Department of Particle Physics and Astrophysics, Weizmann Institute of Science, 76100 Rehovot, Israel \and
\label{inst17}The Oskar Klein Centre, Department of Astronomy, Stockholm University, AlbaNova, SE-10691 Stockholm, Sweden \and
\label{inst18}Astronomical Observatory, University of Warsaw, Al. Ujazdowskie 4, 00-478 Warszawa, Poland \and
\label{inst19add}{Guangxi Key Laboratory for Relativistic Astrophysics, School of Physical Science and Technology, Guangxi University, Nanning 530004, P.R. China} \and
\label{inst19}Center for Astrophysics | Harvard \& Smithsonian, 60 Garden Street, Cambridge, MA 02138-1516, USA
\newpage \and
\label{inst20}The NSF AI Institute for Artificial Intelligence and Fundamental Interactions, USA \and
\label{inst36}Department of Physics, Tsinghua University, Beijing 100084, P.R. China \and
\label{inst21}Cardiff Hub for Astrophysics Research and Technology, School of Physics \& Astronomy, Cardiff University, Queens Buildings, The Parade, Cardiff, CF24 3AA, UK \and
\label{inst22}Astrophysics Research Centre, School of Mathematics and Physics, Queen's University Belfast, Belfast BT7 1NN, UK \and
\label{inst23}UKIRT Observatory, Institute for Astronomy, 640 N.\ A'ohoku Place, University Park, Hilo, Hawai'i 96720, USA 
\and
\label{inst24}Instituto de Alta Investigaci\'{o}n, Universidad de Tarapac\'{a}, Casilla 7D, Arica, Chile \and
\label{inst25}Instituto de Astrof\'isica, Facultad de Ciencias Exactas, Universidad Andres Bello, Av. Fern\'andez Concha 700, Santiago, Chile \and
\label{inst26}Department of Physics, University of Warwick, Gibbet Hill Road, Coventry CV4 7AL, UK 
\and
\label{inst27}Space Science Data Center-ASI, Via del Politecnico SNC, 00133 Roma, Italy 
\and
\label{inst28}Max-Planck-Institut f{\"u}r Extraterrestrische Physik, Giessenbachstra\ss e 1, 85748, Garching, Germany \and
\label{inst29}Manipal Centre for Natural Sciences, Manipal Academy of Higher Education, Manipal - 576104, Karnataka, India\and
\label{inst30}INAF - Osservatorio Astronomico d'Abruzzo, Via M. Maggini snc, 64100 Teramo, Italy \and
\label{inst31}Physics and Astronomy Department, Johns Hopkins University, Baltimore, MD 21218, USA \and
\label{inst32}Center for Astrophysics and Cosmology, University of Nova Gorica, Vipavska 11c, 5270 Ajdov\v{s}\v{c}ina, Slovenia \and
\label{inst33}Institut d'Estudis Espacials de Catalunya (IEEC), 08860 Castelldefels (Barcelona), Spain \and
\label{inst34}Millennium Institute of Astrophysics, Nuncio Monse\~{n}or Sotero Sanz 100, Of. 104, Providencia, Santiago, Chile \and
\label{inst39}{Center for Fundamental Physics, School of Mechanics and opticelectrical Physics, Anhui University of Science and Technology, Huainan, Anhui 232001, P.R. China} \and
\label{inst35}Purple Mountain Observatory, Chinese Academy of Sciences, 10 Yuan Hua Road, Nanjing, Jiangsu 210023, P.R. China \and
\label{inst37}Institute for Frontier in Astronomy and Astrophysics, Beijing Normal University, Beijing 102206, P.R. China \and
\label{inst38}Advanced Institute of Natural Sciences, Beijing Normal University, Zhuhai 519087, P.R. China
}

   \date{Received June xx, 2024; accepted June xx, 2024}
 
  \abstract
  {We present optical and near-infrared observations of two Type Ibn supernovae (SNe), SN\,2018jmt and SN\,2019cj. Their light curves have rise times of about 10 days, reaching an absolute peak magnitude of $M_g$(SN\,2018jmt) = $-$19.07 $\pm$ 0.37 and $M_V$(SN\,2019cj) = $-$18.94 $\pm$ 0.19 mag, respectively. The early-time spectra of SN\,2018jmt are dominated by a blue continuum, accompanied by narrow (600$-$1000 km~s$^{-1}$) He {\sc i} lines with P-Cygni profile. At later epochs, the spectra become more similar to those of the prototypical SN Ibn 2006jc. At early phases, the spectra of SN\,2019cj show flash ionisation emission lines of C {\sc iii}, N {\sc iii} and He {\sc ii} superposed on a blue continuum. These features disappear after a few days, and then the spectra of SN\,2019cj evolve similarly to those of SN\,2018jmt. The spectra indicate that the two SNe exploded within a He-rich circumstellar medium (CSM) lost by the progenitors a short time before the explosion. We model the light curves of the two SNe Ibn to constrain the progenitor and the explosion parameters. The ejecta masses are consistent with either that expected for a canonical SN Ib ($\sim$ 2 M$_{\odot}$) or those from a massive WR star ($>$ $\sim$ 4 M$_{\odot}$), with the kinetic energy on the order of $10^{51}$ erg. The lower limit on the ejecta mass ($>$ $\sim$ {2 M}$_{\odot}$) argues against a scenario involving a relatively low-mass progenitor (e.g., $M_{ZAMS}$ $\sim$ 10 M$_{\odot}$). We set a conservative upper limit of $\sim$~0.1 M$_{\odot}$ for the $^{56}$Ni masses in both SNe. From the light curve modelling, we determine a two-zone CSM distribution, with an inner, flat CSM component, and an outer CSM with a steeper density profile. The physical properties of SN\,2018jmt and SN\,2019cj are consistent with those expected from the core collapse of relatively massive, stripped-envelope (SE) stars.
  }


   \maketitle
\nolinenumbers  
\section{Introduction}
Supernovae (SNe) interacting with a circumstellar medium (CSM) provide an opportunity to probe the latest evolutionary stages of massive stars before their explosion.
In general, the interaction of the SN ejecta with the CSM produces quite complex spectral line profiles, sometimes characterised by multiple width components, but usually dominated by a relatively narrow emission feature. The narrow emission component is thought to originate in the unshocked, photoionised CSM, which is expanding slowly (from tens to several hundreds km s$^{-1}$) \citep{Dessart2024arXiv240504259D}. 
As a consequence, measuring the width of these narrow lines gives an indication of the velocity of the pre-SN stellar wind \citep[e.g., ][]{Chevalier1994ApJ...420..268C, Fransson2002ApJ...572..350F, Pastorello2016MNRAS.456..853P, Smith2017hsn..book..403S}.

In general, interacting core-collapse (CC) SNe are divided into three observational types, depending on the identification in the spectra of individual narrow features \citep{Gal-Yam2017hsn..book..195G}: the H-rich SNe IIn \cite[e.g.,][]{Schlegel1990MNRAS.244..269S,Filippenko1997ARA&A..35..309F,Fraser2020RSOS....700467F}, the He-rich and H-poor SNe Ibn \cite[e.g., ][]{Pastorello2007Natur.447..829P, Pastorello2008MNRAS.389..113P, Hosseinzadeh2017ApJ...836..158H, Maeda2022ApJ...927...25M}, and C/N/O-rich and H/He-poor SNe Icn \cite[e.g., ][]{Fraser2021arXiv210807278F, Gal-Yam2022Natur.601..201G, Perley2022ApJ...927..180P, Pellegrino2022ApJ...938...73P}.

The spectral properties of SNe Ibn are explained in terms of interaction between the SN ejecta and the H-deprived, He-rich CSM. The spectra of SNe Ibn are characterised by narrow \Hei~emission lines, with full-width at half maximum (FWHM) velocities ranging from several hundred to a few thousand \kms~\citep[see e.g., ][]{Pastorello2016MNRAS.456..853P, Hosseinzadeh2017ApJ...836..158H, Wang2021ApJ...917...97W}. However, in a few cases \citep[e.g., SN\,2005la, SN\,2011hw, SN\,2022pda;][Cai et al., in preparation]{Pastorello2008MNRAS.389..131P, Pastorello2015MNRAS.449.1921P} weak H lines were observed in the spectra of some SNe Ibn, suggesting the presence of some residual H in the CSM, forming a subclass of SNe Ibn. Most SNe Ibn show relatively fast-evolving light curves, with a typical rise time of $\leq$ 15 days, a post-peak decline rate of $\sim$ 0.05$-$0.15 mag day$^{-1}$, and a peak absolute magnitude of $\approx-$19 mag \citep[see the statistic study on SNe Ibn by][]{Hosseinzadeh2017ApJ...836..158H}. However, some outliers were observed in the SN Ibn family, such as ASASSN-14ms, which has a highly luminous peak at $M_{\rm{V}}$ $\sim$ $-$20.5 mag \citep{Vallely2018MNRAS.475.2344V, Wang2021ApJ...917...97W}, while OGLE-2012-SN-006 and OGLE-2014-SN-131 shows an unprecedented long-lasting light curve evolution \citep{Pastorello2015MNRAS.449.1941P,Karamehmetoglu2017AA...602A..93K}.

Although the first discovery of this family was SN\,1999cq \citep{Matheson2000AJ....119.2303M}, the label `SNe Ibn' was introduced in the study of SN 2006jc, which is considered to be the prototype of Type Ibn events \citep{Pastorello2007Natur.447..829P}. 
SNe Ibn are a rare group of stellar explosions, with only 66 events discovered so far\footnote{Based on a query conducted on the Transient Name Server (\url{https://www.wis-tns.org/}) on 2024 Jun 14.}.
{Based on data from the Zwicky Transient Facility \citep[ZTF;][]{Bellm2019PASP..131a8002B} Bright Transient Survey, \citet{Perley2020ApJ...904...35P} estimated a detection rate of 0.66\% for Type Ibn SNe within the ZTF transient sample. Additionally, \citet{Maeda2022ApJ...927...25M} estimated that the observed volumetric rate of SNe Ibn is approximately 1\% of all CC SNe.}

Studies in the literature suggest that the progenitors of SNe Ibn can either be H-poor, massive Wolf-Rayet (WR) stars which have experienced major mass-loss events shortly before the terminal CC, or lower-mass He stars in binary systems, where the interaction with the companion favours the mass loss from the primary \citep[e.g., ][]{Maund2016ApJ...833..128M, Hosseinzadeh2019ApJ...871L...9H, Maeda2022ApJ...927...25M,Wang2024MNRAS.530.3906W}. Unfortunately, to date, there have been no direct detection of SN Ibn progenitors. 
{\citet{Sun2020MNRAS.491.6000S} reported the detection of a point source at the location of the SN\,2006jc explosion with the Hubble Space Telescope (HST)\footnote{\url{https://science.nasa.gov/mission/hubble/}}, providing evidence that it is the progenitor's binary companion.}

In general, SNe Ibn are observed in active star-forming regions within their host galaxies \citep[e.g., ][]{Kuncarayakti2013AJ....146...30K, Taddia2015A&A...580A.131T, Pastorello2015MNRAS.449.1941P}. However, as a remarkable exception, PS1-12sk was detected on the outskirts of an elliptical galaxy, challenging the massive star explosion scenario for SNe Ibn \citep{Sanders2013ApJ...769...39S}. Therefore, there are still many open questions regarding SNe Ibn, for example, the homogeneity of the progenitors, the origins of CSM, and the sources powering the SN light curves. 

In this paper, we present our photometric and spectroscopic observational data of two Type Ibn SNe\,2018jmt and 2019cj, which were observed in the framework of the extended Public European Southern Observatory (ESO) Spectroscopic Survey of Transient Objects \citep[ePESSTO,][]{Smartt2015A&A...579A..40S}. The paper is organised as follows: the information on the discovery, distance, and extinction of the two objects are reported in Section \ref{section:Basic_sample_information}. In Section \ref{section:Observations_and_data_reduction}, we present the observations and the data reduction techniques. Photometric and spectroscopic analyses are presented in Sections \ref{section:Photometry} and \ref{section:Spectroscopy}, respectively. Finally, the discussion and conclusions are presented in Section \ref{section:Discussion_and_conclusions}.

\section{Basic sample information}
\label{section:Basic_sample_information}

\subsection{SN~2018jmt}

The discovery of SN\,2018jmt, attributed to the Mobile Astronomical System of the Telescope-Robots \citep[MASTER;][]{Lipunov2012ASInC...7..275L, Gorbovskoy2013ARep...57..233G}, is dated 2018 December 08.28 (epoch corresponding to MJD~=~58460.28; UT dates are used throughout the paper). 
The object was observed in an unfiltered image, with a magnitude of 16.5 \citep{Chasovnikov2018TNSTR1888....1C}.
However, an earlier detection was reported by the All-Sky Automated Survey for Supernovae \citep[ASAS-SN; ][]{Shappee2014AAS...22323603S, Kochanek2017PASP..129j4502K, Jayasinghe2019MNRAS.485..961J} {on 2018 December 05.29 (MJD~=~58457.29), at a magnitude $g$~=~18.32$\pm$0.20 mag\footnote{\url{https://asas-sn.osu.edu/photometry}}.}
{The last non-detection by ASAS-SN was on 2018 December  02.31 (MJD~=~58454.31) in the $g$-band, with an estimated limit of 18.57 mag.}
Soon after its discovery, the SN was classified as a Type Ibn event by the ePESSTO collaboration \citep{Castro-Segura2018TNSCR2064....1C}. The SN coordinates are RA~=~$06^\mathrm{h}54^\mathrm{m}47^\mathrm{s}.100$, Dec~=~$-59^\circ30'10''.80$ (J2000.0). 
The location of the SN within the host galaxy is shown in Fig.~\ref{fig:location}.

SN\,2018jmt is possibly associated with the host galaxy PGC 370943 \citep[2MASSJ06544633-5930163,][]{Skrutskie2006AJ....131.1163S}. Due to the lack of distance information, we inferred the kinematic distance of the host galaxy from the most prominent narrow \Hei ($\lambda_{0}$ ~=~ 5875.6 \AA) line in the SN spectra. We measured the central wavelength of the narrow \Hei lines and obtained the redshift of $z$~=~0.032$\pm$0.001. Adopting a standard cosmology with $H_0$~=~73~$\pm$~5~\kms~$\mathrm{Mpc}^{-1}$, $\Omega_M$~=~0.27, $\Omega_{\Lambda}$~=~0.73 \citep{Spergel2007ApJS..170..377S}, we estimated a luminosity distance $d_L$ ~=~ 134.7 $\pm$ 13.6\,Mpc ($\mu$ ~=~ 35.65 $\pm$ 0.22\,mag) for SN\,2018jmt. 
Regarding the interstellar reddening, we adopt $E\left (B-V \right ) _{\rm Gal} $ ~=~ 0.105\,mag for the Galactic reddening contribution \citep{Schlafly2011ApJ...737..103S}, retrieved via the NASA/IPAC Extragalactic Database (NED)\footnote{\url{https://ned.ipac.caltech.edu}}, assuming a reddening law with $R_V$~=~3.1 \citep{Cardelli1989ApJ...345..245C}. However, the host galaxy extinction cannot be firmly constrained due to the low signal-to-noise ratios (S/N) in our early spectra. Therefore, we adopt $E\left (B-V \right ) _{\rm Total} $ ~=~ 0.105\,mag as the total reddening towards SN\,2018jmt.

\subsection{SN~2019cj}

Although the discovery of SN\,2019cj was officially announced by ASAS-SN on 2019 January 03.26 (MJD~=~58486.26) with a Sloan $g$-band brightness of 18.3 mag \citep[AB,][]{Nicholls2019TNSTR..20....1N}, an earlier detection was obtained by the Asteroid Terrestrial-impact Last Alert System \citep[ATLAS;] []{Tonry2018PASP..130f4505T, Smith2020PASP..132h5002S, Shingles2021TNSAN...7....1S} survey on 2018 December 31.42 (MJD~=~58483.42), with the object having an orange ($o$) band magnitude of $o$~=~19.75$\pm$0.27 mag\footnote{\url{https://fallingstar-data.com/}}.
{The last non-detection by ASAS-SN was on 2018 December 29.30 (MJD~=~58481.30) in the $g$-band, with an estimated limit of 18.68 mag.}
Soon after the discovery, SN\,2019cj was classified as a Type II or Type Ibn SN \citep{Pignata2019TNSCR..42....1P} by ePESSTO. Its coordinates are RA~=~$04^\mathrm{h}56^\mathrm{m}22^\mathrm{s}.977$, Dec~=~$-46^\circ02'13''.68$ (J2000.0).

\begin{figure*}
    \includegraphics[width=\columnwidth]{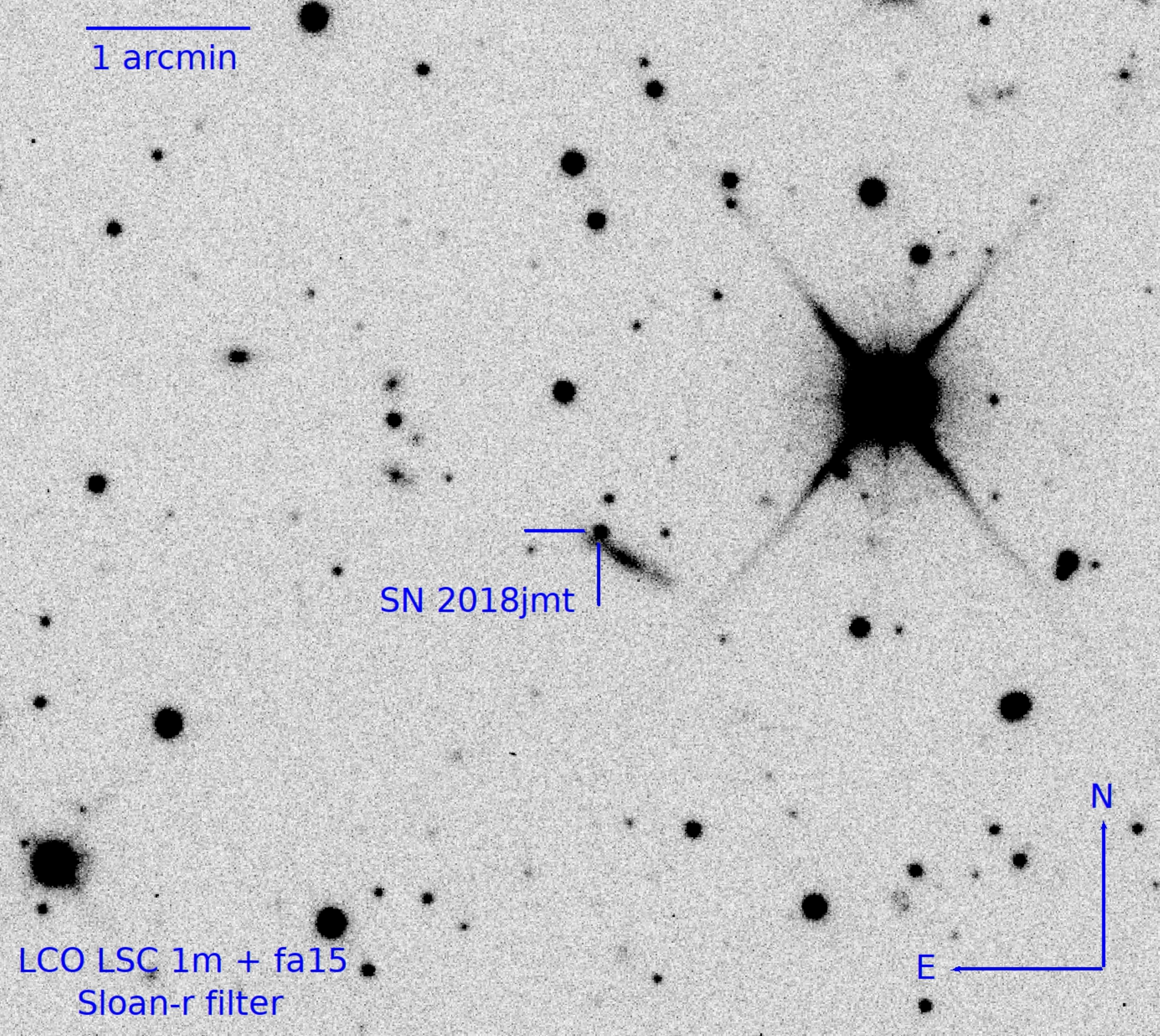}
    \includegraphics[width=\columnwidth]{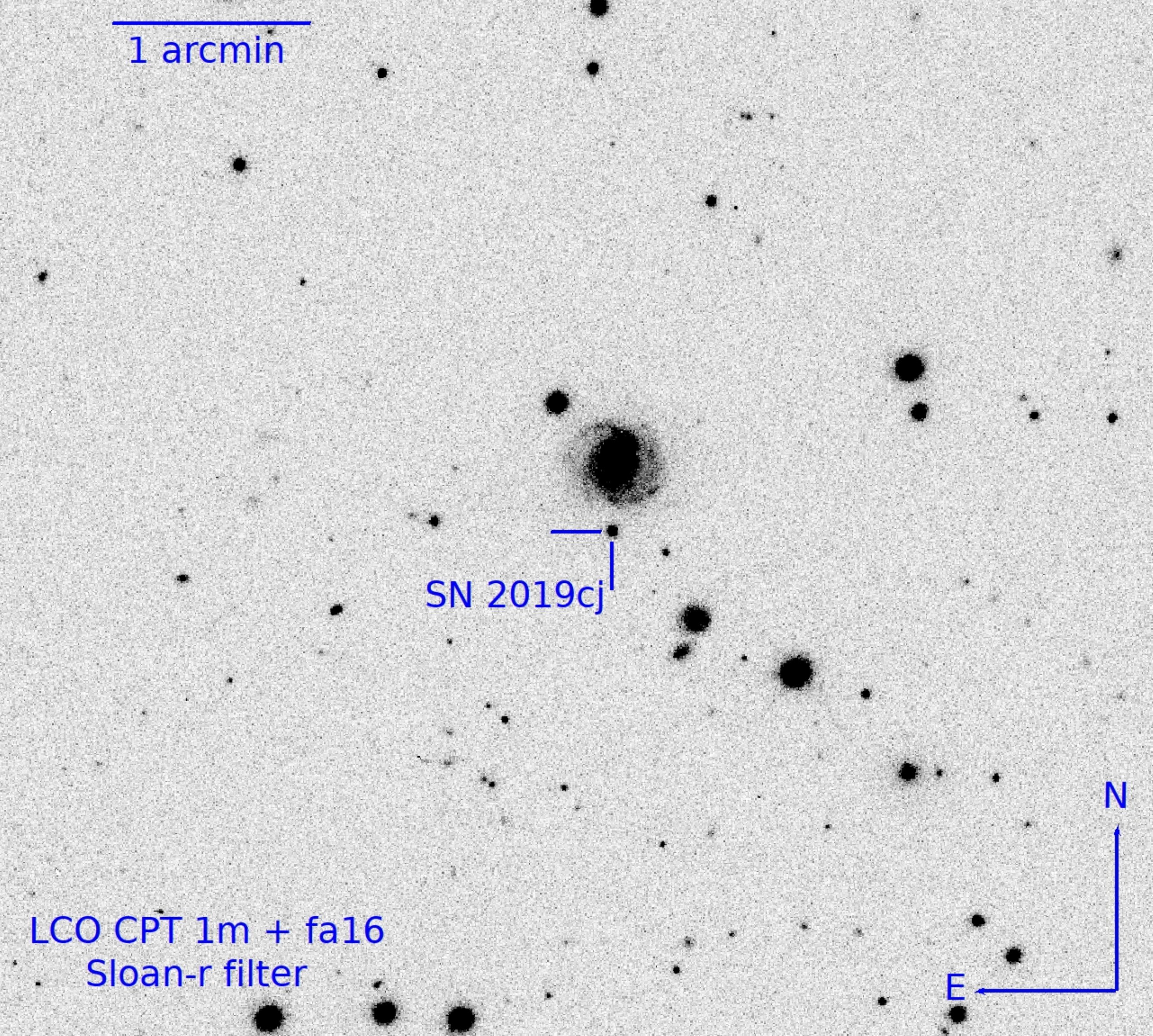}
    \caption{Images of the locations of SN\,2018jmt and SN\,2019cj, taken on 2018 December 20 and 2019 January 11, respectively, by the LCO telescopes with the $r$-filter. The orientation and scale of the images are reported.}
    \label{fig:location}
\end{figure*}

SN\,2019cj is located 21.08" south and 1.43" east from the center of its predicted host galaxy, AM 0454-460 (PGC 130531), which is a face-on Sc-type galaxy \citep{Loveday1996MNRAS.278.1025L, Moustakas2023ApJS..269....3M}. The location of the supernova is shown in Fig.~\ref{fig:location}.
Adopting the recessional velocity of $v$~=~13313 $\pm$ 49 \kms \citep[hence a redshift $z$~=~ 0.0444~$\pm$~0.0002;][]{Mould2000ApJ...529..786M}, corrected for the Virgo Cluster, the Great Attractor, and the Shapley supercluster influence, and adopting the same standard cosmological model, we obtain a luminosity distance of $d_L$ ~=~ 188.7 $\pm 13.9$\,Mpc, hence a distance modulus $\mu_L$ ~=~ 36.38 $\pm$ 0.16\,mag.
The Galactic reddening towards SN\,2019cj is small, $E\left (B-V \right ) _{\rm Gal} $ ~=~ 0.016 \,mag \citep{Schlafly2011ApJ...737..103S}, assuming the same $R_V$ as for SN\,2018jmt. 
The remote location of this SN in the outskirts of AM 0454-460, along with the low S/N in our early spectra, suggests negligible extinction from the host galaxy.
For this reason, we assume the total line-of-sight reddening, $E\left (B-V \right ) _{\rm tot} $ ~=~ 0.016 \,mag, is only due to the Galactic contribution.

\section{Observations and data reduction}
\label{section:Observations_and_data_reduction}

\subsection{Photometric data}
\label{subsection:Photometric_data}

We conducted multi-band optical (Sloan $griz$, Johnson-Cousins $UBV$) and near-infrared (NIR; $JHK$) follow-up campaigns of SNe\,2018jmt and 2019cj starting shortly after their classification. 
The telescopes and instruments utilised were the following:
the 3.58m New Technology Telescope (NTT) equipped with the ESO Faint Object Spectrograph and Camera 2 (EFOSC2) and the Son of Isaac (SOFI), hosted on La Silla (Chile) of the ESO.
Additionally, we obtained a single epoch (2019-02-14) of SN 2018jmt photometry as part of GREAT program \citep{Chen2018ApJ...867L..31C} using the Gamma-Ray Burst Optical/Near-Infrared Detector \citep[GROND,][]{Greiner2008PASP..120..405G}, a 7-channel imager that collects multi-colour photometry simultaneously with $g'r'i'z'JHK$s bands, mounted at the 2.2m MPG telescope at ESO La Silla Observatory, Chile. 
Furthermore, we collected additional data through the Las Cumbres Observatory \citep[LCO,][]{Brown2013PASP..125.1031B} global network. This network includes the 1-m class telescopes identified as fa06, fa14 and fa16, and hosted in the South African Astronomical Observatory (SAAO) in Sutherland, South Africa; the fa03 and fa15 telescopes of the Cerro Tololo Interamerican Observatory site, Chile, and the fa11 and fa12 telescopes at the Siding Spring Observatory in New South Wales, Australia. 

The raw images were first pre-reduced by applying bias and overscan corrections, flat-fielding, and trimming, which are standard correction steps performed in \textsc{iraf}\footnote{\url{https://iraf-community.github.io/}} \citep{Tody1986SPIE..627..733T, Tody1993ASPC...52..173T}. 
The NTT data were retrieved by the ePESSTO collaboration.
The raw images from ePESSTO were pre-reduced using the dedicated \texttt{PESSTO} pipeline \citep[see][]{Smartt2015A&A...579A..40S}.
The GROND raw images were reduced with the \texttt{GROND} pipeline \citep{Kruhler2008ApJ...685..376K},
which applies de-bias and flat-field corrections, stacks images and provides astrometry calibration.
If multiple exposures were taken with the same instrument and in the same night, we combined them into stacked science frames to increase the S/N.

Subsequent photometric data reduction steps  were  carried out with a dedicated pipeline called {\sl ecsnoopy}\footnote{{\sl ecsnoopy} is a package for SN photometry using PSF fitting and/or template subtraction developed by E. Cappellaro. A package description can be found at \url{http://sngroup.oapd.inaf.it/snoopy.html.}},
which consists of several photometric packages, including {\sc daophot}\footnote{\url{http://www.star.bris.ac.uk/~mbt/daophot/}} for magnitude measurement \citep[][]{Stetson1987PASP...99..191S}, {\sc sextractor}\footnote{\url{www.astromatic.net/software/sextractor/}}~for source extraction \citep[][]{Bertin1996A&AS..117..393B},  and {\sc hotpants}\footnote{\url{https://github.com/acbecker/hotpants/}} for template subtraction \citep{Becker2015ascl.soft04004B}. 
The {\sl ecsnoopy} pipeline was used for astrometric calibration, image combination, and point-spread-function (PSF) fitting photometry, with the subtraction of a reference image (referred to as the "template") when required.
We directly adopted the simple PSF-fitting technique for SN\,2019cj, due to its location in the outskirts of the host galaxy, while template subtraction was necessary for SN\,2018jmt to remove the background contamination from the host galaxy\footnote{The $UBVgri$ template images were taken through LCO-fa15 in November 2023, i.e. about 5 years after the discovery.}. The PSF-fitting technique consists of constructing a PSF model by selecting for each image bright and isolated stars in the SN field. The sky background was then subtracted by fitting a low-order polynomial (e.g., a second or third order) in the SN proximity. The modelled source was subtracted from the original images, and the fitting process was repeated to minimise the residuals. When the SN was not detected, an upper limit to the object brightness was estimated. 

Photometric calibration of instrumental magnitudes was performed by adopting instrumental zero points (ZPs) and color terms (CTs) inferred through observations of standard stars on photometric nights.
Specifically, Johnson-Cousins filter photometry was calibrated using standard stars from the \citet{Landolt1992AJ....104..340L} catalogue, while the Sloan data were calibrated using standards catalogued by Pan-STARRS \citep{Chambers2016arXiv161205560C}, as the fields of both SN\,2018jmt and SN\,2019cj were not sampled by the Sloan Digital Sky Survey (SDSS) \citep{Abdurro'uf2022ApJS..259...35A}. To correct the instrumental ZPs on non-photometric nights and improve the photometric calibration accuracy, we compared the average magnitudes of local sequences of standard stars in the fields of the two SNe to those obtained on photometric nights. With the corrected ZPs, we fine-tuned the SN apparent magnitudes on all nights. 

The instrumental magnitude errors were computed through artificial star experiments, in which fake stars (with a similar magnitude as the SN) were placed near the SN location. The simulated frame is then processed with the PSF fit and the magnitudes measured. The dispersion of individual artificial star experiments was combined (in quadrature) with the PSF fit and the ZP correction, providing the final errors for the photometric data. 

NIR raw images required some preliminary processing procedures, such as flat-fielding, distortion corrections and the subtraction of the background contamination. To construct sky images for each filter, we median-combined several dithered science frames.
We then combined sky-subtracted frames to increase the S/N. Note that these steps were performed through the \texttt{PESSTO} pipeline for the NTT-SOFI raw data. 
Seeing measurements, astrometry, PSF-fitting, and ZP corrections were carried out using {\sl ecsnoopy} and are similar to those discussed for the optical frames.
Finally, reference stars from the Two Micron All Sky Survey \citep[2MASS\footnote{\url{http://irsa.ipac.caltech.edu/Missions/2mass.html/}};][]{Skrutskie2006AJ....131.1163S} catalogue were used to calibrate the NIR instrumental magnitudes.

We also collected photometric data from the public ATLAS and ASAS-SN sky surveys for transients. The orange and cyan($c$) band light curves were directly produced by the ATLAS data-release server \footnote{\url{https://fallingstar-data.com/forcedphot/}} \citep{Shingles2021TNSAN...7....1S}, while some $g$-band data were obtained from the ASAS-SN Sky Patrol\footnote{\url{https://asas-sn.osu.edu}} \citep{Hart2023arXiv230403791H}. 
The Transiting Exoplanet Survey Satellite ($TESS$) \citep{Ricker2015JATIS...1a4003R} space telescope, which is operated by the National Aeronautics and Space Administration (NASA), is equipped with four wide field-of-view optical cameras.
\citet{Vallely2021MNRAS.500.5639V} presented the early-time light curves of a sample of SNe, including SN,2018jmt, and provided a highly accurate determination of the time of explosion with negligible uncertainty.
The final photometric data of SN\,2018jmt and SN\,2019cj are listed in Tables \ref{table:SN2018jmt_lightcurves_data} - \ref{table:SN2019cj_lightcurves_data} in Appendix \ref{appendix:lightcurves_data}, while their apparent light curves are shown in Fig.~\ref{fig:light_curve}. 

\subsection{Spectroscopic data}
\label{subsection:Spectroscopic_data}

Spectroscopic observations of the two SNe Ibn were carried out using the following telescopes:
The 3.58m NTT equipped with EFOSC2; the 4.1m Southern Astrophysical Research Telescope (SOAR) at Cerro Pach\'on, Chile, equipped with the Goodman High Throughput Spectrograph (GHTS); the 11m Southern African Large Telescope (SALT) at the SAAO equipped with the Robert Stobie Spectrograph (RSS); the 2m Faulkes telescope with the FLOYDS spectrograph, hosted by the Siding Spring Observatory, which is also part of the LCO global network. 

All raw spectral data were reduced following the standard steps in \textsc{iraf}\footnote{\url{http://iraf.noao.edu/}} \citep{Tody1986SPIE..627..733T, Tody1993ASPC...52..173T} or with dedicated pipelines such as \texttt{PySALT}\footnote{\url{http://pysalt.salt.ac.za/}} \citep{Crawford2010SPIE.7737E..25C}, \texttt{PESSTO} \citep{Smartt2015A&A...579A..40S} and \texttt{FLOYDS}\footnote{\url{https://lco.global/documentation/data/floyds-pipeline/}}. The pre-reduction steps, such as bias, overscan, flat-fielding correction, and trimming, are similar to those described for the imaging data. Then, the one-dimensional (1D) spectra were optimally extracted from the 2D images. Wavelength calibrations were performed using arc lamps, while flux calibrations were performed using spectrophotometric standard stars taken on the same nights. Subsequently, the strongest telluric absorption bands, such as O$_2$ and H$_2$O, were removed from the SN spectra using the spectra of the standard stars. Finally, the accuracy of flux calibration for all spectra was checked against the coeval photometric data. The information on the instrumentation used for the spectroscopic observations is reported in Tables \ref{2018jmtSpecInfo} and \ref{2019cjSpecInfo} (Appendix \ref{SpecInfo}).

\section{Photometry}
\label{section:Photometry}

\subsection{Apparent light curves}

\begin{figure*}
    \includegraphics[width=0.49\textwidth]{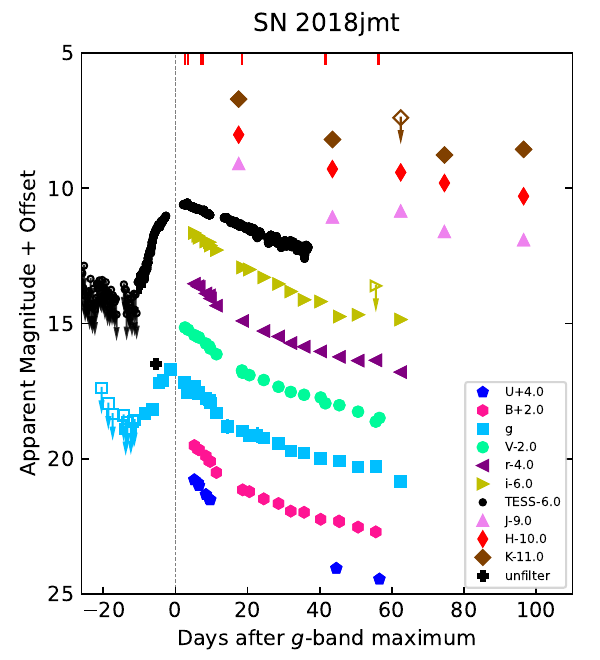}
    \includegraphics[width=0.49\textwidth]{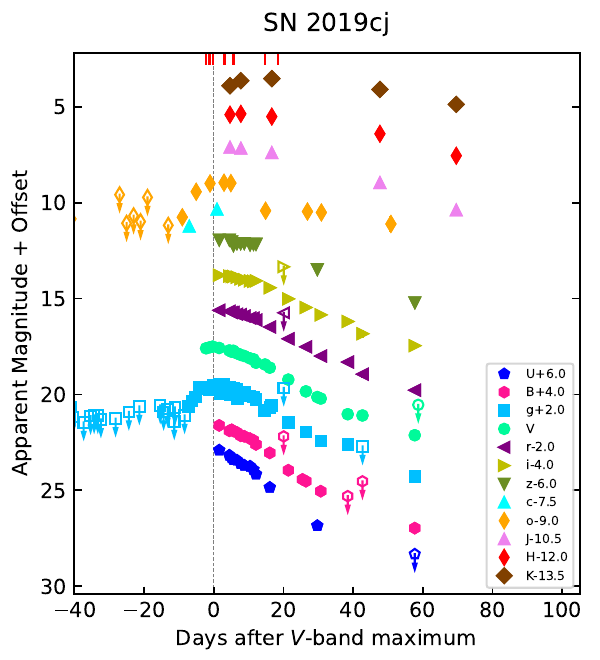}
    \caption{Multi-band light curves of SN\,2018jmt (left) and SN\,2019cj (right). A dashed vertical line is used to visually represent the reference epoch, which corresponds to the $g$/$V$-band maximum light. 
    The epochs of our spectra are marked with vertical solid red lines on the top. The upper limits are indicated by empty symbols with down-arrows. For clarity, the light curves are shifted by constant amounts reported in the legends. In most cases, the errors associated with the magnitudes are smaller than the plotted symbol sizes. }
    \label{fig:light_curve}
\end{figure*}

\begin{figure}
    \includegraphics[width=\columnwidth]{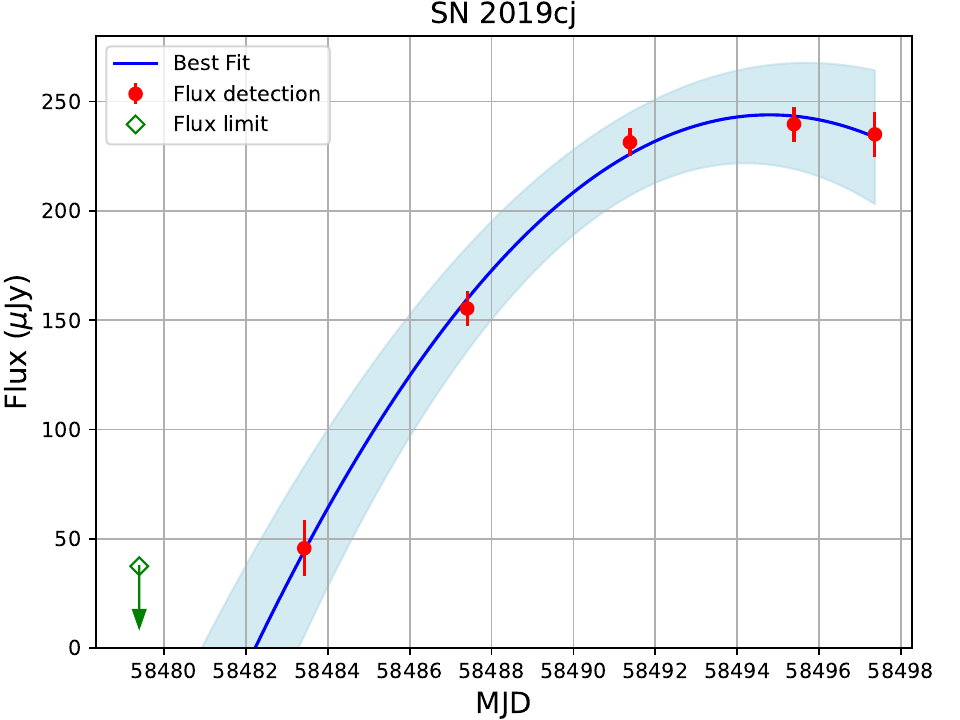}
    \caption{Estimate of the explosion epoch for SN\,2019cj. The ATLAS-$o$ light curves of SN\,2019cj (detections use red dot markers, limits use green diamond markers) are shown in flux space (expressed in $\mu$Jy).
    The early light curves are fitted with a 2nd-order polynomial, represented by a blue solid line. The blue shaded region around the fitted curve represents the 3-$\sigma$ uncertainty in the fitting process.}
    \label{fig:explosion_time}
\end{figure}

\begin{table}
    \centering
    \caption{Decline rates of the light curves of SN\,2018jmt and SN\,2019cj.}
    \begin{tabular}{cccc}
    \hline \hline
        \multicolumn{4}{c}{SN\,2018jmt} \\
        \hline
        Filter & $\gamma_{0-15}^\ddag$ & $\gamma_{15-60}^\ddag$ & $\gamma_{60-100}^\ddag$\\ \hline
        U & 17.37$\pm$0.57 & 3.32$\pm$1.12 & -\\
        B & 15.43$\pm$1.14 & 4.17$\pm$0.21 & -\\
        g & 14.02$\pm$1.54 & 4.09$\pm$0.15 & -\\
        V & 11.10$\pm$0.38 & 4.61$\pm$0.14 & -\\
        r & 12.93$\pm$0.83 & 4.21$\pm$0.24 & -\\
        i &  9.72$\pm$0.75 & 6.00$\pm$0.39 & - \\
        J & - & 7.59$\pm$0.66 & 2.74$\pm$1.35\\
        H & - & 4.88$\pm$1.30 & 2.22$\pm$2.30\\
        K & - & 5.74$\pm$0.99 & -0.95$\pm$2.36\\
        \hline
        \multicolumn{4}{c}{SN\,2019cj} \\
        \hline
        Filter & $\gamma_{0-15}^\ddag$ & $\gamma_{15-48}^\ddag$ & $\gamma_{48-80}^\ddag$\\ \hline
        U & 10.84$\pm$0.67 & 14.62$\pm$0.88 & -\\
        B & 8.68$\pm$0.58 & 13.53$\pm$0.78 & -\\
        g & 7.33$\pm$0.43 & 11.78$\pm$0.78 & -\\
        V & 7.08$\pm$0.50 & 10.97$\pm$0.22 & -\\
        r & 4.99$\pm$0.34 & 9.19$\pm$0.41 & -\\
        i & 3.46$\pm$0.30 & 8.93$\pm$0.30 & -\\
        z & 2.46$\pm$0.46 & - & -\\
        o & 5.11$\pm$9.27 & - & -\\
        J & 1.61$\pm$2.80 & 5.06$\pm$0.36 & 6.51$\pm$0.83\\
        H & -1.45$\pm$2.99 & 2.89$\pm$0.34 & 5.23$\pm$0.72\\
        K & -8.54$\pm$4.46 & 1.82$\pm$0.46 & 3.56$\pm$0.72\\ \hline
    \end{tabular}
    \begin{flushleft}
    $^\ddag$ in mag (100 d)$^{-1}$
    \end{flushleft}
    \label{tab:slope}
\end{table}

We conducted a continuous monitoring of the photometric evolution of SN\,2018jmt and SN\,2019cj for about three months after discovery.
The optical and near-infrared light curves of SNe\,2018jmt and 2019cj are shown in Fig.~\ref{fig:light_curve}. 

The determination of the explosion epoch of a SN is located between the last non-detection and the first detection of the event. 
\citet{Vallely2021MNRAS.500.5639V} reports a relatively accurate explosion time of 58455.01 $\pm$ 0.24 in the $TESS$ $T$-band for SN\,2018jmt, based on a curved power-law fit to the pre-peak TESS light curve. For SN\,2019cj, the last non-detection $t_l$ dates back to MJD~=~58479.4 (in the ATLAS $o$ band), whilst the first detection epoch $t_d$ is MJD~=~58483.4 (in the $o$ band). The midpoint between $t_l$ and $t_d$ provides a rough estimate of the explosion epoch. The maximum error is given by half of the difference between $t_l$ and $t_d$. 
Using this method, we obtain the explosion epoch of MJD~=~ 58481.4$\pm$2.0 for SN\,2019cj.
To improve our estimate of the explosion epoch of SN\,2019cj, we adopted the fireball expansion method. As shown in  Fig.~\ref{fig:explosion_time}, we applied a second-order polynomial fit to the data captured within a 20-day period before and around the peak of the light curve in flux space \citep[e.g.,][]{Gonzalez-Gaitan2015MNRAS.451.2212G}. Following this approach, we estimated the explosion epoch, as the time when the flux reaches 0, to be MJD~=~58482.2 $\pm$ 1.1 for SN\,2019cj, which will be adopted hereafter. 

To estimate the peak magnitude of SN\,2018jmt, a 3rd-order polynomial fit is performed on the $g$-band light curve data within a 4-week period centred around the peak in magnitude space. We obtained a peak magnitude of $g$ ~=~ 17.0 $\pm$ 0.3 on MJD ~=~ 58465.7 $\pm$ 1.2 for SN\,2018jmt. 
Using a similar approach, we estimated the peak magnitude of SN\,2019cj as 17.5 $\pm$ 0.1 on MJD ~=~ 58492.4 $\pm$ 0.2 in the $V$-band.

We also estimated the post-maximum decline rate of SN\,2018jmt and SN\,2019cj in various bands by performing a linear regression on the post-peak data. 
The results are reported in Table~\ref{tab:slope}.
Given the observed change in the slope of the light curves of SN\,2018jmt at approximately +15 d in the optical and +60 d in the NIR, we computed the decline rates in three different time intervals. 
We observe a notable difference in the decline rates among different filters. Specifically, the bluer light curves exhibit a faster decline compared to the redder ones. This trend is particularly evident during the early decline phase ($\gamma_{0-15}$ in Table~\ref{tab:slope}).

From 0 to 15 days, the light curves of SN\,2019cj show a faster decline in the blue bands, while in the NIR, the object is still rising towards its peak.
Later on, at phases beyond 15 days, the light curves show a steeper decline (e.g., 
$\gamma_{15-48}$(B) $\approx 0.14$ mag~d$^{-1}$, 
$\gamma_{15-48}$(r) $\approx 0.09$ mag~d$^{-1}$, 
$\gamma_{15-48}$(K) $\approx 0.02$ mag~d$^{-1}$). 
An increased rate of decline in the optical luminosity at late phases is frequently observed in SNe Ibn \citep[e.g.,][see also other examples in Section \ref{absLC}]{Mattila2008MNRAS.389..141M, Pastorello2015MNRAS.453.3649P}. 

\subsection{Colour evolution}

\begin{figure}
    \centering
    \includegraphics[width=1\linewidth]{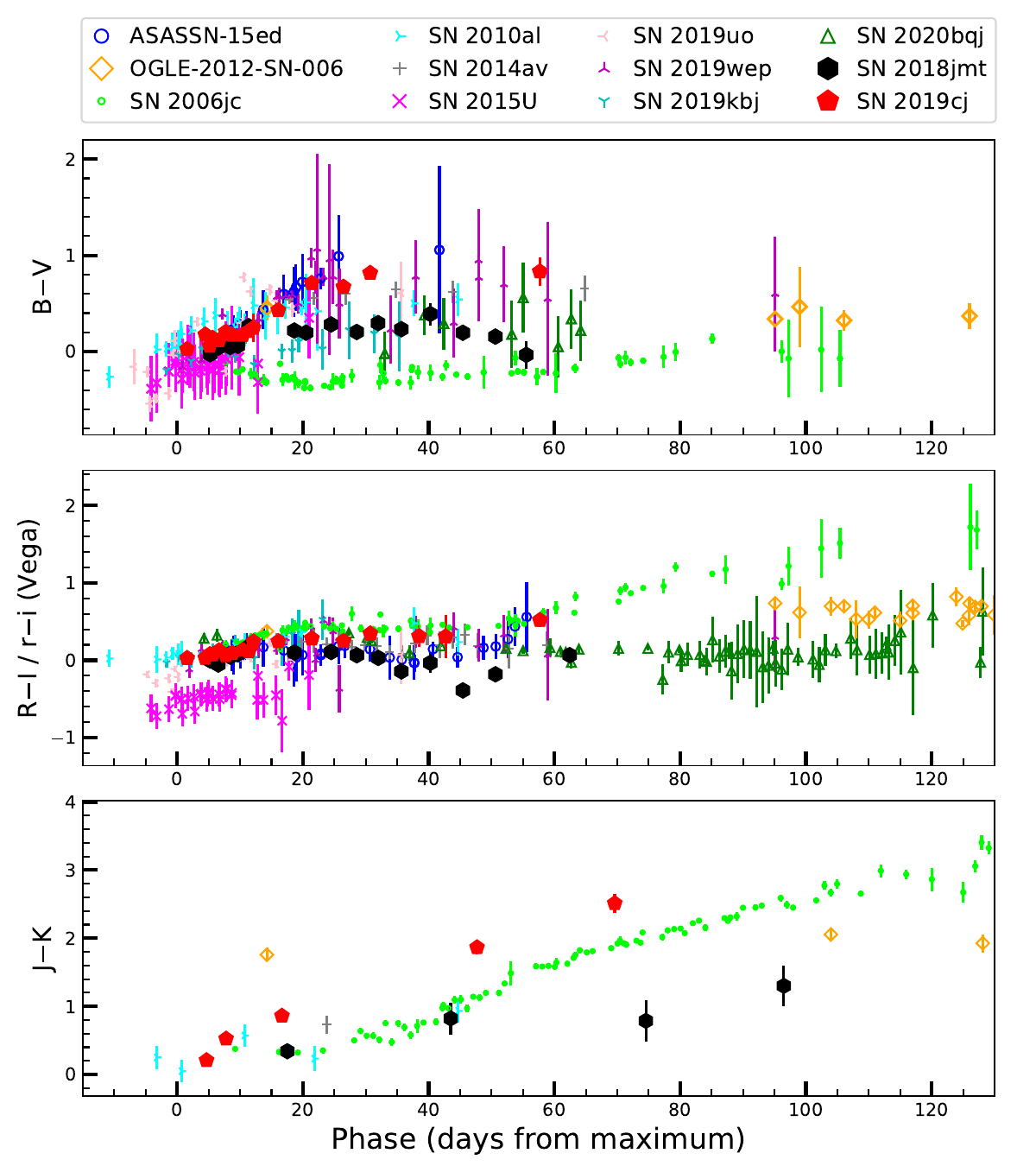}
    \caption{Colour evolution of SN\,2018jmt and SN\,2019cj, compared with those of a sample of SNe Ibn.
    Top panel: $B~-~V$ colour curves; Middle panel: $R~-~I$~/~$r~-~i$ colour curves; Bottom panel: $J-K$ colour curves. 
    The colour curves are corrected for Galactic and host galaxy extinction.}
    \label{fig:colour_evolution}
\end{figure}

Figure~\ref{fig:colour_evolution} displays the colour evolution of SN~2018jmt and SN~2019cj, along with those of other well-studied SNe Ibn, including ASASSN-15ed \citep{Pastorello2015MNRAS.453.3649P}, OGLE-2012-SN-006 \citep{Pastorello2015MNRAS.449.1941P}, SN\,2006jc \citep{Pastorello2007Natur.447..829P}, SN\,2010al \citep{Pastorello2015MNRAS.449.1921P}, SN\,2014av \citep{Pastorello2016MNRAS.456..853P}, SN\,2015U \citep{Pastorello2015MNRAS.454.4293P,Shivvers2016MNRAS.461.3057S}, SN\,2019uo \citep{Gangopadhyay2020ApJ...889..170G}, 
SN\,2019wep \citep{Gangopadhyay2022ApJ...930..127G}, SN\,2019kbj \citep{Ben-Ami2023ApJ...946...30B}, 
and SN\,2020bqj \citep{Kool2021AA...652A.136K}.

At an early stage, $\sim$5 days from maximum, SN\,2018jmt exhibits intrinsic $B~-~V$ and $r~-~i$ colours that are both close to 0 mag. At around +10 days, the object undergoes a transition towards red colours, with $B~-~V~\sim~0.3$ mag and $r~-~i~\sim~0.1$ mag. This is followed by a period, from +10 to +40 days, during which the $B~-~V$ colour is slowly increasing to {0.4} mag, while the $r~-~i$ colour is nearly constant. 
Later on, the colours of SN\,2018jmt become bluer again.
Similarly, the $r~-~i$ colour reaches its minimum value of approximately $-$0.4 mag at around +45 days. After that, the $r~-~i$ colour starts to rise again to +0.1 mag at +62 days. 
The $B~-~V$ colour behaviour of SN\,2019wep, SN\,2015U, and SN\,2019uo is similar to that of SN\,2018jmt, becoming initially redder and then turning bluer again.
The $R~-~I$~/~$r~-~i$ colour behaviour of SN\,2010al, SN\,2019kbj, and SN\,2014av, like SN\,2018jmt, shows a trend towards redder in the early stages, followed by a transition to bluer.
The $J-K$ colour evolution follows a similar trend as the $B~-~V$ and $r~-~i$ ones, becoming gradually redder from 0.3 mag to 1.3 mag up to 100 days, like SN\,2010al.

For SN\,2019cj, the $B~-~V$ colour evolves steadily from approximately 0 mag near the maximum to $\approx$0.8 mag at around +30 days (as shown in Fig.~\ref{fig:colour_evolution}).
Beyond 30 days, the $B~-~V$ colour remains fairly constant. The evolution trend of $B~-~V$  colour is similar to ASASSN-15ed, SN~2010al and SN~2014av.
Similarly, from the maximum to about 1 month after peak, the $r~-~i$ colour increases from approximately 0 mag to 0.3 mag. Between 30 and 45 days, also the $r~-~i$ colour remains constant, although - at around 60 days - $r~-~i$ rises again towards redder colours ({0.5} mag). 
The evolution trend of $R~-~I$~/~$r~-~i$  colour is similar to that of SNe~2015U, 2019uo, and OGLE-2012-SN-006, although the timescales can be significantly different among individual SNe.
In contrast to the optical colours, the $J-K$ colour rises monotonically from 0.2 to 2.5 mag over the entire 2 months of follow-up, like SN 2006jc.

\subsection{Absolute light curves}
\label{absLC}

\begin{figure}
    \centering
    \includegraphics[width=1\linewidth]{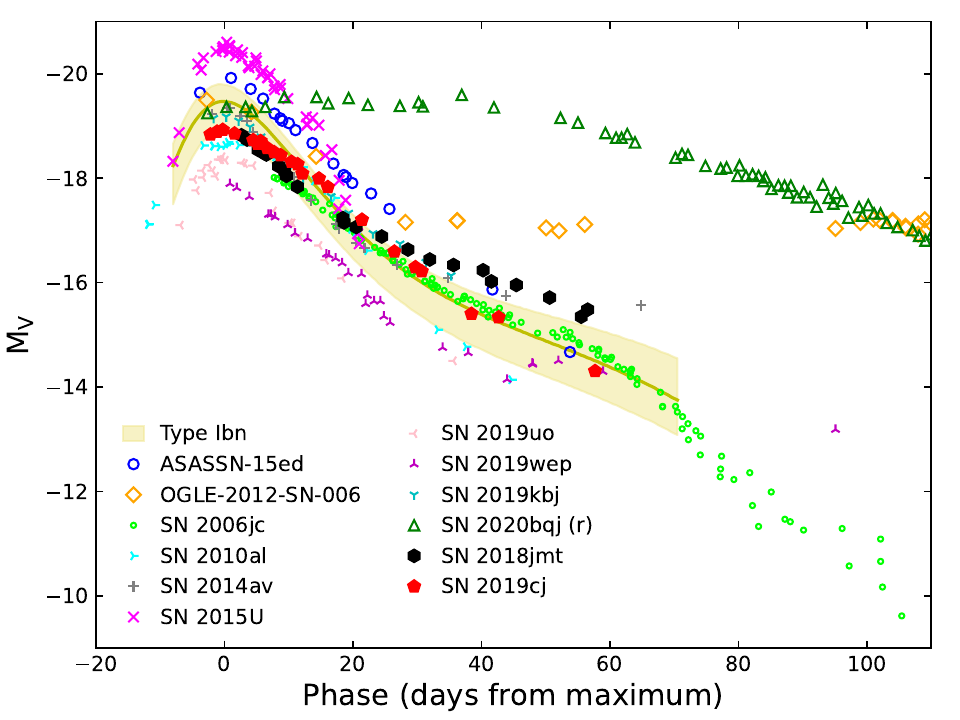}
    \caption{Absolute $V$-band light curve of SN\,2018jmt and SN\,2019cj compared to other SNe Ibn and the SN Ibn template presented by \citet{Hosseinzadeh2017ApJ...836..158H}.}
    \label{fig:Absolute_Magnitude}
\end{figure}

Taking into account the distance and reddening estimates reported in Section \ref{section:Basic_sample_information}, we calculate for SN\,2018jmt the absolute magnitude at maximum to be $M_g$ ~=~ $-19.03 \pm 0.37$ mag, while the $V$-band peak absolute magnitude of SN~2019cj is $M_V$ ~=~ $-18.94 \pm 0.19$ mag.  
A comparison of the absolute $V$-band magnitudes for a subset of the Type Ibn SN sample is presented in Fig.~\ref{fig:Absolute_Magnitude}. 
When $V$-band observations were not available, we included observations in adjacent bands for the comparison. For example, in the case of SN 2020bqj, we utilised observations in the $r$-band.
Upon comparing the $V$-band light curves of SN\,2018jmt and SN\,2019cj with those of other Type Ibn SNe, we find that they generally follow the behaviour of the template presented by \cite{Hosseinzadeh2017ApJ...836..158H} around the maximum light.
SNe Ibn are relatively luminous, with most of them having absolute $V$-band magnitudes around $-19$ mag, but all falling within the range of $-18$ mag to $-21$ mag.

The heterogeneity of SNe Ibn is more evident in the post-peak decline, as many objects display an almost linear post-peak optical drop, while a few others may show a light curve with double-phase declines: an initially faster luminosity drop followed by a much slower decline. The most extreme case is OGLE-2012-SN-006, which experienced an early decline slope of $\approx$ 8 mag (100 d)$^{-1}$, followed by an extended phase characterised by a nearly flat light curve ($\sim$0.1 mag (100 d)$^{-1}$ between 25 and 130 days, and 2 mag (100 d)$^{-1}$ thereafter). 
Another example is SN 2020bqj, which exhibits a plateau between $-19.1$ and $-19.3$ mag in the $r$-band that persists for 40 days, followed by a linear decline lasting over 90 days at a rate of 4 mag (100 d)$^{-1}$.

\subsection{Pseudo-bolometric light curves} \label{subsection:pseudoBLC}
\begin{figure}
    \centering
    \includegraphics[width=1\linewidth]{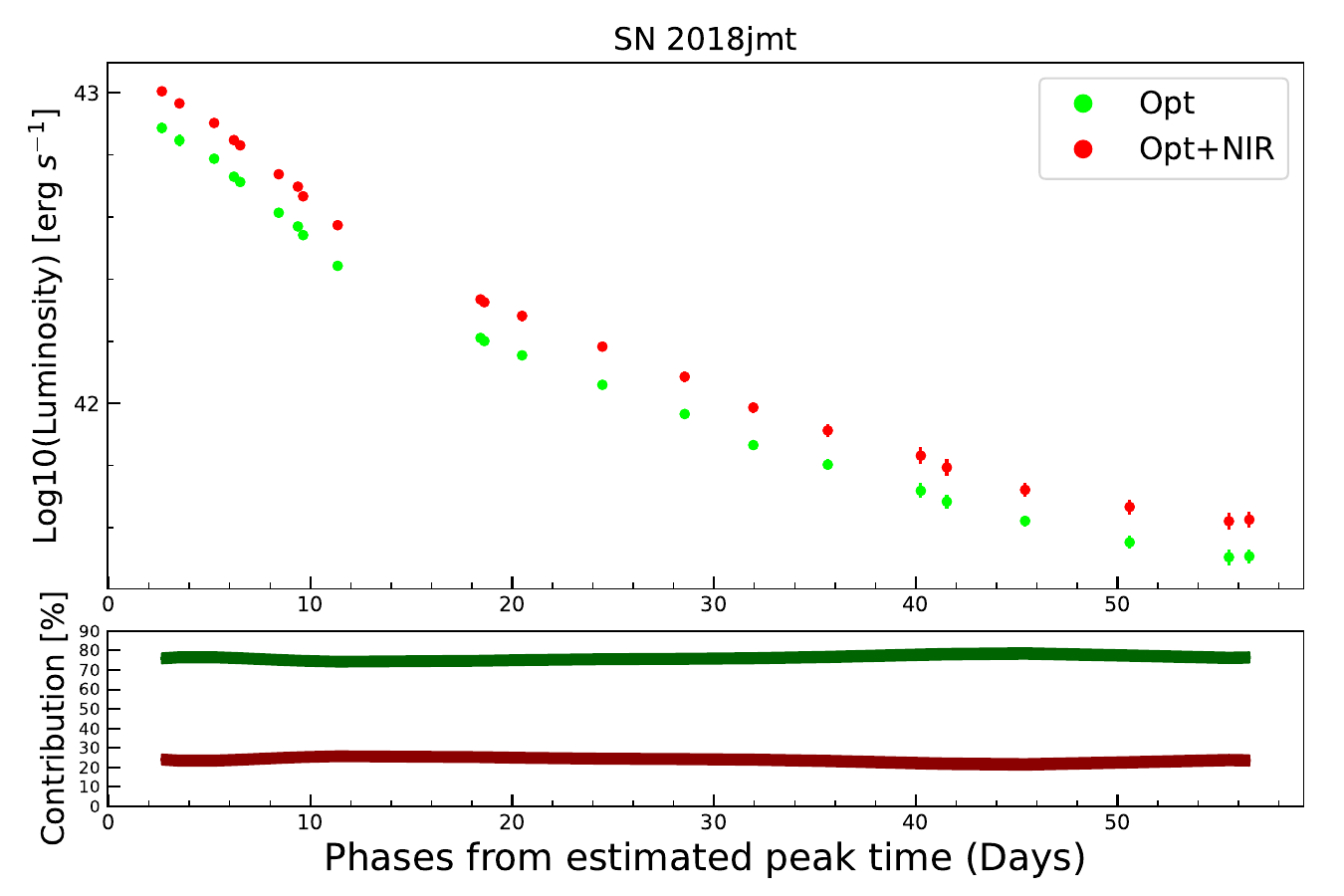}
    \includegraphics[width=1\linewidth]{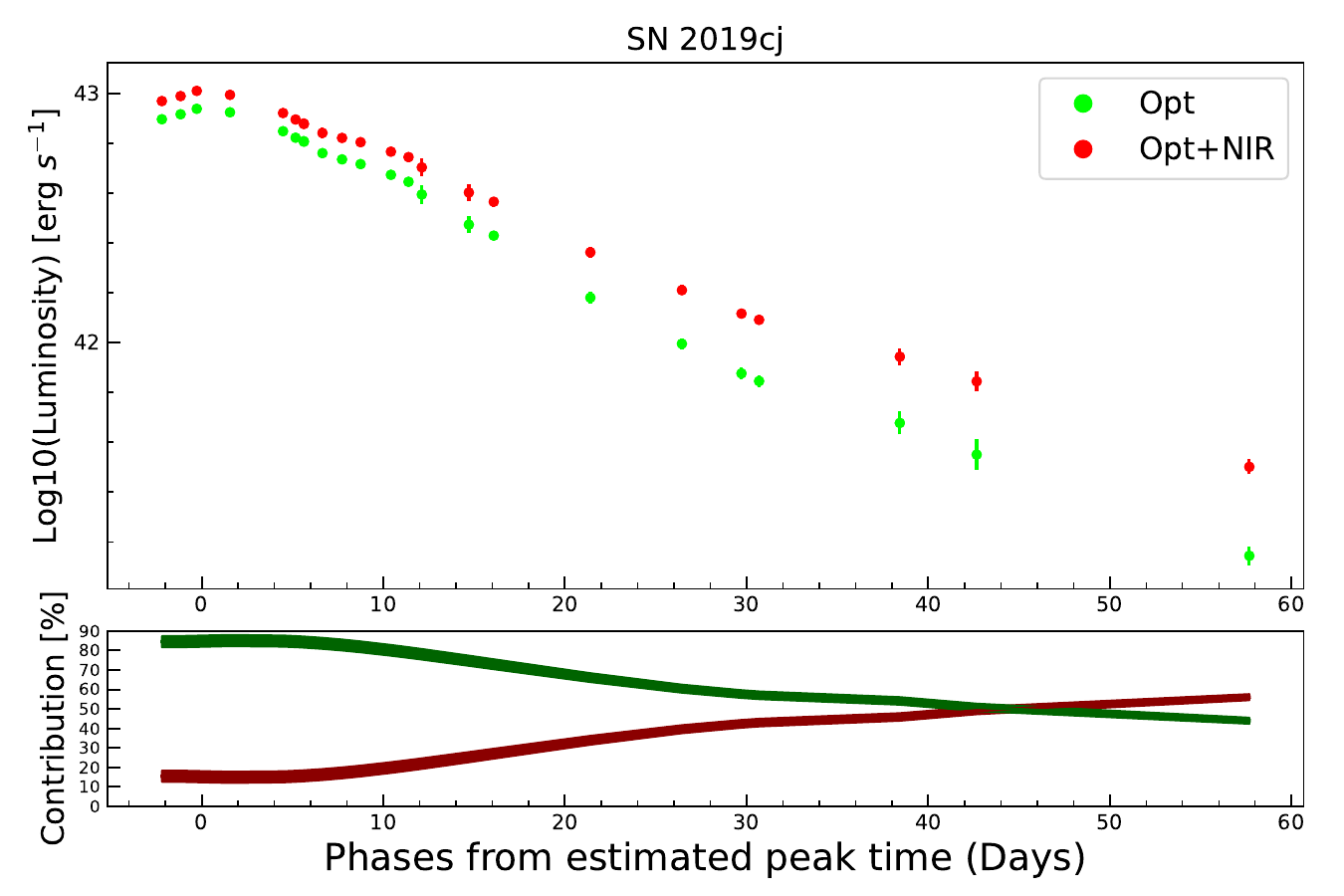}
    \includegraphics[width=1\linewidth]{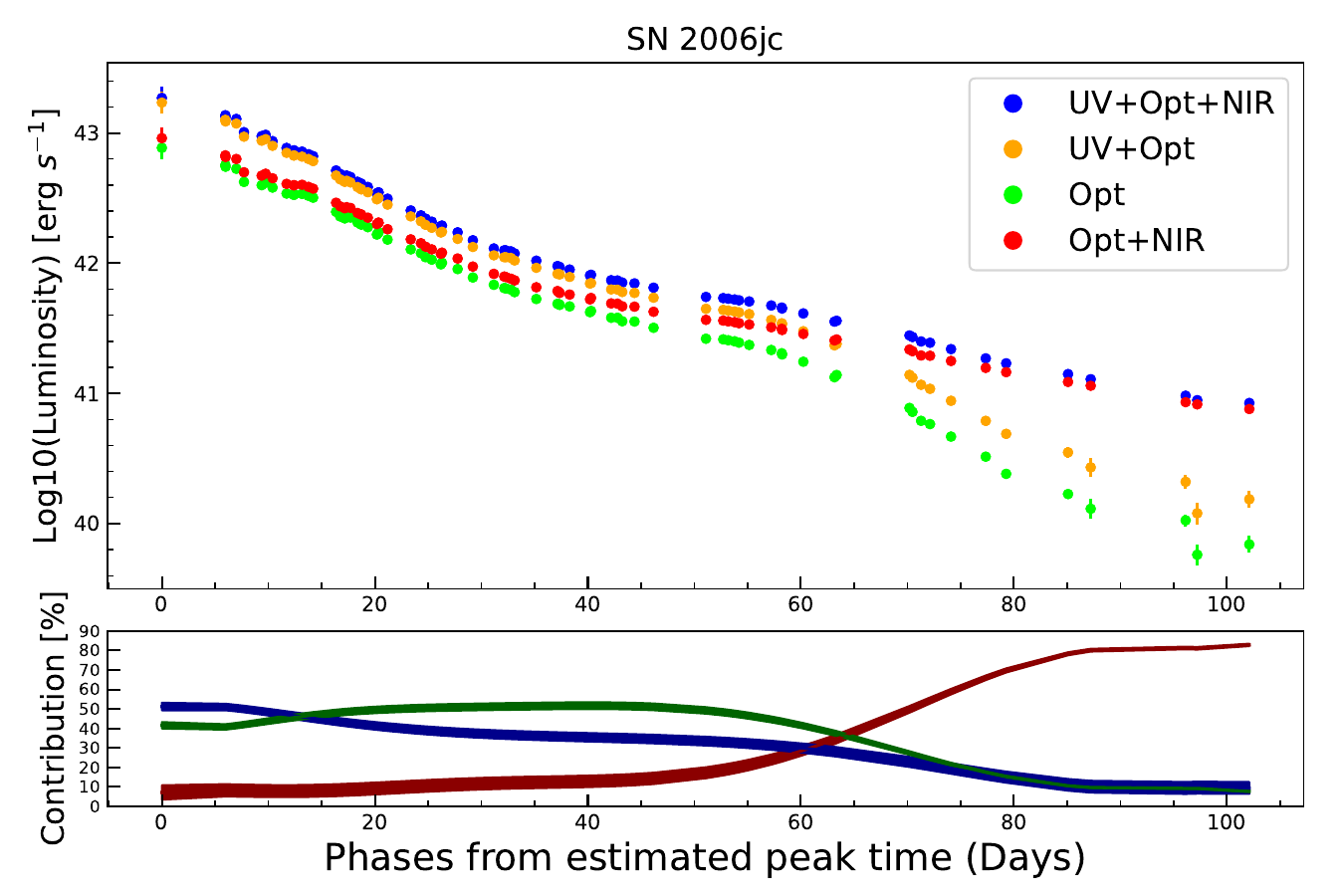}
    \caption{Top panels: Pseudo-bolometric light curves of SNe\,2018jmt, 2019cj and 2006jc, along with the light curves constructed using the UV + optical, just optical, and optical + NIR photometry. Bottom panels: Evolution of the contribution of the individual electromagnetic regions with time computed as a percentage of the pseudo-bolometric luminosity. In the comparison, we adopted the following colour codes: {UV = dark blue, Optical = dark green, and NIR = dark red. }
    }
    \label{fig:pseudo_bolometric}
\end{figure}

We calculated the "optical" pseudo-bolometric light curves of SNe\,2018jmt and 2019cj by integrating the flux contributions from individual optical bands. 
In our analysis, we made the assumption that the flux outside the integration limits is zero. 
When photometric data were not available for certain epochs in a particular filter, we estimated the flux contribution from the missing bands by interpolating between epochs with available data or extrapolating from earlier or later available epochs, assuming a consistent colour evolution.
We also computed pseudo-bolometric light curves, including the contribution of ultraviolet (UV) and NIR photometry when available.
The pseudo-bolometric light curves for SN\,2018jmt, SN\,2019cj, and the prototypical SN 2006jc are displayed in the top panels of Fig.~\ref{fig:pseudo_bolometric}. The relative flux contribution of each electromagnetic domain to the overall pseudo-bolometric light curve is shown in the lower panels of Fig.~\ref{fig:pseudo_bolometric}.

Public $TESS$ data of SN\,2018jmt would potentially help tracking the evolution of its bolometric luminosity in the pre-maximum phase. However, the information on the colour or spectroscopic evolution at those early phases is not available. Besides, there are known issues regarding the photometric calibration across multiple sectors for $TESS$ \citep{Vallely2021MNRAS.500.5639V}. 
In \citet{Vallely2021MNRAS.500.5639V}, a method was employed where the flux offset was selected to match the linear extrapolations from the last $\sim$ 2 days of the earlier sector and the first $\sim$ 2 days of the later sector for flux calibration across different sectors.
In the case of SN 2018jmt, this approach may introduce a significant bias when comparing flux, and thus a photometric correction, between pre- and post-maximum phases because the peak lies in the sector gap. As a result, linear extrapolation can be a poor approximation for the light curve in the gap.
Because of these two factors, there would be major uncertainties on the bolometric correction to apply. 
For this reason, we only use traditional broad-band observations to compute a pseudo-bolometric light curve for SN\,2018jmt.

The lack of UV observations for SNe\,2018jmt and 2019cj limits our ability to accurately determine the real bolometric luminosity at peak, when the UV contribution is expected to be significant. Therefore, we can only provide a lower limit to the maximum luminosity for the two objects, i.e., $L > 1.01 \times 10^{43}$ erg~$\mathrm{s}^{-1}$.
Throughout the evolution of SN\,2018jmt , only minor changes are registered in the relative contribution of the optical and NIR luminosity. The flux contribution of the optical bands dominates the bolometric luminosity, as it accounts for approximately 76\% of the overall emission. We note, however, that during phases later than +60 days, the object fades below the detection thresholds in the optical, while it remains visible in the NIR domain up to 100 days. The NIR light curves also show an evident flattening, suggesting a dramatic increase in the luminosity contribution of the NIR bands over the optical ones. 

It is worth noting that the contribution from near-infrared (NIR) emission to the pseudo-bolometric light curve of SN\,2019cj is relatively small around the time of maximum brightness, as it accounts for $\sim$ 15\% of the total luminosity. However, as time progresses, the NIR contribution becomes progressively larger.
Interestingly, a similar behaviour was also observed in SN 2006jc, where the NIR contribution increased with time and became more dominant at later phases (bottom panel of Fig.~\ref{fig:pseudo_bolometric}), and was attributed to the contribution to warm dust in a cool dense shell \citep{Mattila2008MNRAS.389..141M, Smith2008ApJ...680..568S}.
Unfortunately, we were not able to obtain photometry data for SNe 2018jmt and 2019cj beyond 60 days after their discovery.

\begin{figure*}
    \centering
    \includegraphics[width=0.48\linewidth]{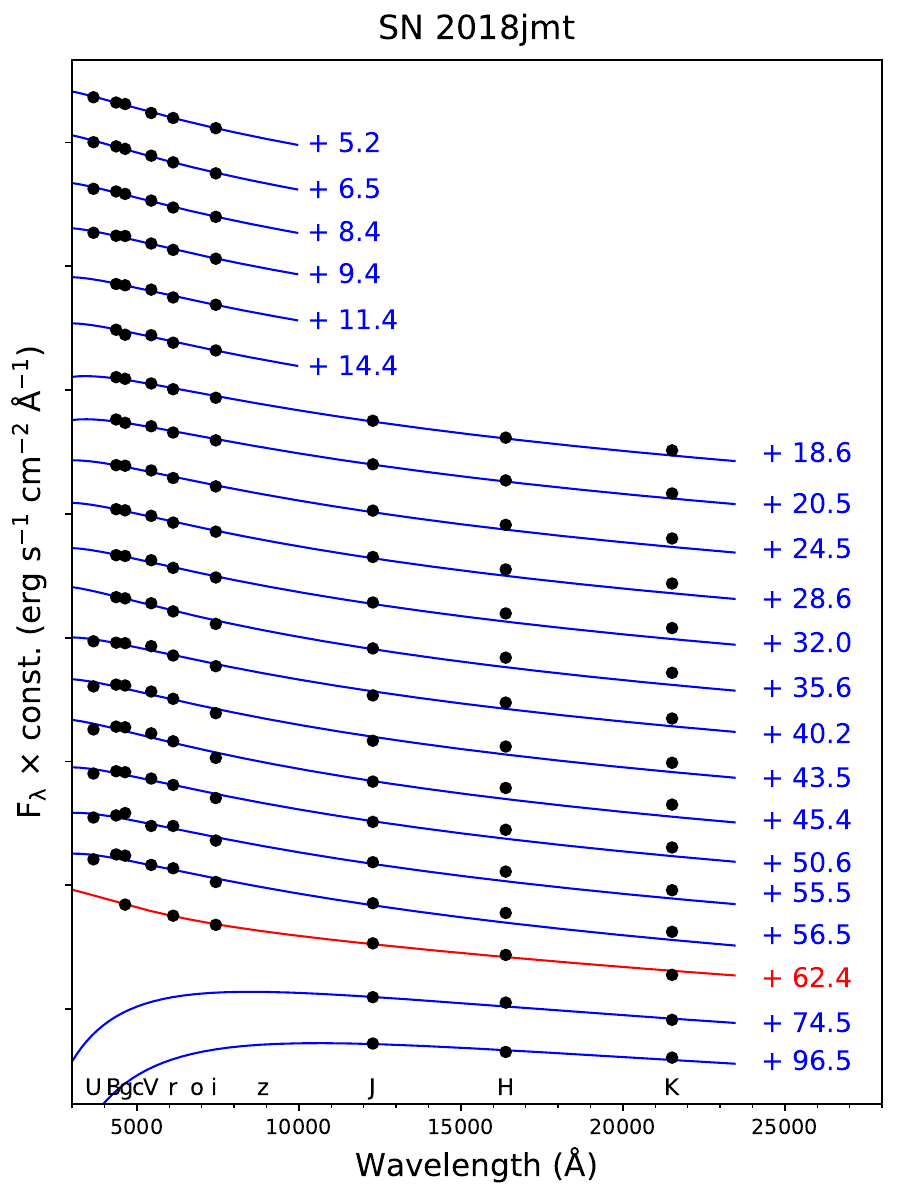}
    \includegraphics[width=0.48\linewidth]{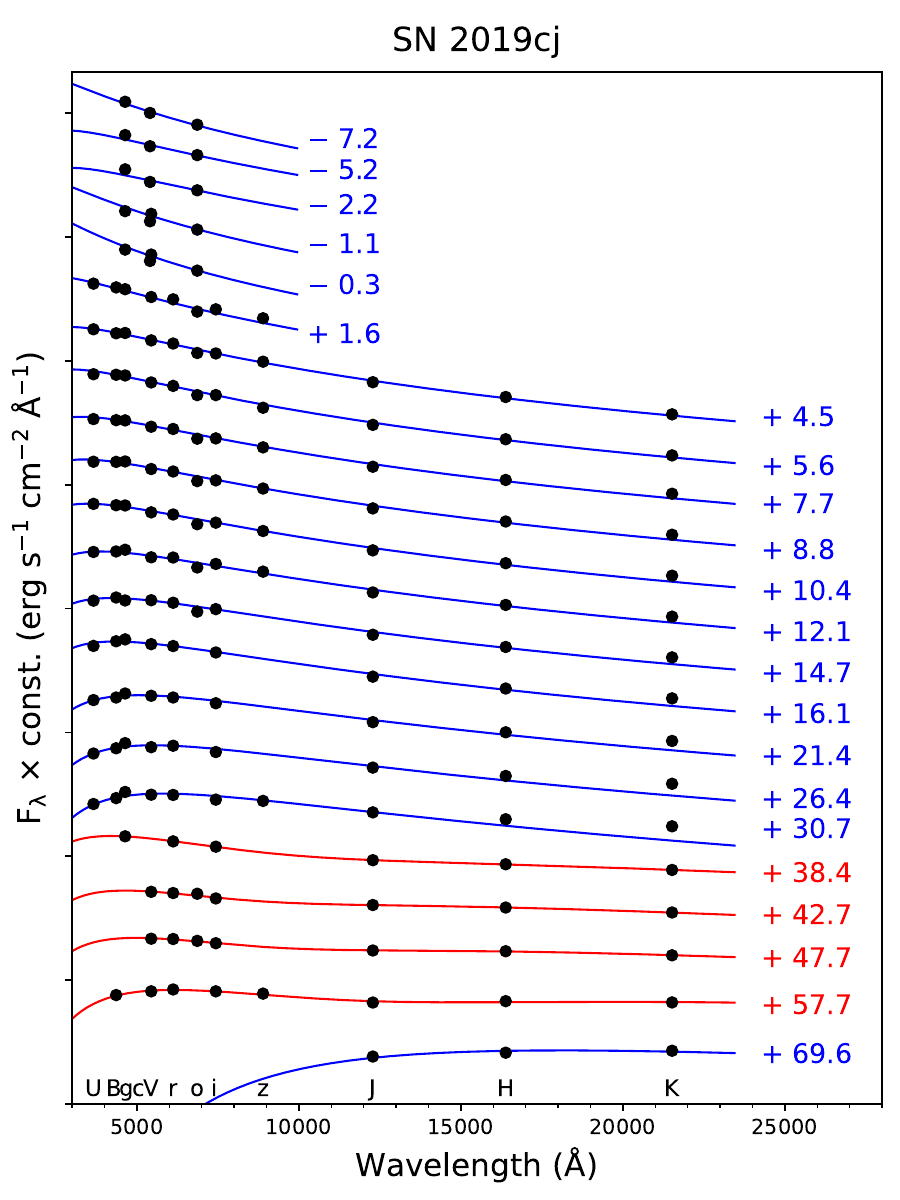}
    \caption{{SED evolution of SNe~2018jmt (left panel) and 2019cj (right panel). The lines represent the best fitted blackbody functions, which are overplotted on each SED. Blue lines are the best fits of the single blackbody, while red lines are the best fits of the two-component blackbodies. Epochs reported to the right of each SED are relative to their maximum light. SEDs have been shifted vertically by an arbitrary constant for clarity.}
    }
    \label{fig:SED}
\end{figure*}

\begin{figure*}
    \centering
    \includegraphics[width=0.8\linewidth]{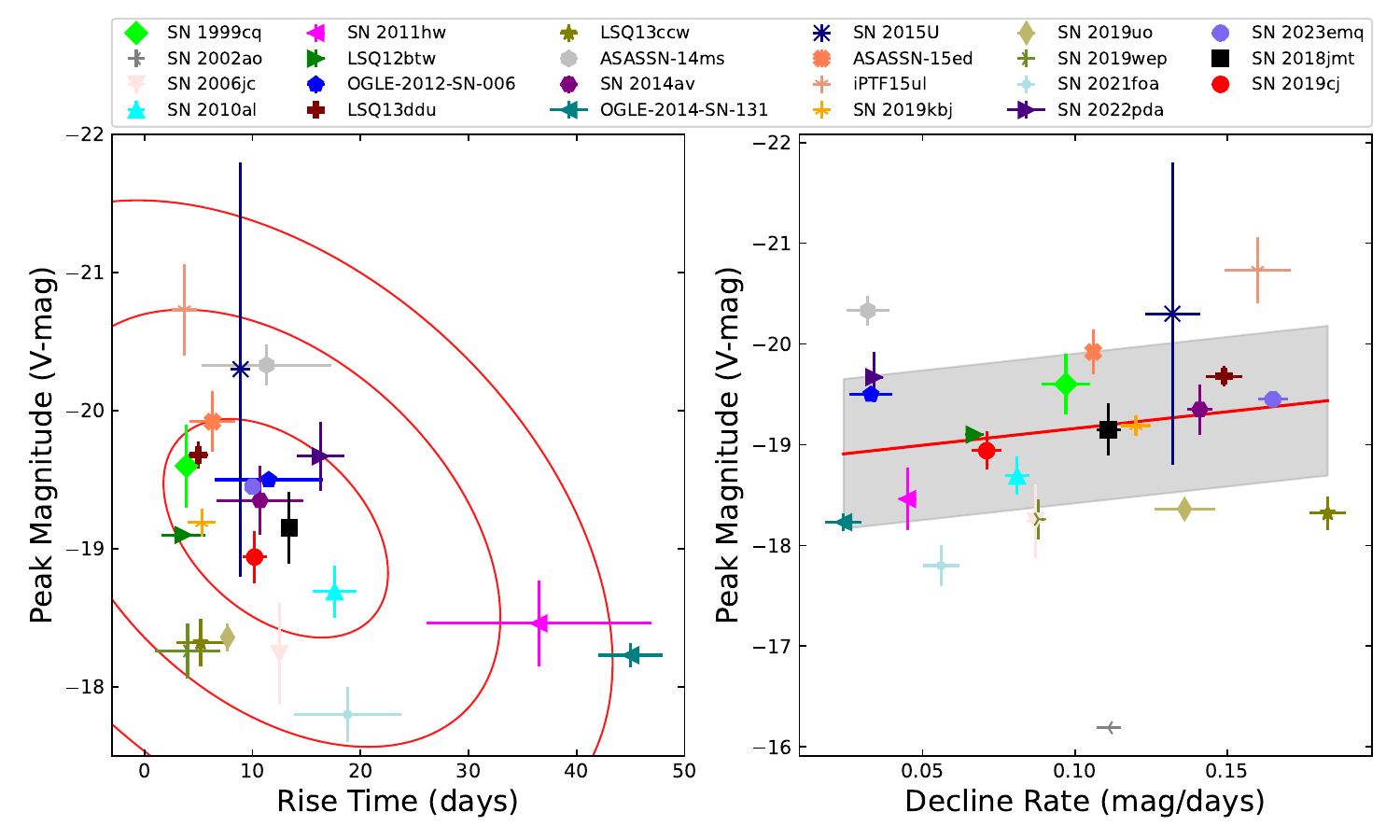}
    \caption{Phase-space diagrams showing peak magnitudes vs. rise time (left), and peak magnitudes vs. decline rates (right) for a sample of SNe Ibn, including SNe 2018jmt and 2019cj.
    In the left panel, three red ellipses represent the $1\sigma$, $2\sigma$, and $3\sigma$ confidence intervals, indicating regions where approximately 68.27\%, 95.45\%, and 99.73\% of the points are expected to lie, respectively. These ellipses are centered at the mean values of the points and are oriented according to the principal components of the covariance matrix. In the right panel, linear fitting was applied to the observed data, and the 95\% confidence interval was calculated using a standard deviation multiplier of 1.96 to determine the shaded region.
    Data for comparison objects are taken from \citet{Matheson2000AJ....119.2303M, Pastorello2007Natur.447..829P, Pastorello2008MNRAS.389..131P, Mattila2008MNRAS.389..141M, Pastorello2008MNRAS.389..113P, Sanders2013ApJ...769...39S, Morokuma2014CBET.3894....1M, Gorbikov2014MNRAS.443..671G, Pastorello2015MNRAS.454.4293P, Pastorello2015MNRAS.449.1954P, Pastorello2015MNRAS.449.1941P, Pastorello2015MNRAS.449.1921P, Pastorello2015MNRAS.453.3649P, Pastorello2016MNRAS.456..853P, Karamehmetoglu2017AA...602A..93K, Hosseinzadeh2017ApJ...836..158H, Vallely2018MNRAS.475.2344V, Wang2020ApJ...900...83W, Clark2020MNRAS.492.2208C, Gangopadhyay2020ApJ...889..170G, Prentice2020MNRAS.499.1450P, Karamehmetoglu2021AA...649A.163K, Kool2021AA...652A.136K, Gangopadhyay2022ApJ...930..127G, Reguitti2022AA...662L..10R, Pursiainen2023ApJ...959L..10P, Ben-Ami2023ApJ...946...30B, Wang2024MNRAS.530.3906W}; Cai et al., in preparation.}
    \label{fig:RiseTime_DeclineRate_PeakMagnitude}
\end{figure*}

{
\subsection{Spectral energy distribution (SED) analysis}

In order to compare in a meaningful way SN\,2018jmt, SN\,2019cj and the prototypical Type Ibn SN\,2006jc, we constructed their pseudo-bolometric light curves based on the observed wavelength range (see details in Section \ref{subsection:pseudoBLC}). Instead, to better estimate the full bolometric light curves of SNe\,2018jmt and 2019cj, we fitted the broad-band photometry with a blackbody curve, extrapolating the luminosity contribution of the blackbody tails outside the observed range. To do so, we performed blackbody fits on the spectral energy distributions (SEDs) of SN\,2018jmt and SN\,2019cj at different epochs, following the descriptions in \citet{Cai2018MNRAS.480.3424C}. The resulting full bolometric light curves are used to model SNe\,2018jmt and 2019cj, as presented in Section \ref{subsection:LCmodelling}.

In Fig.~\ref{fig:SED}, we show the SED evolution of SNe\,2018jmt and 2019cj with their best blackbody fits. During the early and middle phases of evolution, the SEDs of both SNe\,2018jmt and 2019cj are well-fitted by a single blackbody. At late-time epochs, a single blackbody is unable to well reproduce the NIR fluxes and hence a second blackbody component is needed. The optical domain is represented by a "hot" blackbody associated to the SN photosphere, in contrast with the "warm" blackbody which emerges at late phases. Specifically, as shown in the left panel of Fig.~\ref{fig:SED}, the SEDs of SN\,2018jmt are well-fitted by a single blackbody until epoch $+$56.5 d. At the epoch $+$62.4 d, the SED clearly reveals a NIR flux excess over a single blackbody model, hence it was fitted with two-component (hot+warm components) blackbody functions. Although the NIR flux excess is likely attributed to the newly formed dust, there is also a relevant contribution from the emission lines in the SN spectra, which can bring deviations from a thermal continuum (see the late spectra of SN\,2018jmt in the top panel of Fig.~\ref{fig:spectral}). 

The onset of dust formation in SN\,2006jc, starting at $\sim$55~d past maximum, is marked by a sharp decline in the optical light curves, coincidental with a relative increase in the NIR fluxes \citep[see e.g.,][]{DiCarlo2008ApJ...684..471D, Mattila2008MNRAS.389..141M, Anupama2009MNRAS.392..894A}. Unfortunately, since the lack of simultaneous observations in optical bands at late epochs from $+$62.4 to $+$96.5 d, our limited NIR observations cannot give us a stringent constraint on the possible dust formation for SN\,2018jmt. Assuming the late-time NIR emission is purely due to dust condensation, it is possible to obtain an estimate of the dust mass for SN\,2018jmt using the methods adopted and described in \citet{Fox2011ApJ...741....7F, Gan2021ApJ...914..125G, Wang2024NatAs...8..504W}. We adopt the species of dust grains made of graphite and silicate with the same size distribution ($a=0.1$~$\mu$m). The inferred dust masses are about $3 \times 10^{-6}$ \msun~(graphite dust) and $3 \times 10^{-5}$ \msun~(silicate dust), respectively. It is important to note that these values have to be considered as upper limits, due to our simplifying assumptions on dust formation. The SED evolution of SN\,2019cj is similar to that observed in SN\,2018jmt (see the right panel of Fig.~\ref{fig:SED}), we hence adopted the same approach and estimated upper limits on the dust masses in SN\,2019cj of the order of several $10^{-4}$ \msun~for both dust species. 

}

\subsection{Observational parameter correlations}

To better characterise the light curve evolution of SN\,2018jmt and SN\,2019cj in the context of SN Ibn variety, we present in  Fig.~\ref{fig:RiseTime_DeclineRate_PeakMagnitude} the locations of a large Type Ibn SN sample in the diagrams of absolute peak magnitude versus rise time, and absolute peak magnitude versus decline rate.
Note that the rise time in the $V$-band for SN\,2018jmt was actually estimated using the $g$-band. Furthermore, the $V$-band absolute peak magnitude was extrapolated based on the initial decline rate observed in the $V$-band. All of this adds uncertainty to the estimates for this object.

The two objects fall in the $1\sigma$ confidence interval in the left panel, and their rise times and peak magnitudes are comparable to the median values observed in the Type Ibn SN sample (median rise time = 9.6 days, median peak magnitude = $-19.19$ mag). In the right panel, both SN\,2018jmt and SN\,2019cj fall within the 95\% confidence interval (shaded area).
SN\,2018jmt and SN\,2019cj exhibit characteristics similar to other SNe Ibn in our sample, suggesting that they adhere to the typical characteristics of Type Ibn SNe and likely share a similar origin.

\subsection{Light curve modelling} \label{subsection:LCmodelling} 

In this section, we present the bolometric light curve modelling of SNe 2018jmt and 2019cj, adopting the (1D spherical) model framework of \citet{Maeda2022ApJ...927...25M}, under the assumption that their optical emissions are entirely powered by the SN-CSM interaction. 
We assume a broken power law for the density structure of the SN ejecta, with the outer power slope fixed to be $n$~=~$7$ ($\rho_{\rm ej} \propto v^n$), while the inner part is represented by a flat distribution. This setup allows the ejecta mass ($M_{\rm ej}$) and the kinetic energy of the expansion ($E_{\rm K}$), used as the input parameters, to determine the properties of the SN ejecta. The CSM density distribution is given as $\rho_{\rm CSM}$($r$) ~=~ $10^{-14} D'$($r$/$5\times10^{14} \ {\rm cm}$)$^{-s}$ g cm$^{-3}$, specified by the parameters $D'$ and $s$, if a single power-law is assumed. 

\begin{figure}
    \includegraphics[width=0.95\linewidth]{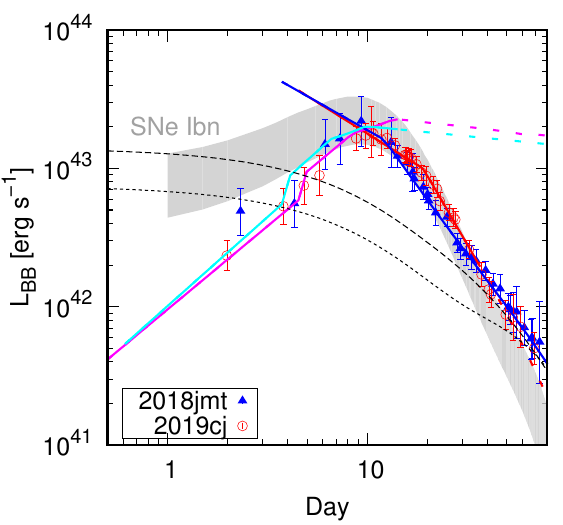}
    \caption{Models for the bolometric light curves of SNe 2018jmt (blue triangles) and 2019cj (red circles). {The models shown here assume $M_{\rm ej}$ ~=~ $2M_\odot$, resulting in $E_{\rm K}$ ~=~ ${1.6}$ and ${1.9} \times 10^{51}$ ergs for SNe 2018jmt and 2019cj, respectively. The CSM parameters are as follows; for the outer components: ($s, D'$) ~=~ (${2.6, 4.2}$) for SN\,2018jmt (blue), and (${2.8, 4.4}$) for SN\,2019cj (red); for the inner components: ($s, D'$) ~=~ (${0.0, 1.0}$) for SN\,2018jmt (cyan), and (${0.1, 0.8}$) for SN\,2019cj (magenta). The thick-dashed curve is the expected contribution from the $^{56}$Ni/Co decay with $M$($^{56}$Ni) ~=~ {0.15}$M_\odot$ taking into account the optical depth to the decay $\gamma$-rays adopting $M_{\rm ej}$~=~$2M_\odot$ and $E_{\rm K}$~=~${1.6} \times 10^{51}$ erg (for SN\,2018jmt), which sets the upper limit for $M$($^{56}$Ni). The same curve but with $M$($^{56}$Ni) ~=~ {0.08}$M_\odot$ is shown by the thin-dashed line, for the case adopting $M_{\rm ej}$ ~=~ $4M_\odot$ and $E_{\rm K}$~=~${2.3} \times 10^{51}$ erg.}
    }
    \label{fig:model_lc}
\end{figure}

\begin{figure}
    \includegraphics[width=0.9\linewidth]{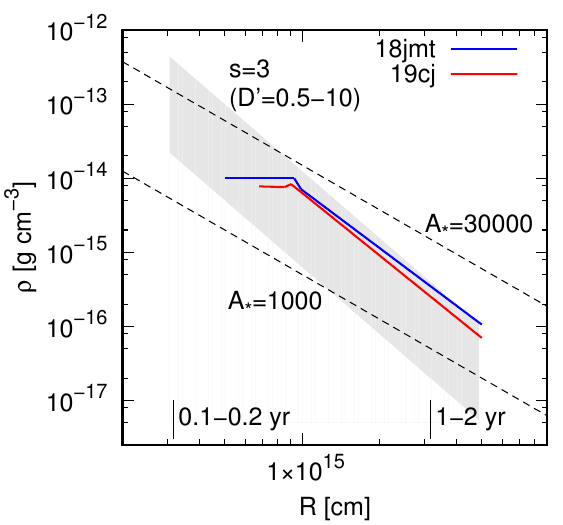}
    \caption{The CSM radial-density distribution derived for SNe 2018jmt (blue) and 2019cj (red). The typical range found for a sample of SNe Ibn is shown by the grey-shaded area. For comparison, the CSM distribution by a steady-state mass loss is shown by the dashed lines for the CSM density parameter of $A_*$ = $30,000$ and $1,000$ (corresponding to $D'$ = $6$ and $0.2$ with $s$~=~$2.0$). On the bottom, the look-back time in the mass-loss history is indicated, assuming $v_{\rm w}$ = $500$-$1,000$ km s$^{-1}$. 
    }
    \label{fig:model_csm}
\end{figure}

Fig. \ref{fig:model_lc} shows the results of the light curve models, and Fig. \ref{fig:model_csm} shows the derived CSM density distribution. These models assume $M_{\rm ej}$ ~=~ 2 $M_\odot$, and the $E_{\rm K}$ required to fit the light curve is derived to be ${1.6}$ and ${1.9} \times 10^{51}$ ergs, for SNe 2018jmt and 2019cj, respectively. We introduced a two-component CSM for both SNe (see below), represented by the inner and outer components having different sets of CSM parameters. 
For the outer components, ($s, D'$) ~=~ (${2.6 \pm 0.1, 4.2 \pm 0.3}$) for SN\,2018jmt, and (${2.8 \pm 0.1, 4.4 \pm 0.3}$) for SN\,2019cj; for the inner components, ($s, D'$) ~=~ (${0.0 \pm 0.5, 1.0 \pm 0.3}$) for SN\,2018jmt, and (${0.1 \pm 0.5, 0.8 \pm 0.2}$) for SN\,2019cj. {The uncertainties here only account for the errors in the bolometric luminosities and the distances; these errors should be treated as lower limits, since the systematic uncertainties linked to the assumptions in the emission model are difficult to quantify and are not included here. The relative errors both in $s$ and $D'$ are larger for the inner components, reflecting the larger errors in the BB fits in the pre-peak epochs.} Overall, the inferred physical properties are within the range derived for a sample of SNe Ibn \citep{Maeda2022ApJ...927...25M}, both for the SN ejecta and CSM. This is expected since the observational properties of the two SNe are similar to other SNe Ibn. 
{Note that the CSM densities of SNe 2018jmt and 2019cj are on the highest side of the SNe Ibn analyzed by \citet{Maeda2022ApJ...927...25M}, but it can partly be an artifact; they used quasi-bolometric LCs and therefore might underestimate the CSM densities for the samples of SNe Ibn shown here. }

As discussed by \citet{Maeda2022ApJ...927...25M}, the parameters of the CSM ($s$ and $D'$) can be well constrained from the post-peak light curves. The ejecta properties are on the other hand not uniquely determined. In the post-peak light curves (i.e., after the shock enters into the outer CSM component), the kink where the decline rate accelerates can be interpreted as a transition phase of the shock from the cooling regime to the adiabatic regime, and $E_{\rm K}$ is determined from this transition for a given $M_{\rm ej}$. However, the outer ejecta structure degenerates in terms of a combination of $M_{\rm ej}$ and $E_{\rm K}$, and for $n$~=~$7$ adopted here, this scaling is given as $E_{\rm K} \propto \sqrt{M_{\rm ej}}$ \citep[e.g.,][]{Moriya2013MNRAS.435.1520M}. Models with more massive ejecta can reproduce essentially the same light curve if the ejecta properties follow this relation; for example, the light curves of SNe 2018jmt and2019cj can be fit with $M_{\rm ej}$~=~$4~ M_\odot$ if $E_{\rm K}$ is increased to $\sim$ (${2.3 - 2.7}$) $\times 10^{51}$ erg. 
The same applies to the less-massive ejecta case, but with another constraint: the ejecta mass cannot be too low, otherwise the reverse shock reaches the inner part of the ejecta too early (which is the argument used by \citealt{Nagao2023A&A...673A..27N} in constraining the ejecta properties for SNe Icn). From the light curve calculations, we find the lower limits for the ejecta masses as {$M_{\rm ej}$ > ${1.6~M_\odot}$ for SN\,2018jmt and > ${1.8~M_\odot}$ for SN\,2019cj}.

Under the model framework, the CSM density distribution is uniquely determined. Similar CSM density structures are derived for SNe 2018jmt and 2019cj, both within the range found for a sample of SNe Ibn. The slopes of the outer components are similar to other SNe Ibn \citep[$s \sim 2.5-3$;][]{Maeda2022ApJ...927...25M}, indicating the increase of the mass-loss rate in the last few years toward the explosion. 
With the (outer) CSM parameters used to fit their post-peak light curves, we find that the diffusion time scale becomes too short (a few days) compared to the rise time (about 10 days). This indicates that the inner part of the CSM density distribution could deviate from the extrapolation from the outer part of the CSM distribution. We are thus motivated to introduce the inner CSM component separately from the outer component. By adopting the flatter density distribution for the inner part (Fig. \ref{fig:model_csm}), the pre-peak light curve evolution is well reproduced (Fig. \ref{fig:model_lc}). 

The two-component CSM, with the flat part inside and the steep part outside, has been inferred for a few SNe Ibn following a similar analysis \citep{Maeda2022ApJ...927...25M}. We note that this may likely be a common property of SNe Ibn; this analysis requires intensive photometric data in the pre-peak phase, and the flat inner part has been frequently inferred when such data are available. This highlights the importance of discovering SNe Ibn soon after the explosion and coordinating the high-cadence and intensive follow-up observations immediately after the discovery, as demonstrated in the present work for SNe 2018jmt and 2019cj. 

Most of the SNe Ibn decline rapidly after the peak without requiring an additional energy input from the $^{56}$Co decay; this sets the upper limits for the masses of $^{56}$Ni ejected in SNe Ibn, which have been found to be lower than the typical amount estimated for canonical SNe Ib/Ic \citep{Drout2011ApJ...741...97D, Lyman2016MNRAS.457..328L, Ouchi2021ApJ...922..141O, Maeda2022ApJ...927...25M}. 
We performed this test in Fig. \ref{fig:model_lc}, where the two light curves powered by the hypothetical $^{56}$Ni/$^{56}$Co decay for the cases with $M_{\rm ej}$ = 2 and 4 $M_\odot$ for SN\,2018jmt are shown. Similar limits are obtained for SN\,2019cj. The upper limits thus obtained, $M$($^{56}$Ni) = {0.15 or 0.08} $M_\odot$, are below the mean $^{56}$Ni production found in a sample of canonical SNe Ib/c (\citealt{Meza2020A&A...641A.177M}, but see \citealt{Ouchi2021ApJ...922..141O}). However, the limits here are not very strong and within the diversity of canonical SNe Ib/c; longer-term follow-up observations until later epochs might have placed a stronger constraint, as were found for some SNe Ibn.

\section{Spectroscopy}
\label{section:Spectroscopy}

\begin{figure*}
    \centering
    \includegraphics[width=13.8cm, height=9.8cm]{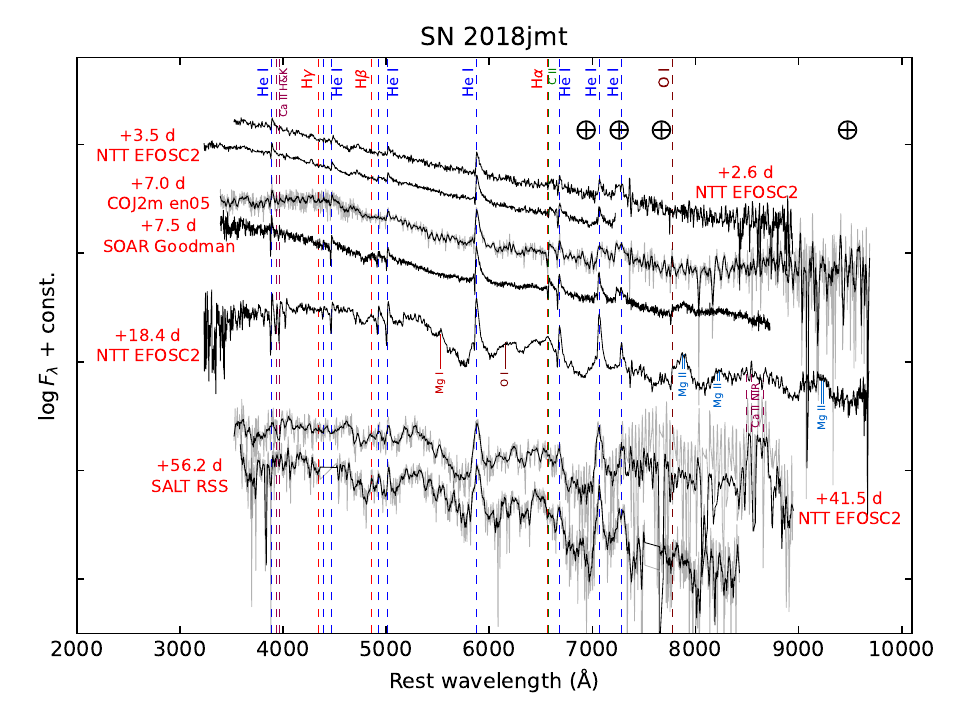}
    \includegraphics[width=13.8cm, height=9.8cm]{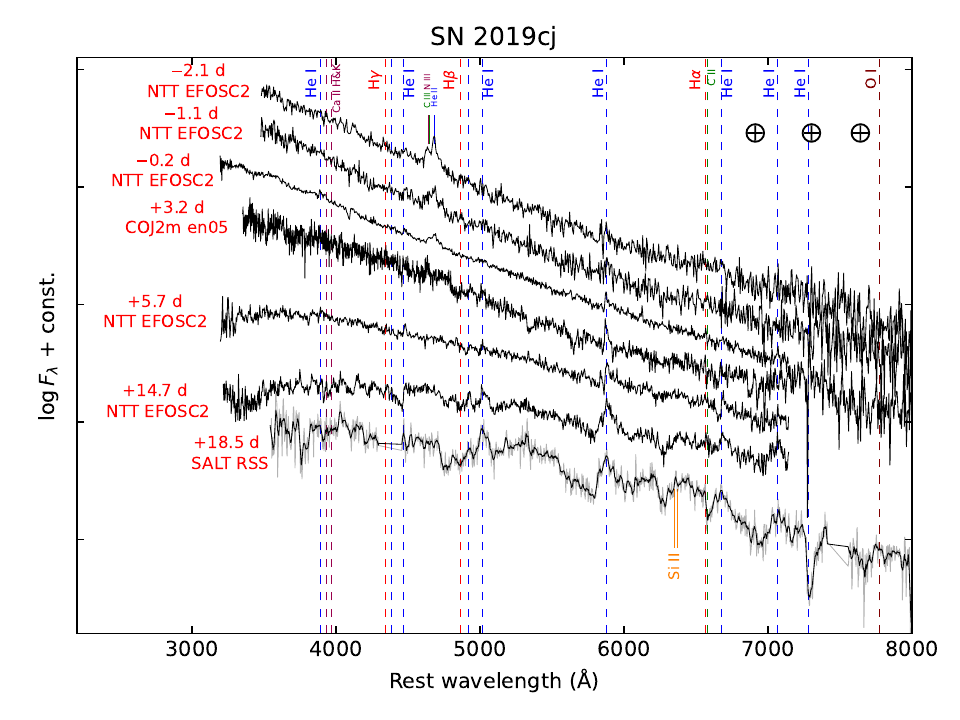}
    \caption{Spectral sequences of SN\,2018jmt (top) and SN\,2019cj (bottom).
    The position of the principal transitions from H and He I are highlighted by the dashed vertical lines. The $\oplus$ symbols mark the position of the strongest telluric absorption bands. All spectra have been corrected for redshift and extinction. In some cases, spectra with lower S/N have been smoothed using a Savitzky-Golay filter (indicated by the gray line).}
    \label{fig:spectral}
\end{figure*}

\begin{figure*}
    \centering
    \includegraphics[width=0.8\linewidth]{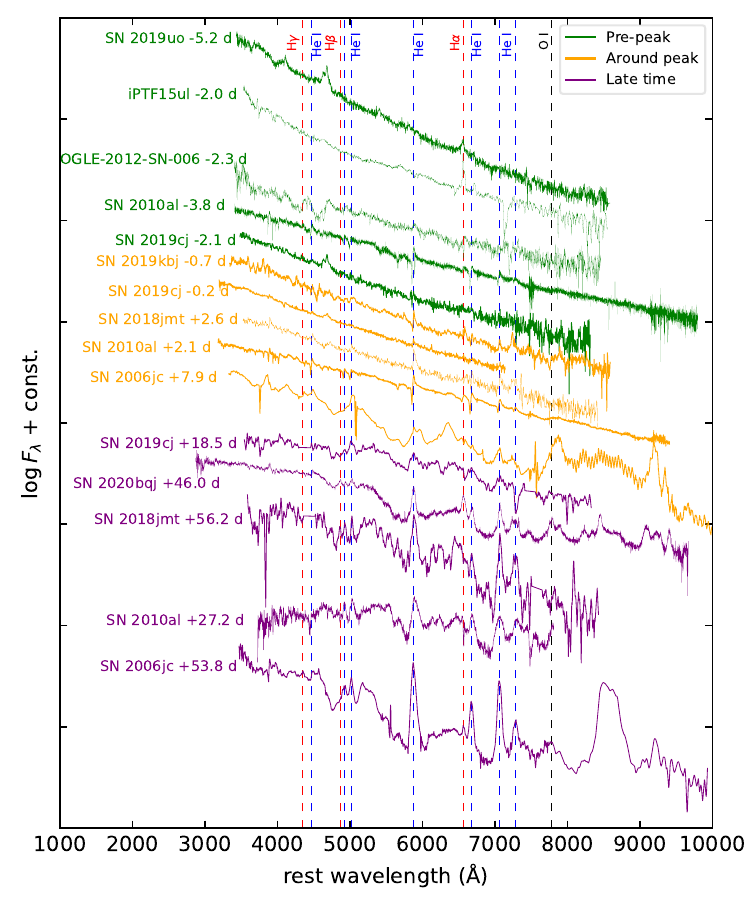}
    \caption{Comparison of pre-peak, around peak, and late-time spectra of SN\,2018jmt and SN\,2019cj with those of other SNe Ibn at similar phases. The H, \Hei, and \Oi lines are marked with dashed vertical lines. All spectra have been corrected for the respective redshift and extinction. The pre-peak spectra are shown in green, those taken near the maximum light are in orange, and the post-maximum spectra are in purple.
    The most significant \Hei lines are indicated by vertical dashed blue lines, while Balmer lines are marked with red dashed lines. The \Oi $\lambda$7774 line is represented by a dashed black line.
    }
    \label{fig:IBN_spectral}
\end{figure*}

\begin{figure*}
    \centering
    \includegraphics[width=0.8\linewidth]{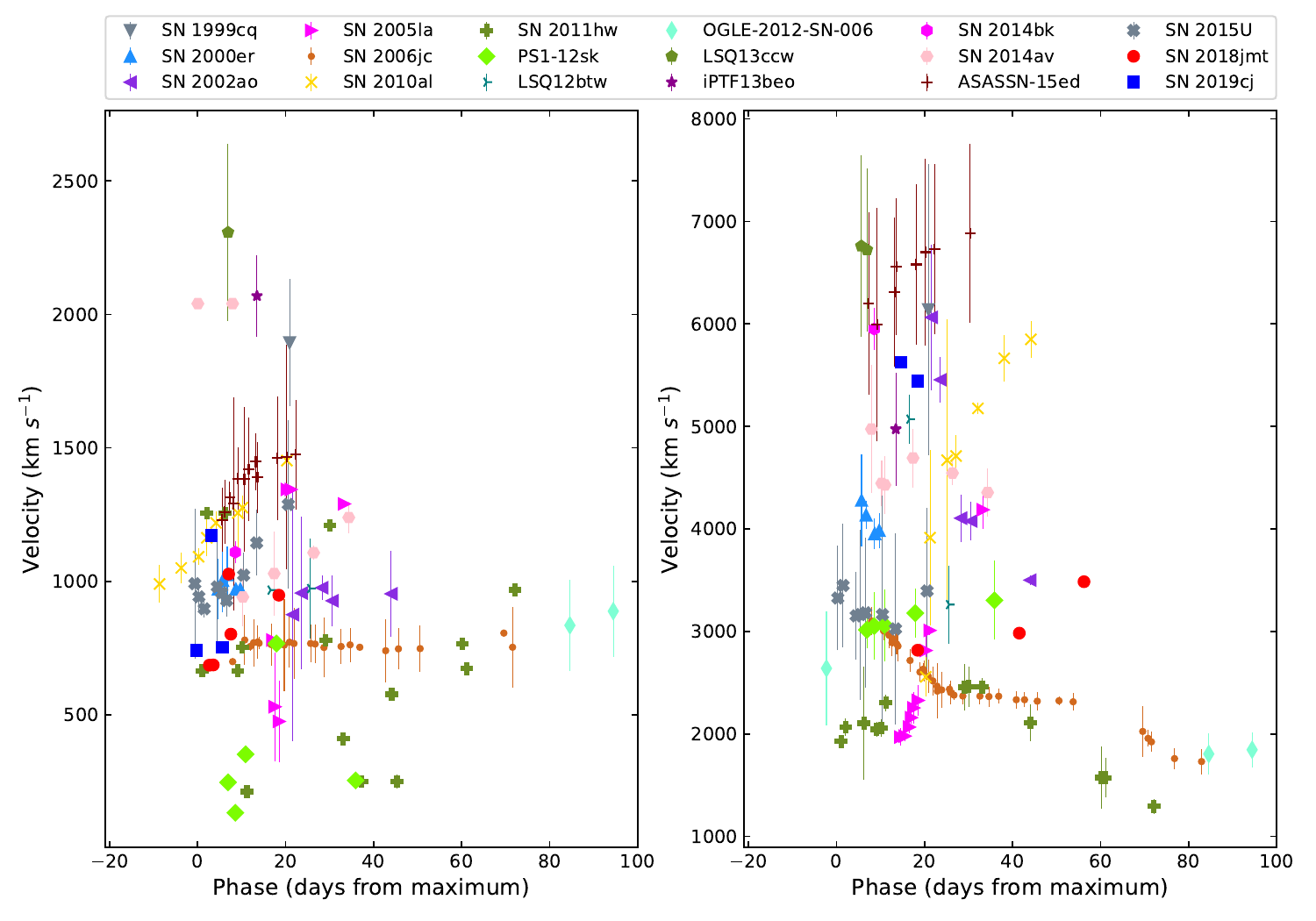}
    \caption{Velocity evolution of the \Hei emission lines. In the left panel, the velocity evolution of the narrow \Hei line components is shown. On the right panel, the velocity of the intermediate/broad emission components is displayed.
    The data of comparison SNe Ibn are taken from \citet{Pastorello2016MNRAS.456..853P}.
    }
    \label{fig:velocity_evolution}
\end{figure*}

\subsection{Spectral sequence of SN~2018jmt}

SN\,2018jmt was monitored in optical spectroscopy from its discovery until approximately 60 days after maximum brightness.
Top panel of Fig.~\ref{fig:spectral} displays, in chronological order, our spectra of SN\,2018jmt. The first spectrum shown corresponds to the classification spectrum \citep{Castro-Segura2018TNSCR2064....1C}. 
During the entire observational period, the spectra exhibit a modest evolution.
The \Hei lines are the most prominent features observed in the spectra of SN\,2018jmt. Some of these lines exhibit complex profiles, with narrow features overlaid on broader line components (e.g. in the +18 d spectrum).

The earlier spectra of SN\,2018jmt reveal a blue continuum. By fitting a blackbody model to the first four spectra (taken at 2.6, 3.5, 7.0, and 7.5 days after the maximum light), the photospheric temperature ranges from 12,000 to 9,000 K. 
Notably, significant P-Cygni profiles are observed in the \Hei 5876 \AA~line, with the minimum being blueshifted by about 600~-~1000 km~s$^{-1}$.
Several \Hei lines are also identified at 4471, 4921, 5016, 6678, 7065, and 7281 \AA. 

Furthermore, we note the presence of a weak and narrow H$\alpha$ with a P-Cygni profile
(the minimum of the P-Cygni absorption is blue-shifted by about 150~-~300 km~s$^{-1}$).
It is worth mentioning that other Balmer lines, which are prominent in Type IIn SNe, are only marginally detected in SN\,2018jmt.

In the subsequent three spectra (from +18 to +56 d), the dominant feature is a blue pseudo-continuum that extends up to approximately 5600 \AA.
The nature of this blue pseudo-continuum has been extensively discussed by \citet{Stritzinger2012ApJ...756..173S} and \citet{Smith2012MNRAS.426.1905S}. They propose that it arises from a combination of numerous narrow and intermediate-width Fe lines, similar to those observed in SN~2005ip (with a \textit{v}$_{\mathrm{FWHM}}\sim$ 150~-~200 km s$^{-1}$) and SN~2006jc \citep[with a \textit{v}$_{\mathrm{FWHM}}\approx$2000~-~2500 km~s$^{-1}$;][]{Chugai2009MNRAS.400..866C}.
This blend of Fe lines can explain several features observed in the spectrum of SN\,2018jmt. They account for the apparent discontinuity in the continuum at around 5600~\AA, the broad `W'-shaped feature between 4600 and 5200 \AA~(although some \Hei lines may also contribute to it), and the broad bump observed between 6100 and 6600 \AA. 

At 18.4 days after maximum, prominent \Hei lines in emission are observed at $\lambda$3889, $\lambda$4388 (weak), $\lambda$4471, $\lambda$$\lambda$4921, 5016, $\lambda$5876 (possibly blended with the \Nai doublet), $\lambda$6678, $\lambda$7065, and $\lambda$7281. These lines exhibit an emission component that largely dominates over the blue-shifted absorption. The most prominent \Hei emission features display a distinct double-component profile.
By deconvolving the $\lambda$7065 line into Lorentzian and Gaussian components, we deduce the presence of a broader component with a FWHM velocity of $\sim$ 2800 km~s$^{-1}$, which shows marginal evolution over time. Additionally, a narrower line with a FWHM velocity of $\sim$ 950 km~s$^{-1}$ is superimposed on the broader component.
From this spectrum, there is a marginal detection of the H$\alpha$ line (possibly blended with \Cii $\lambda$6578) and possibly even H$\beta$.
We also identify \Mgi$\lambda$5528, \Caii H\&K $\lambda$$\lambda$3934,3969, as well as \Oi $\lambda$6158. We searched for the presence of lines typical of thermonuclear SNe, such as \Sii and \SiII ions, but we were unable to securely identify them. It is likely that \Feii lines are responsible for the majority of the broad absorption blends observed at $\lambda$~<~5600 \AA.
Additionally, broad bumps are detected around 7900~\AA~(likely attributed to \Mgii $\lambda$$\lambda$7877,7896), 8200~\AA~(possibly \Mgii $\lambda\lambda$8214,8235 lines), 8500 \AA~(due to Ca II NIR triplet), and 9200 \AA~(attributed to \Mgii $\lambda\lambda$9218,9244 lines).

At later epochs, the emission lines become stronger, allowing for a more robust identification of the spectral features. Using our latest spectra taken at 41.5 and 56.2 days after maximum light, we accurately identify the most prominent features, as shown in Fig.~\ref{fig:spectral}. However, it should be noted that the S/N in these spectra is relatively low.
We continue to observe the prominent \Hei lines, now exhibiting a FWHM velocity of $\sim$ 3200 km s$^{-1}$.
Once again, we tentatively identify H$\alpha$ and H$\beta$~(v$_{\mathrm{FWHM}}$~$\sim$~600 km s$^{-1}$)~emission lines, although alternative identifications cannot be completely ruled out, such as \Cii ($\lambda$~6578) and \Nii ($\lambda$ 4803).

\subsection{Spectral sequence of SN~2019cj} \label{spec_2019cj}

The spectroscopic monitoring campaign of SN\,2019cj began 4 days after its discovery and lasted for 20 days.
Information regarding the spectroscopic observations can be found in Table \ref{2019cjSpecInfo}, while the sequence of available spectra for SN\,2019cj is shown in Fig. \ref{fig:spectral}, bottom panel.

The first spectrum, obtained 8.1 days after the explosion (i.e., 2.1 days before the maximum light), exhibits a blue continuum. By fitting a blackbody to the continuum, we infer a temperature of $T_{\mathrm{bb}}$~=~16800$\pm$2400 K.
The most prominent emission feature is observed in the blue region of the spectrum, specifically around 4660 \AA, and it displays a double-peaked profile. The redder component of the emission is most likely to be \Heii $\lambda$4686. On the other hand, the bluer emission is possibly attributed to either \Ciii $\lambda$4648 or \Niii $\lambda$4640, or a combination of both.
We note that these lines, which were also identified by \citet{Silverman2010CBET.2223....1S} and \citet{Cooke2010ATel.2491....1C}, are commonly seen in WR winds. They are also frequently observed as "flash spectroscopy" features in very early spectra of many CC SNe \citep[e.g.,][]{Gal-Yam2014Natur.509..471G, Bostroem2023ApJ...956L...5B, Bruch2023ApJ...952..119B, ZHANG20232548, Jacobson2024arXiv240302382J, Jacobson2024arXiv240419006J}. A similar characteristic was also detected in the early spectra of other SNe Ibn, such as SN 2010al \citep{Pastorello2015MNRAS.449.1921P} and SN 2019uo \citep{Gangopadhyay2020ApJ...889..170G}.
A low-contrast feature is observed around 5830-5890 \AA, and it is likely due to the \Hei $\lambda$5876, exhibiting a weak P-Cygni profile.

Between the second and third spectra (taken 1.1 and 0.2 days before maximum light), we still observe a dominant blue continuum, with a photospheric temperature ($T_{\mathrm{bb}}$) decreasing from 16700$\pm$2700~K to 14300$\pm$1900 K.
The blend of lines from highly ionised elements, detected in the first spectrum at 4600$-$4700 \AA, gradually weakens, although remaining the most prominent feature, and still displaying a double-peaked profile.
In the spectrum taken at maximum light, other lines start to emerge, in particular, \Hei $\lambda$5876, which exhibits a very narrow P-Cygni profile. The position of the minimum of the blue-shifted absorption component suggests a velocity of the He-rich material around 740 km~s$^{-1}$.
In the fourth spectrum (+3.2 days), $T_{\mathrm{bb}}$ has decreased to 11900$\pm$1200 K, while the P-Cygni \Hei lines become progressively more prominent. From the position of the absorption minimum, we infer an expansion velocity of approximately 1200 km~s$^{-1}$. The previously observed feature at around 4600-4700 \AA~has now completely vanished.
Furthermore, another weak P-Cygni feature can be seen in the spectrum, specifically the \Hei $\lambda$6678 line.
The subsequent spectrum (+5.7 days) does not exhibit any significant changes or evolution, except a slightly redder continuum ($T_{\mathrm{bb}}$ = 8800$\pm$1400 K).
The spectrum obtained at 14.7 days after maximum displays significant changes. The continuum has shifted towards redder colours, with a photospheric temperature of 7400$\pm$1000 K.
The most prominent line observed is \Hei $\lambda$5876, which is mainly in emission and exhibits a FWHM of approximately 5600 km s$^{-1}$. Additionally, several new emission lines are detected, including \Hei $\lambda$$\lambda$4921, 5016, and $\lambda$7065.
In the last spectrum taken on day 18.5, most features observed in the previous spectrum are confirmed, with a notable blue pseudo-continuum. The lines of \Hei $\lambda$5016, $\lambda$5876 (possibly blended with  \ion{Na}{i} D), $\lambda$6678, and $\lambda$7065 are now seen as prominent emission features.
We can confidently rule out any \Sii features. The absorption observed around 6300 \AA~could potentially be attributed to \SiII $\lambda$6355, while alternative identifications include \Mgii $\lambda$6346 and/or [\Oi] $\lambda\lambda$6300, 6364.

{The transition of the spectral features from the `flash' features to those of a classical Type Ibn SN might be consistent with the two-component CSM structure inferred through the LC modeling. When the shock is in the inner component (in the pre-maximum phase), with sufficiently dense materials to create the recombination lines and not yet swept up, they might create the highly-ionized emission lines as irradiated by the high-energy photons from inside. Once the shock enters into the outer component (in the post-maximum phase), the CSM density above the shock wave rapidly decreases, and thus they are not able to produce the recombination lines anymore. Note that the pre-maximum spectra to test the same prediction for SN 2018jmt are not available. However, lacking detailed spectral modeling, this picture is only speculative; this needs to be verified by future efforts in both theoretical modeling and advanced observations.}

\subsection{Comparison of Type Ibn SN spectra}

In Fig. \ref{fig:IBN_spectral}, we present a comparison of the pre-peak, around-peak, and late-time spectra of SN\,2018jmt and SN\,2019cj with those of other SNe Ibn at similar phases.
On top of Fig. \ref{fig:IBN_spectral}, an early spectrum of SN\,2019cj is compared with pre-peak spectra of other Type Ibn SNe, including SN 2019uo, iPTF15ul, OGLE-2012-SN-006, and SN 2010al.
The pre-peak spectra of these SNe exhibit notable differences. 
SN\,2019uo, iPTF15ul, and OGLE-2012-SN-006 display evident H$\alpha$ lines, while in SN\,2010al and SN\,2019cj, this feature is very faint or missing.
The \Hei $\lambda$5876 line is quite prominent in OGLE-2012-SN-006, SN 2010al, and SN\,2019cj, whereas it is missing or weak in SN 2019uo and iPTF15ul.
One of the strongest features visible in the early spectra SNe 2019uo and 2019cj is observed at around 4660 \AA, showing a double-peak profile, likely a blend of \Heii and \Niii/\Ciii (see Sect. \ref{spec_2019cj}). The situation is different in the OGLE-2012-SN-006 spectrum, where this line is not securely detected, but a broad absorption feature is present at a similar position, likely due to \Oii lines \citep{Pastorello2015MNRAS.449.1941P}. We remark that  \Oii lines have never been observed before in SNe Ibn, although these lines are ubiquitously observed in the early spectra of super-luminous stripped-envelope (SE) SNe \citep{Quimby2011Natur.474..487Q}.

In the middle of Fig. \ref{fig:IBN_spectral}, a spectrum around the peak of SN\,2018jmt and SN\,2019cj is compared with spectra of SNe 2019kbj, 2010al, and 2006jc taken at similar phases.
The spectra of SNe Ibn share a very similar blue continuum with narrow P-Cygni profiles of \Hei lines. However, there are some subtle differences. 
Specifically, H$\alpha$ is still observable as a weak emission in SN 2019kbj and SN 2010al, while its presence is not secure in the spectra of SN\,2018jmt and SN\,2019cj. In this small sample, SN~2006jc is somewhat of an exception, as its spectrum shows a larger number of lines with a broader width. 
While the phase of SN 2006jc is measured relative to its peak time, which is relative to the presumed time of the maximum light \citep[MJD = 54012.29, according to][]{Maund2016ApJ...833..128M}, the properties of the spectrum suggest a somewhat older evolutionary phase.

At the bottom of Fig.~\ref{fig:IBN_spectral}, a late spectrum of SN\,2018jmt and one of SN\,2019cj is compared with those of SNe 2020bqj, 2010al, and 2006jc at similar phases.
The spectra of the five objects exhibit a remarkable similarity in terms of the blue pseudo-continuum and the presence of prominent broader spectral lines.
The wide velocity range observed in these lines can be attributed to various gas regions where they originate, such as the unperturbed CSM, shocked shells, shocked or unshocked supernova ejecta, or a combination of different emitting regions \citep{Pastorello2016MNRAS.456..853P}.

\subsection{He I line velocity evolution} 
\label{subsec:velocity evolution}

The velocity evolution of the spectral lines allows us to constrain the properties of the stellar wind and understand the nature of the emitting regions. SNe that interact with the CSM, such as Type Ibn and Type IIn events, exhibit lines with multiple-width components. These components are believed to originate from different gas regions \citep{Chugai1997Ap&SS.252..225C}.

In SNe Ibn, the presence of multiple components in the spectral lines indicates that the emitting material is expanding at different velocities. 
When a clear P-Cygni profile is identified, the velocity of the He-rich expanding material can be determined by measuring the position of the minimum point of the blue-shifted absorption.
If this component is undetected, the velocity is estimated by measuring the FWHM of the strongest \Hei emission lines, which are obtained by deblending the full line profile using Gaussian fits.
The evolution of velocities in \Hei emission lines is illustrated in Fig.~\ref{fig:velocity_evolution}.

The study of the narrowest line profiles provides insights into the velocity of the unshocked CSM hence offers key information about the mass-loss history of the progenitors of SNe Ibn in the latest stages of its life. 
The temporal evolution of the velocity of the narrow \Hei line components is shown in the left panel of Fig.~\ref{fig:velocity_evolution}.
In most cases, including SN\,2018jmt and SN\,2019cj, the narrow \Hei components in our spectral sample exhibit velocities of 800-1000 km~s$^{-1}$.
It is worth noting that, due to the limited spectral resolution of the spectra, the measurements for the two SNe should be considered as upper limits in some cases.
However, it is worth noting that the narrowest components observed in the spectra of our Type Ibn supernova sample span a wide range of velocities. Objects such as PS1-12sk (approximately 250 km~s$^{-1}$) and the two transitional Type Ibn/IIn SNe, SNe 2005la (about 500 km~s$^{-1}$) and 2011hw (200-250 km~s$^{-1}$), display the lowest velocities for the unperturbed CSM.
On the other extreme, LSQ13ccw, iPTF13beo, and SN 1999cq exhibit narrow components with P-Cygni profiles having velocities of approximately 1900-2300 km~s$^{-1}$. Such a large range of CSM velocities, as identified by \citet{Pastorello2016MNRAS.456..853P}, suggests that Type Ibn SNe may arise from different progenitor types and/or different explosion mechanisms.

The evolution of the expansion velocities for the intermediate/broad components of the \Hei lines is shown in right panel of Fig.~\ref{fig:velocity_evolution}. These components exhibit velocities that are 4-6 times higher than those measured for the narrow components.
Unlike the narrow features, the broader components of the \Hei lines experience a significant evolution over time. This evolution is influenced by the velocity of the ejecta and the density of the interacting material. The velocities of these broader components can provide insights into the gas interface between two shock fronts, which in turn depend on the speed of the expanding SN ejecta. 
An increasing velocity of the intermediate/broad \Hei components is observed in SN 2010al, ASASSN-15ed, SN\,2018jmt, SN 2005la, and PS1-12sk.
In some cases, the \Hei lines become narrower with time. For instance, in SN 2002ao, the width of the \Hei intermediate component decreases by a factor of 2 within three weeks. This trend is also observed in SNe 2014av, 2000er, 2002ao, 2019cj and 2006jc.
This apparent decline in the velocity of the shocked gas regions is possibly attributed to an increased density of the CSM. 

\section{Summary and discussion}
\label{section:Discussion_and_conclusions}

The high-cadence $TESS$ light curve of SN\,2018jmt presented by \protect\citet{Vallely2021MNRAS.500.5639V} reveals no evidence for a rapidly evolving shock break-out peak. The subsequent light-curve rise time to maximum light is 13.4 $\pm$ 0.3 days, slightly longer than the 10.2 days estimated in this paper for SN\,2019cj (in the $T$-band and $V$-band, respectively). At maximum, SN\,2018jmt and SN\,2019cj exhibit a similar luminosity, reaching a peak magnitude of $M_{g}$~$\sim$~-19~mag and $M_V$~$\sim$ -19 mag, implying a very similar bolometric luminosity of about 10$^{43}$ erg $\mathrm{s}^{-1}$.
The post-peak decline in the light curve of SN 2018jmt is initially steep ($\gamma_{0-15}$(r)$~\approx~0.13$ mag d$^{-1}$) and then slows down ($\gamma_{15-60}$(r)$~\approx~0.04$ mag d$^{-1}$), whereas in the case of SN 2019cj, it starts off slow ($\gamma_{0-15}$(r)$~\approx~0.05$ mag d$^{-1}$) and then becomes steep ($\gamma_{15-48}$(r)$~\approx~0.09$ mag d$^{-1}$), as reported in Table \ref{tab:slope}. This is consistent with the decline rate range of $\gamma_{0-15}$(R): 0.05 $\sim$ 0.24 mag d$^{-1}$ observed in the SN Ibn group \citep{Hosseinzadeh2017ApJ...836..158H}.

The spectra of SN\,2018jmt evolve from a distinct blue continuum in the early phases to being dominated by narrow \Hei lines ($v \sim$ 600~-~1000 km~s$^{-1}$), while $T_{\mathrm{bb}}$ ranges from 12,000 to 9,000~K. A weak and narrow H$\alpha$ line with a P-Cygni profile ($v \sim$ 150~-~300 km~s$^{-1}$) is present, while other Balmer lines are either absent or weak. In the subsequent stages, the spectra exhibit a blue pseudo-continuum with a narrower line superimposed on the broader component, which eventually transitions into a broad line ($v_{FWHM} \sim$~3200 km~s$^{-1}$). 
At the early stages of SN 2019cj, an intriguing feature observed in the spectra is the potential identification of flash ionisation signatures formed within a He-rich CSM.
The most prominent line in the subsequent spectra of SN\,2019cj was the \Hei line at 5876 \AA, initially displaying a P$-$Cygni profile ($v \sim$ 740~-~1200 km~s$^{-1}$), and later transitioning into broad features in emission ($v_{FWHM} \sim$ 5600 km~s$^{-1}$).

The high intrinsic luminosity, the blue colours persisting for a long time, the emission-line spectra, and the fast-declining light curve without any apparent flattening to the $^{56}$Co tail suggest that the observables of SNe 2018jmt and 2019cj are primarily due to ejecta$-$CSM interaction.
In particular, as there is no spectroscopic evidence of dust formation (e.g., blue-shifted line emission peaks), the faint late-time luminosity of the two SNe (see Fig.~\ref{fig:model_lc}) can be explained assuming that the contribution of the synthesised $^{56}$Ni/$^{56}$Co to the light curve is very small. 

From light curve modelling (see Fig. \ref{fig:model_csm}), we may determine the CSM configuration of SNe 2018jmt and 2019cj. The CSM distribution is constrained within a range from ${5} \times 10^{14}$ to ${5} \times 10^{15}$ cm. The inner and outer radii correspond to look-back times of approximately 0.1 to 0.2 years and 1 to 2 years, respectively, assuming a mass-loss history with $v_{w}$ = 500 to 1000 km s$^{-1}$. Notably, the CSM distribution exhibits a steeper trend compared to steady-state mass loss, characterised by a power-law index of approximately s = 3.
Furthermore, the derived CSM density is remarkably high for the outer components, with D' = {4.2} for SN\,2018jmt and {4.4} for SN\,2019cj. At a distance of approximately $5 \times 10^{14}$ cm, this corresponds to $D \sim ({4.2} \text{ and } {4.4}) \times 10^{-14}$ g cm$^{-1}$ or $A_{*} \sim {21,000} \text{ and } {22,000}$; $\dot{M}\sim {0.21} \text{ and } {0.22 } ~M_\odot$ yr$^{-1}$ for $v_{\rm w}$ = 1,000 km s$^{-1}$, or $\dot{M}\sim {0.105} \text{ and {0.11} }M_\odot$ yr$^{-1}$ for $v_{\rm w}$ = 500 km s$^{-1}$.
These two objects exhibit an inner flat CSM component and an outer steep CSM component at a radius of approximately $(0.8-1) \times 10^{15}$ cm. 
A common feature in SNe Ibn is the possible existence of the inner flat part CSM component, as observed in SNe 2010al, 2011hw, and LSQ12btw \citep{Maeda2022ApJ...927...25M}. Considering the time scale of the two-component CSM transition, 0.3 $-$ 0.6 years, it is possible that this new component corresponds to an eruptive pre-SN mass-loss event as observed in SN 2006jc \citep{Pastorello2007Natur.447..829P}. In this latter object, the outburst occurred approximately two years prior to the SN explosion. 

\subsection{Host environment metallicity}

\citet{Pastorello2015MNRAS.449.1954P} conducted a characterisation study of the host galaxies of SNe Ibn, revealing that all of them were found in spiral galaxies, with the exception of PS1-12sk, which originated in the outskirts of an elliptical galaxy \citep{Sanders2013ApJ...769...39S}.
In order to explore the possible connection of SNe Ibn with the evolution of very massive WR stars, we study the environments of our two Type Ibn SNe.
SN\,2018jmt exploded within an edge-on disk galaxy, most likely a spiral galaxy, while SN\,2019cj occurred in the outskirts of a late-type (Sc-type) spiral galaxy.

The oxygen abundance for the host galaxies at the SN location can be calculated using the luminosity-metallicity relation of \citet{Pilyugin2004A&A...425..849P}.
The oxygen abundance at the location of SN 2018jmt is approximately 8.54 dex, while for SN 2019cj, it is 8.62 dex. These values are nearly solar, assuming a Solar metallicity of 12 + log(O/H) = 8.69 dex \citep[see e.g.,][]{Asplund2009ARA&A..47..481A, vonSteiger2016ApJ...816...13V, Vagnozzi2019Atoms...7...41V}.
\citet{Pastorello2015MNRAS.449.1954P} estimated an average metallicity of 12 + log(O/H) = 8.63 $\pm$ 0.42 at the SN positions for Ibn SNe. \citet{Taddia2015A&A...580A.131T}, using a smaller sample, found a slightly lower average oxygen abundance of 12 + log(O/H) = 8.45 $\pm$ 0.10.
The discovery of SNe Ibn in environments spanning a wide range of metallicities led \citet{Pastorello2016MNRAS.456..853P} to suggest that metallicity has a marginal influence on the evolutionary path of the progenitors of SNe Ibn. 

\subsection{Progenitor and explosion scenarios}

In our attempts to model the light curves of SNe 2018jmt and 2019cj, we constrained the physical parameters as follows:
\begin{itemize}
    \item Ejecta: $M_{\rm ej}$ ranges between 1 $\msun$ to 4 $\msun$, and $E_{\rm K}$ is of the order of $10^{51}$ erg, although there is some degeneracy in the above values.
    Adopting an average value of $M_{\rm ej}$ = 2 $M_\odot$ for the two objects, $E_{\rm K} \sim {1.6} \times 10^{51}$ erg for SN\,2018jmt and ${1.9} \times 10^{51}$ erg for SN\,2019cj are required to fit the light curves. 
    Our analysis provides lower limits for the ejecta masses as 
    {$M_{\rm ej}$ > ${1.6~M_\odot}$ for SN\,2018jmt and > ${1.8~M_\odot}$ for SN\,2019cj}.
    \item CSM: We adopt $M_{\rm ej}$ = 2 $\msun$ and a two-zone CSM distribution, with a flat-density inner component (s $\sim$ {0.1}, $D'$ $\sim$ 0.9) and a steeper density outer component (s $\sim$ 2.7, $D'$ $\sim$ {4.3}).
    Specifically, for the outer components, we obtained ($s, D'$) = (${2.6 \pm 0.1, 4.2 \pm 0.3}$) for SN\,2018jmt and (${2.8 \pm 0.1, 4.4 \pm 0.3}$) for SN\,2019cj. For the inner components, we infer ($s, D'$) = (${0.0 \pm 0.5, 1.0 \pm 0.3}$) for SN\,2018jmt and (${0.1 \pm 0.5, 0.8 \pm 0.2}$) for SN\,2019cj.
    \item $^{56}$Ni production: While the light curves  of the two SNe can be comfortably reproduced with a pure CSM-interaction model, without necessarily invoking a $^{56}$Ni production, we could constrain an upper limit for the ejected $^{56}$Ni mass from the late luminosity. Assuming ejected masses of $M_{\rm ej}$ = 2 $\msun$ and 4 $\msun$, respectively, we obtained {0.15} $\msun$and {0.08} $\msun$as upper limits for the $^{56}$Ni amounts (the above values are virtually identical for SNe 2019cj and 2018jmt). 
\end{itemize}

The above CSM properties (D$'$) are quite close to the upper limit expected for Type Ibn SNe. According to \citet{Maeda2022ApJ...927...25M},  when the mass-loss rate significantly exceeds D$'$ $\sim$ 4, the entire helium envelope is ejected. Further mass loss could then lead to the formation of a C/O$-$rich CSM and would result in the emergence of SNe Icn. 

\subsubsection{Thermonuclear SNe from He white dwarfs?}

SNe 2018jmt and 2019cj exhibit a rise time of approximately 10 days, thus belonging to the well-populated sample of fast-evolving SNe Ibn. The evolutionary timescales of these SNe~Ibn resemble those observed in other classes of transients.  
SN 2002bj \citep{Poznanski2010Sci...327...58P}, in particular, is a fast-evolving, He-rich transient that was tentatively interpreted as a helium shell detonation on a white dwarf \citep[an example of the so-called Type .Ia SNe; see][]{Bildsten2007ApJ...662L..95B}.
These transients are expected to be relatively faint, with peak magnitudes ranging from $-$15 to $-$18 in the $V$-band \citep{Perets2010Natur.465..322P, Kasliwal2010ApJ...723L..98K, Perets2011ApJ...730...89P, Fesen2017ApJ...848..130F}. 
More importantly, they exhibit a rapid evolution, typically with a rise time of 1 to 6 days, with dimmer objects usually experiencing a faster rise. 
The sample of Type .Ia SN candidates includes $^{56}$Ni masses ranging from very small values (0.02 $\msun$in the case of SN\,2010X; \citealp{Kasliwal2010ApJ...723L..98K}) to $\sim$0.2 $\msun$for SN\,2002bj \citep{Kasliwal2010ApJ...723L..98K}. The upper limits for the $^{56}$Ni mass estimated for SN\,2018jmt and SN\,2019cj are alone not sufficient to rule out the possibility of a  Type .Ia SN interpretation.

However, the ejecta/CSM parameters estimated for SNe 2018jmt and 2019cj are inconsistent with those expected in a very low progenitor mass scenario. For instance, interpreting them as thermonuclear SNe from He white dwarfs is an improbable scenario, given the relatively high ejected masses and the overall CSM parameters. Another argument against a thermonuclear explosion of white dwarfs is the lack of the \Sii spectral lines typical of SNe Ia, and also the Si~II features are not securely detected. For all these reasons, we believe that the two SNe Ibn are explosions associated with much more massive, envelope-stripped stars.

\subsubsection{Core-collapse SNe from moderate-mass He stars in binary systems}

Important constraints on the progenitor's nature can be inferred by studying the circumstellar wind, in particular, the composition and the velocity of the CSM. SNe 2018jmt and 2019cj exhibit emission-line spectra with faint or absent H features, while the prominent lines of He~I suggest wind velocities of 700~$-$~1000 km s$^{-1}$. The broadening of the He~I emission components with time suggests the presence of an intermediate-width component, and a growing intensity of the shocked region emission.
The initial wind velocity is quite consistent with that expected in WR winds, although similar velocities were also observed in the Type Ibn SN~2015G \citep{Shivvers2017MNRAS.471.4381S}, whose progenitor was proposed to be a moderate-mass He star in a binary system \citep{Sun2020MNRAS.491.6000S}. 

Ejected masses higher than {1.6}~$-$~{1.8} $\msun$ are consistent with those expected in canonical stripped-envelope CC SNe, or even in some giant, non-terminal eruptions of very massive stars \citep{Karamehmetoglu2021AA...649A.163K}. While a significant amount of $^{56}$Ni is synthesised in a SE CC SN explosion, giant eruptions are expected to produce no $^{56}$Ni. Unfortunately, for SN\,2018jmt and 2019cj, we could only pose upper limits on the $^{56}$Ni masses ($\leq {0.08 - 0.15}$ $\msun$), which are lower than the average $^{56}$Ni production observed in canonical SE CC SNe, although similar amounts were occasionally observed in SNe Ibc \citep[e.g.][]{Richmond1996AJ....111..327R, Hunter2009A&A...508..371H}. $^{56}$Ni masses of a few $\times 10^{-3}$ $\msun$ were also observed in faint H-rich  CC SNe \citep[e.g.][]{Spiro2014MNRAS.439.2873S}. 
However, we need to remark that the zero $^{56}$Ni mass case, supportive of a non-terminal eruption, cannot be ruled out. For this reason, we can conclude that the $^{56}$Ni mass constraints alone do not allow us to discriminate between CC SNe and giant eruption scenarios. 

From Fig.~\ref{fig:RiseTime_DeclineRate_PeakMagnitude}, we note that most SNe Ibn cluster in a small region of the phase-space diagrams, possibly suggesting some homogeneity in the explosion scenarios and the progenitor masses,  hence the involvement of moderate-mass stars rather than massive progenitors. 
As mentioned above, \citet{Sun2020MNRAS.491.6000S} proposed that SNe Ibn may originate from lower-mass progenitors in interacting binary systems, and the pre-SN
eruptions occasionally observed before the explosion of some Type Ibn SNe could also be triggered by binary interaction.

In brief, a plausible scenario for SNe 2018jmt and 2019cj is that they arose from the explosion of relatively massive stars producing partially stripped CC SNe. This conclusion is also supported by the inspection of the latest spectrum of SN 2018jmt (+56 d), which exhibits some similarity to the spectra of a CC SN in the transition towards the nebular phase. In fact, while the [O I] $\lambda\lambda$6300, 6364, was not securely identified, the strengthening of the He I $\lambda$7281 line vs. the He I $\lambda$7065 can likely be attributed to the emerging of the [Ca II] $\lambda\lambda$7291, 7324 doublet, a classical feature of CC SNe in the nebular phase. 

\subsubsection{The explosion of massive Wolf-Rayet stars}

Another plausible scenario is that SNe 2018jmt and 2019cj mark the endpoints of the lives of higher-mass WR stars. \citet{Maeda2022ApJ...927...25M} suggested that at least a fraction of SNe Ibn is produced by the explosion of envelope-stripped WRs with zero-age main sequence masses (M$_{ZAMS}$) exceeding 18 $\msun$. This interpretation would have some evident advantages. Invoking a massive WR progenitor would comfortably explain the eruptive pre-SN mass loss events, as well as the CSM composition and velocity of typical SNe Ibn. The observed properties of SNe 2018jmt and 2019cj are quite similar to those of classical SNe Ibn \citep{Pastorello2016MNRAS.456..853P, Hosseinzadeh2017ApJ...836..158H}. 

In this scenario, the binding energy of the helium or carbon-oxygen (C+O) core is estimated to be around $10^{51}$ erg \citep{Maeda2022ApJ...927...25M}. 
Consequently, if the canonical explosion energy (approximately $10^{51}$ erg, as constrained by the light-curve analysis) is achieved during the supernova explosion following neutron star (NS) formation, a significant amount of fallback onto the NS, which may or may not lead to black hole formation, is expected. Due to this fallback, the ejection of $^{56}$Ni will be minimal or nonexistent \citep{Woosley1995ApJS..101..181W, Zampieri1998ApJ...502L.149Z, Maeda2007ApJ...666.1069M, Moriya2010ApJ...719.1445M}.
According to \citet{Valenti2009Natur.459..674V}, the absence of [O~I] 6300, 6364 lines in the late spectra is also in agreement with the expectations of the fall-back SN scenario.
During their evolution, these high-mass stars develop large cores with high luminosity. While the relationship between core nature and final evolution is yet to be fully understood \citep{Fuller2017MNRAS.470.1642F, Fuller2018MNRAS.476.1853F}, the substantial luminosity could contribute to heightened activity in the final stages just before CC, resulting in a significant increase in mass-loss rate leading up to the SN event. According to \citet{Heger2003ApJ...591..288H} and \citet{Langer2012ARA&A..50..107L}, it is reasonable to expect that massive stars with initial masses greater than 18 $\msun$ lose mass through strong stellar winds also without the need for a binary companion, leaving a C+O core surrounded by a He-rich CSM.

WR stars much more massive than $\sim$18 $\msun$ can also produce SN-like phenomena with properties compatible with those observed in Type Ibn SNe.
Pulsational pair-instability (PPI) arises from stars with He-core masses of 30~$-$~64 $\msun$ \citep{Woosley2007Natur.450..390W, Woosley2017ApJ...836..244W}, causing intense nuclear flashes during which the H-envelope and portions of the He core are expelled. The frequency and duration of these pulses depend on the He-core mass, with more energetic pulses resulting in longer intervals. The collisions among the ejected shells may generate luminous, interacting events with SN-like observable properties. 
SNe 2018jmt and 2019cj may share some of the PPI SN characteristics, such as ejecta mass, ejecta velocity, and metallicity. \cite{Karamehmetoglu2021AA...649A.163K} proposed several PPI models for a Type Ibn SN with ejecta masses up to 2.65 $\msun$. However, PPI SNe are expected to occur in metal-poor environments, which is not the case for SNe 2018jmt and 2019cj.
The progenitors of PPI SNe are also expected to experience recurrent outbursts, as suggested by \citet{Woosley2007Natur.450..390W} and \citet{Woosley2017ApJ...836..244W}.This is a potential problem for invoking the PPI SN scenario for SNe 2018jmt and 2019cj, as they both appear to be single SN-like events without previously detected eruptions. A PPI model was also proposed for SN 2006jc, which exhibited a luminous precursor two years before the alleged terminal explosion. However, subsequent investigations revealed that the progenitor of SN 2006jc was inconsistent with an extremely massive star, thus challenging the PPI scenario. Instead, the eruptive history is more likely to be explained by more conventional binary interaction \citep{Sun2020MNRAS.491.6000S}.

Regardless of the physical mechanisms triggering the progenitor's mass loss, studying the pre-SN eruptions, as done for SN\,2006jc \citep{Pastorello2007Natur.447..829P}, is a key step to constrain the properties of the progenitor stars and the terminal explosion scenario. 
Indeed, pre-SN outbursts were observed for a handful of Type Ibn SNe, including SN\,2011hw \citep{Dintinjana2011CBET.2906....1D}, SN\,2019uo \citep{Strotjohann2021ApJ...907...99S}, SN\,2022pda (Cai et al., in preparation), and SN\,2023fyq \citep{Brennan2024A&A...684L..18B}. 
For the latter, spectra obtained when the progenitor was quiescent and later in outburst revealed complex \Hei line profiles, characterised by a relatively narrow P-Cygni component, whose minimum is blue-shifted by about 1700 km s$^{-1}$, superposed on a very broad base (extended up to 10$^4$ km s$^{-1}$). The spectra published by \citet{Brennan2024A&A...684L..18B} indicate the presence of a high-velocity progenitor's wind and a highly asymmetric CSM distribution.

Unfortunately, information on the pre-SN variability of the progenitor star is an exception in SNe Ibn, either due to the lack of archival observations of the progenitor sites, or because these SNe are simply located in distant galaxies and pre-SN outbursts are below the instrumental detection thresholds.

\subsection{Concluding remarks} 

With the available dataset for SNe 2018jmt and 2019cj, we cannot securely constrain the mass of their progenitors and the explosion mechanism. However, several clues tend to favour a scenario according to which SNe Ibn are terminal CC SNe from massive stars. Whether the progenitors are massive WRs or lower-mass He stars in binaries is still disputed.

{\citet{Maeda2022ApJ...927...25M} argued that the progenitors of SNe Ibn are WR stars with a mass exceeding 18 \msun. The volumetric rate of SNe Ibn is approximately 1\% of the CC SN population \citep{Maeda2022ApJ...927...25M}. Although this proportion falls below the fraction of massive stars with M$_{ZAMS}$ $\geq$ 18 \msun to those with M$_{ZAMS}$ $\geq$ 8 \msun, many of these potential SNe may have remained undetected in the optical due to a significant portion of their emission being UV radiation. 
Therefore, conducting high-cadence UV surveys is crucial for detecting the population of UV-emitting transients, including SNe Ibn.
Future facilities, such as the Ultraviolet Transient Astronomy Satellite (ULTRASAT)\footnote{\url{https://www.weizmann.ac.il/ultrasat/}} space mission \citep{Shvartzvald2024ApJ...964...74S} and the Ultraviolet Explorer (UVEX) \footnote{\url{https://www.uvex.caltech.edu}} mission \citep{Kulkarni2021arXiv211115608K} will be devoted to conducting wide-field, high-cadence surveys of the sky in the UV, which will play a critical role in studying highly energetic and fast-evolving transient objects.
Additionally, there is a lack of modelling of pre-peak light curves. Hence, it is imperative to consider these limitations in future observational and theoretical endeavours.
The assistance of next-generation instruments, like the Chinese Space Station Telescope (CSST)\footnote{\url{http://nao.cas.cn/csst/}} and the Vera C. Rubin Observatory\footnote{\url{https://www.lsst.org/}} will play a vital role in increasing the sampling frequency and refining current models of Type Ibn SNe.}

\bibliographystyle{aa} 
\bibliography{Ibnref.bib} 

\begin{thebibliography}{144}
\expandafter\ifx\csname natexlab\endcsname\relax\def\natexlab#1{#1}\fi

\bibitem[{{Abdurro'uf} {et~al.}(2022){Abdurro'uf}, {Accetta}, {Aerts}, {Silva
  Aguirre}, {Ahumada}, {Ajgaonkar}, {Filiz Ak}, {Alam}, {Allende Prieto},
  {Almeida}, {Anders}, {Anderson}, {Andrews}, {Anguiano}, {Aquino-Ort{\'\i}z},
  {Arag{\'o}n-Salamanca}, {Argudo-Fern{\'a}ndez}, {Ata}, {Aubert},
  {Avila-Reese}, {Badenes}, {Barb{\'a}}, {Barger}, {Barrera-Ballesteros},
  {Beaton}, {Beers}, {Belfiore}, {Bender}, {Bernardi}, {Bershady}, {Beutler},
  {Bidin}, {Bird}, {Bizyaev}, {Blanc}, {Blanton}, {Boardman}, {Bolton},
  {Boquien}, {Borissova}, {Bovy}, {Brandt}, {Brown}, {Brownstein}, {Brusa},
  {Buchner}, {Bundy}, {Burchett}, {Bureau}, {Burgasser}, {Cabang}, {Campbell},
  {Cappellari}, {Carlberg}, {Wanderley}, {Carrera}, {Cash}, {Chen}, {Chen},
  {Cherinka}, {Chiappini}, {Choi}, {Chojnowski}, {Chung}, {Clerc}, {Cohen},
  {Comerford}, {Comparat}, {da Costa}, {Covey}, {Crane}, {Cruz-Gonzalez},
  {Culhane}, {Cunha}, {Dai}, {Damke}, {Darling}, {Davidson}, {Davies},
  {Dawson}, {De Lee}, {Diamond-Stanic}, {Cano-D{\'\i}az}, {S{\'a}nchez},
  {Donor}, {Duckworth}, {Dwelly}, {Eisenstein}, {Elsworth}, {Emsellem},
  {Eracleous}, {Escoffier}, {Fan}, {Farr}, {Feng}, {Fern{\'a}ndez-Trincado},
  {Feuillet}, {Filipp}, {Fillingham}, {Frinchaboy}, {Fromenteau}, {Galbany},
  {Garc{\'\i}a}, {Garc{\'\i}a-Hern{\'a}ndez}, {Ge}, {Geisler}, {Gelfand},
  {G{\'e}ron}, {Gibson}, {Goddy}, {Godoy-Rivera}, {Grabowski}, {Green},
  {Greener}, {Grier}, {Griffith}, {Guo}, {Guy}, {Hadjara}, {Harding},
  {Hasselquist}, {Hayes}, {Hearty}, {Hern{\'a}ndez}, {Hill}, {Hogg},
  {Holtzman}, {Horta}, {Hsieh}, {Hsu}, {Hsu}, {Huber}, {Huertas-Company},
  {Hutchinson}, {Hwang}, {Ibarra-Medel}, {Chitham}, {Ilha}, {Imig}, {Jaekle},
  {Jayasinghe}, {Ji}, {Johnson}, {Jones}, {J{\"o}nsson}, {Katkov}, {Khalatyan},
  {Kinemuchi}, {Kisku}, {Knapen}, {Kneib}, {Kollmeier}, {Kong}, {Kounkel},
  {Kreckel}, {Krishnarao}, {Lacerna}, {Lane}, {Langgin}, {Lavender}, {Law},
  {Lazarz}, {Leung}, {Leung}, {Lewis}, {Li}, {Li}, {Lian}, {Liang}, {Lin},
  {Lin}, {Lin}, {Lintott}, {Long}, {Longa-Pe{\~n}a}, {L{\'o}pez-Cob{\'a}},
  {Lu}, {Lundgren}, {Luo}, {Mackereth}, {de la Macorra}, {Mahadevan},
  {Majewski}, {Manchado}, {Mandeville}, {Maraston}, {Margalef-Bentabol},
  {Masseron}, {Masters}, {Mathur}, {McDermid}, {Mckay}, {Merloni},
  {Merrifield}, {Meszaros}, {Miglio}, {Di Mille}, {Minniti}, {Minsley},
  {Monachesi}, {Moon}, {Mosser}, {Mulchaey}, {Muna}, {Mu{\~n}oz}, {Myers},
  {Myers}, {Nadathur}, {Nair}, {Nandra}, {Neumann}, {Newman}, {Nidever},
  {Nikakhtar}, {Nitschelm}, {O'Connell}, {Garma-Oehmichen}, {Luan Souza de
  Oliveira}, {Olney}, {Oravetz}, {Ortigoza-Urdaneta}, {Osorio}, {Otter},
  {Pace}, {Padilla}, {Pan}, {Pan}, {Parikh}, {Parker}, {Peirani}, {Pe{\~n}a
  Ram{\'\i}rez}, {Penny}, {Percival}, {Perez-Fournon}, {Pinsonneault},
  {Poidevin}, {Poovelil}, {Price-Whelan}, {B{\'a}rbara de Andrade Queiroz},
  {Raddick}, {Ray}, {Rembold}, {Riddle}, {Riffel}, {Riffel}, {Rix}, {Robin},
  {Rodr{\'\i}guez-Puebla}, {Roman-Lopes}, {Rom{\'a}n-Z{\'u}{\~n}iga}, {Rose},
  {Ross}, {Rossi}, {Rubin}, {Salvato}, {S{\'a}nchez}, {S{\'a}nchez-Gallego},
  {Sanderson}, {Santana Rojas}, {Sarceno}, {Sarmiento}, {Sayres}, {Sazonova},
  {Schaefer}, {Schiavon}, {Schlegel}, {Schneider}, {Schultheis}, {Schwope},
  {Serenelli}, {Serna}, {Shao}, {Shapiro}, {Sharma}, {Shen}, {Shetrone}, {Shu},
  {Simon}, {Skrutskie}, {Smethurst}, {Smith}, {Sobeck}, {Spoo}, {Sprague},
  {Stark}, {Stassun}, {Steinmetz}, {Stello}, {Stone-Martinez},
  {Storchi-Bergmann}, {Stringfellow}, {Stutz}, {Su}, {Taghizadeh-Popp},
  {Talbot}, {Tayar}, {Telles}, {Teske}, {Thakar}, {Theissen}, {Tkachenko},
  {Thomas}, {Tojeiro}, {Hernandez Toledo}, {Troup}, {Trump}, {Trussler},
  {Turner}, {Tuttle}, {Unda-Sanzana}, {V{\'a}zquez-Mata}, {Valentini},
  {Valenzuela}, {Vargas-Gonz{\'a}lez}, {Vargas-Maga{\~n}a}, {Alfaro},
  {Villanova}, {Vincenzo}, {Wake}, {Warfield}, {Washington}, {Weaver},
  {Weijmans}, {Weinberg}, {Weiss}, {Westfall}, {Wild}, {Wilde}, {Wilson},
  {Wilson}, {Wilson}, {Wolf}, {Wood-Vasey}, {Yan}, {Zamora}, {Zasowski},
  {Zhang}, {Zhao}, {Zheng}, {Zheng}, \& {Zhu}}]{Abdurro'uf2022ApJS..259...35A}
{Abdurro'uf}, {Accetta}, K., {Aerts}, C., {et~al.} 2022, \apjs, 259, 35

\bibitem[{{Anupama} {et~al.}(2009){Anupama}, {Sahu}, {Gurugubelli}, {Prabhu},
  {Tominaga}, {Tanaka}, \& {Nomoto}}]{Anupama2009MNRAS.392..894A}
{Anupama}, G.~C., {Sahu}, D.~K., {Gurugubelli}, U.~K., {et~al.} 2009, \mnras,
  392, 894

\bibitem[{{Asplund} {et~al.}(2009){Asplund}, {Grevesse}, {Sauval}, \&
  {Scott}}]{Asplund2009ARA&A..47..481A}
{Asplund}, M., {Grevesse}, N., {Sauval}, A.~J., \& {Scott}, P. 2009, \araa, 47,
  481

\bibitem[{{Becker}(2015)}]{Becker2015ascl.soft04004B}
{Becker}, A. 2015, {HOTPANTS: High Order Transform of PSF ANd Template
  Subtraction}, Astrophysics Source Code Library

\bibitem[{{Bellm} {et~al.}(2019){Bellm}, {Kulkarni}, {Graham}, {Dekany},
  {Smith}, {Riddle}, {Masci}, {Helou}, {Prince}, {Adams}, {Barbarino},
  {Barlow}, {Bauer}, {Beck}, {Belicki}, {Biswas}, {Blagorodnova}, {Bodewits},
  {Bolin}, {Brinnel}, {Brooke}, {Bue}, {Bulla}, {Burruss}, {Cenko}, {Chang},
  {Connolly}, {Coughlin}, {Cromer}, {Cunningham}, {De}, {Delacroix}, {Desai},
  {Duev}, {Eadie}, {Farnham}, {Feeney}, {Feindt}, {Flynn}, {Franckowiak},
  {Frederick}, {Fremling}, {Gal-Yam}, {Gezari}, {Giomi}, {Goldstein},
  {Golkhou}, {Goobar}, {Groom}, {Hacopians}, {Hale}, {Henning}, {Ho}, {Hover},
  {Howell}, {Hung}, {Huppenkothen}, {Imel}, {Ip}, {Ivezi{\'c}}, {Jackson},
  {Jones}, {Juric}, {Kasliwal}, {Kaspi}, {Kaye}, {Kelley}, {Kowalski},
  {Kramer}, {Kupfer}, {Landry}, {Laher}, {Lee}, {Lin}, {Lin}, {Lunnan},
  {Giomi}, {Mahabal}, {Mao}, {Miller}, {Monkewitz}, {Murphy}, {Ngeow},
  {Nordin}, {Nugent}, {Ofek}, {Patterson}, {Penprase}, {Porter}, {Rauch},
  {Rebbapragada}, {Reiley}, {Rigault}, {Rodriguez}, {van Roestel}, {Rusholme},
  {van Santen}, {Schulze}, {Shupe}, {Singer}, {Soumagnac}, {Stein}, {Surace},
  {Sollerman}, {Szkody}, {Taddia}, {Terek}, {Van Sistine}, {van Velzen},
  {Vestrand}, {Walters}, {Ward}, {Ye}, {Yu}, {Yan}, \&
  {Zolkower}}]{Bellm2019PASP..131a8002B}
{Bellm}, E.~C., {Kulkarni}, S.~R., {Graham}, M.~J., {et~al.} 2019, \pasp, 131,
  018002

\bibitem[{{Ben-Ami} {et~al.}(2023){Ben-Ami}, {Arcavi}, {Newsome}, {Farah},
  {Pellegrino}, {Terreran}, {Burke}, {Hosseinzadeh}, {McCully}, {Hiramatsu},
  {Gonzalez}, \& {Howell}}]{Ben-Ami2023ApJ...946...30B}
{Ben-Ami}, T., {Arcavi}, I., {Newsome}, M., {et~al.} 2023, \apj, 946, 30

\bibitem[{{Bertin} \& {Arnouts}(1996)}]{Bertin1996A&AS..117..393B}
{Bertin}, E. \& {Arnouts}, S. 1996, \aaps, 117, 393

\bibitem[{{Bildsten} {et~al.}(2007){Bildsten}, {Shen}, {Weinberg}, \&
  {Nelemans}}]{Bildsten2007ApJ...662L..95B}
{Bildsten}, L., {Shen}, K.~J., {Weinberg}, N.~N., \& {Nelemans}, G. 2007,
  \apjl, 662, L95

\bibitem[{{Bostroem} {et~al.}(2023){Bostroem}, {Pearson}, {Shrestha}, {Sand},
  {Valenti}, {Jha}, {Andrews}, {Smith}, {Terreran}, {Green}, {Dong},
  {Lundquist}, {Haislip}, {Hoang}, {Hosseinzadeh}, {Janzen}, {Jencson},
  {Kouprianov}, {Paraskeva}, {Meza Retamal}, {Reichart}, {Arcavi}, {Bonanos},
  {Coughlin}, {Dobson}, {Farah}, {Galbany}, {Guti{\'e}rrez}, {Hawley}, {Hebb},
  {Hiramatsu}, {Howell}, {Iijima}, {Ilyin}, {Jhass}, {McCully}, {Moran},
  {Morris}, {Mura}, {M{\"u}ller-Bravo}, {Munday}, {Newsome}, {Pabst}, {Ochner},
  {Gonzalez}, {Pastorello}, {Pellegrino}, {Piscarreta}, {Ravi}, {Reguitti},
  {Salo}, {Vink{\'o}}, {de Vos}, {Wheeler}, {Williams}, \&
  {Wyatt}}]{Bostroem2023ApJ...956L...5B}
{Bostroem}, K.~A., {Pearson}, J., {Shrestha}, M., {et~al.} 2023, \apjl, 956, L5

\bibitem[{{Brennan} {et~al.}(2024){Brennan}, {Sollerman}, {Irani}, {Schulze},
  {Chen}, {Das}, {De}, {Fransson}, {Gal-Yam}, {Gkini}, {Hinds}, {Lunnan},
  {Perley}, {Qin}, {Stein}, {Wise}, {Yan}, {Zimmerman}, {Anand}, {Bruch},
  {Dekany}, {Drake}, {Fremling}, {Healy}, {Karambelkar}, {Kasliwal}, {Kong},
  {Kulkarni}, {Masci}, {Post}, {Purdum}, {Rich}, \&
  {Wold}}]{Brennan2024A&A...684L..18B}
{Brennan}, S.~J., {Sollerman}, J., {Irani}, I., {et~al.} 2024, \aap, 684, L18

\bibitem[{{Brown} {et~al.}(2013){Brown}, {Baliber}, {Bianco}, {Bowman},
  {Burleson}, {Conway}, {Crellin}, {Depagne}, {De Vera}, {Dilday}, {Dragomir},
  {Dubberley}, {Eastman}, {Elphick}, {Falarski}, {Foale}, {Ford}, {Fulton},
  {Garza}, {Gomez}, {Graham}, {Greene}, {Haldeman}, {Hawkins}, {Haworth},
  {Haynes}, {Hidas}, {Hjelstrom}, {Howell}, {Hygelund}, {Lister}, {Lobdill},
  {Martinez}, {Mullins}, {Norbury}, {Parrent}, {Paulson}, {Petry}, {Pickles},
  {Posner}, {Rosing}, {Ross}, {Sand}, {Saunders}, {Shobbrook}, {Shporer},
  {Street}, {Thomas}, {Tsapras}, {Tufts}, {Valenti}, {Vander Horst}, {Walker},
  {White}, \& {Willis}}]{Brown2013PASP..125.1031B}
{Brown}, T.~M., {Baliber}, N., {Bianco}, F.~B., {et~al.} 2013, \pasp, 125, 1031

\bibitem[{{Bruch} {et~al.}(2023){Bruch}, {Gal-Yam}, {Yaron}, {Chen},
  {Strotjohann}, {Irani}, {Zimmerman}, {Schulze}, {Yang}, {Kim}, {Bulla},
  {Sollerman}, {Rigault}, {Ofek}, {Soumagnac}, {Masci}, {Fremling}, {Perley},
  {Nordin}, {Cenko}, {Ho}, {Adams}, {Adreoni}, {Bellm}, {Blagorodnova},
  {Burdge}, {De}, {Dekany}, {Dhawan}, {Drake}, {Duev}, {Graham}, {Graham},
  {Jencson}, {Karamehmetoglu}, {Kasliwal}, {Kulkarni}, {Miller}, {Neill},
  {Prince}, {Riddle}, {Rusholme}, {Sharma}, {Smith}, {Sravan}, {Taggart},
  {Walters}, \& {Yan}}]{Bruch2023ApJ...952..119B}
{Bruch}, R.~J., {Gal-Yam}, A., {Yaron}, O., {et~al.} 2023, \apj, 952, 119

\bibitem[{{Cai} {et~al.}(2018){Cai}, {Pastorello}, {Fraser}, {Botticella},
  {Gall}, {Arcavi}, {Benetti}, {Cappellaro}, {Elias-Rosa}, {Harmanen},
  {Hosseinzadeh}, {Howell}, {Isern}, {Kangas}, {Kankare}, {Kuncarayakti},
  {Lundqvist}, {Mattila}, {McCully}, {Reynolds}, {Somero}, {Stritzinger}, \&
  {Terreran}}]{Cai2018MNRAS.480.3424C}
{Cai}, Y.~Z., {Pastorello}, A., {Fraser}, M., {et~al.} 2018, \mnras, 480, 3424

\bibitem[{{Cardelli} {et~al.}(1989){Cardelli}, {Clayton}, \&
  {Mathis}}]{Cardelli1989ApJ...345..245C}
{Cardelli}, J.~A., {Clayton}, G.~C., \& {Mathis}, J.~S. 1989, \apj, 345, 245

\bibitem[{{Castro-Segura} {et~al.}(2018){Castro-Segura}, {Pursiainen}, {Smith},
  \& {Yaron}}]{Castro-Segura2018TNSCR2064....1C}
{Castro-Segura}, N., {Pursiainen}, M., {Smith}, M., \& {Yaron}, O. 2018,
  Transient Name Server Classification Report, 2018-2064, 1

\bibitem[{{Chambers} {et~al.}(2016){Chambers}, {Magnier}, {Metcalfe},
  {Flewelling}, {Huber}, {Waters}, {Denneau}, {Draper}, {Farrow}, {Finkbeiner},
  {Holmberg}, {Koppenhoefer}, {Price}, {Rest}, {Saglia}, {Schlafly}, {Smartt},
  {Sweeney}, {Wainscoat}, {Burgett}, {Chastel}, {Grav}, {Heasley}, {Hodapp},
  {Jedicke}, {Kaiser}, {Kudritzki}, {Luppino}, {Lupton}, {Monet}, {Morgan},
  {Onaka}, {Shiao}, {Stubbs}, {Tonry}, {White}, {Ba{\~n}ados}, {Bell},
  {Bender}, {Bernard}, {Boegner}, {Boffi}, {Botticella}, {Calamida},
  {Casertano}, {Chen}, {Chen}, {Cole}, {Deacon}, {Frenk}, {Fitzsimmons},
  {Gezari}, {Gibbs}, {Goessl}, {Goggia}, {Gourgue}, {Goldman}, {Grant},
  {Grebel}, {Hambly}, {Hasinger}, {Heavens}, {Heckman}, {Henderson}, {Henning},
  {Holman}, {Hopp}, {Ip}, {Isani}, {Jackson}, {Keyes}, {Koekemoer}, {Kotak},
  {Le}, {Liska}, {Long}, {Lucey}, {Liu}, {Martin}, {Masci}, {McLean}, {Mindel},
  {Misra}, {Morganson}, {Murphy}, {Obaika}, {Narayan}, {Nieto-Santisteban},
  {Norberg}, {Peacock}, {Pier}, {Postman}, {Primak}, {Rae}, {Rai}, {Riess},
  {Riffeser}, {Rix}, {R{\"o}ser}, {Russel}, {Rutz}, {Schilbach}, {Schultz},
  {Scolnic}, {Strolger}, {Szalay}, {Seitz}, {Small}, {Smith}, {Soderblom},
  {Taylor}, {Thomson}, {Taylor}, {Thakar}, {Thiel}, {Thilker}, {Unger},
  {Urata}, {Valenti}, {Wagner}, {Walder}, {Walter}, {Watters}, {Werner},
  {Wood-Vasey}, \& {Wyse}}]{Chambers2016arXiv161205560C}
{Chambers}, K.~C., {Magnier}, E.~A., {Metcalfe}, N., {et~al.} 2016, arXiv
  e-prints, arXiv:1612.05560

\bibitem[{{Chasovnikov} {et~al.}(2018){Chasovnikov}, {Lipunov}, {Kornilov},
  {Gorbovskoy}, {Tiurina}, {Kuznetsov}, {Vladimirov}, {Balanutsa}, {Chazov},
  {Zimnukhov}, {Vlasenko}, {Podesta}, {Podesta}, {Francile}, {Levato}, {Gress},
  {Budnev}, {Gabovich}, {Shumkov}, {Pogrosheva}, {Tlatov}, \&
  {Senik}}]{Chasovnikov2018TNSTR1888....1C}
{Chasovnikov}, A., {Lipunov}, V., {Kornilov}, D.~V., {et~al.} 2018, Transient
  Name Server Discovery Report, 2018-1888, 1

\bibitem[{{Chen} {et~al.}(2018){Chen}, {Inserra}, {Fraser}, {Moriya}, {Schady},
  {Schweyer}, {Filippenko}, {Perley}, {Ruiter}, {Seitenzahl}, {Sollerman},
  {Taddia}, {Anderson}, {Foley}, {Jerkstrand}, {Ngeow}, {Pan}, {Pastorello},
  {Points}, {Smartt}, {Smith}, {Taubenberger}, {Wiseman}, {Young}, {Benetti},
  {Berton}, {Bufano}, {Clark}, {Della Valle}, {Galbany}, {Gal-Yam},
  {Gromadzki}, {Guti{\'e}rrez}, {Heinze}, {Kankare}, {Kilpatrick},
  {Kuncarayakti}, {Leloudas}, {Lin}, {Maguire}, {Mazzali}, {McBrien},
  {Prentice}, {Rau}, {Rest}, {Siebert}, {Stalder}, {Tonry}, \&
  {Yu}}]{Chen2018ApJ...867L..31C}
{Chen}, T.~W., {Inserra}, C., {Fraser}, M., {et~al.} 2018, \apjl, 867, L31

\bibitem[{{Chevalier} \& {Fransson}(1994)}]{Chevalier1994ApJ...420..268C}
{Chevalier}, R.~A. \& {Fransson}, C. 1994, \apj, 420, 268

\bibitem[{{Chugai}(1997)}]{Chugai1997Ap&SS.252..225C}
{Chugai}, N.~N. 1997, \apss, 252, 225

\bibitem[{{Chugai}(2009)}]{Chugai2009MNRAS.400..866C}
{Chugai}, N.~N. 2009, \mnras, 400, 866

\bibitem[{{Clark} {et~al.}(2020){Clark}, {Maguire}, {Inserra}, {Prentice},
  {Smartt}, {Contreras}, {Hossenizadeh}, {Hsiao}, {Kankare}, {Kasliwal},
  {Nugent}, {Shahbandeh}, {Baltay}, {Rabinowitz}, {Arcavi}, {Ashall}, {Burns},
  {Callis}, {Chen}, {Diamond}, {Fraser}, {Howell}, {Karamehmetoglu}, {Kotak},
  {Lyman}, {Morrell}, {Phillips}, {Pignata}, {Pursiainen}, {Sollerman},
  {Stritzinger}, {Sullivan}, \& {Young}}]{Clark2020MNRAS.492.2208C}
{Clark}, P., {Maguire}, K., {Inserra}, C., {et~al.} 2020, \mnras, 492, 2208

\bibitem[{{Cooke} {et~al.}(2010){Cooke}, {Ellis}, {Nugent}, {Howell},
  {Sullivan}, \& {Gal-Yam}}]{Cooke2010ATel.2491....1C}
{Cooke}, J., {Ellis}, R.~S., {Nugent}, P.~E., {et~al.} 2010, The Astronomer's
  Telegram, 2491, 1

\bibitem[{{Crawford} {et~al.}(2010){Crawford}, {Still}, {Schellart}, {Balona},
  {Buckley}, {Dugmore}, {Gulbis}, {Kniazev}, {Kotze}, {Loaring}, {Nordsieck},
  {Pickering}, {Potter}, {Romero Colmenero}, {Vaisanen}, {Williams}, \&
  {Zietsman}}]{Crawford2010SPIE.7737E..25C}
{Crawford}, S.~M., {Still}, M., {Schellart}, P., {et~al.} 2010, in Society of
  Photo-Optical Instrumentation Engineers (SPIE) Conference Series, Vol. 7737,
  Observatory Operations: Strategies, Processes, and Systems III, ed. D.~R.
  {Silva}, A.~B. {Peck}, \& B.~T. {Soifer}, 773725

\bibitem[{{Dessart}(2024)}]{Dessart2024arXiv240504259D}
{Dessart}, L. 2024, arXiv e-prints, arXiv:2405.04259

\bibitem[{{Di Carlo} {et~al.}(2008){Di Carlo}, {Corsi}, {Arkharov}, {Massi},
  {Larionov}, {Efimova}, {Dolci}, {Napoleone}, \& {Di
  Paola}}]{DiCarlo2008ApJ...684..471D}
{Di Carlo}, E., {Corsi}, C., {Arkharov}, A.~A., {et~al.} 2008, \apj, 684, 471

\bibitem[{{Dintinjana} {et~al.}(2011){Dintinjana}, {Mikuz}, {Skvarc}, {Koff},
  {Valenti}, {Pastorello}, {Benetti}, {Tomasella}, {Bufano}, \&
  {Ochner}}]{Dintinjana2011CBET.2906....1D}
{Dintinjana}, B., {Mikuz}, H., {Skvarc}, J., {et~al.} 2011, Central Bureau
  Electronic Telegrams, 2906, 1

\bibitem[{{Drout} {et~al.}(2011){Drout}, {Soderberg}, {Gal-Yam}, {Cenko},
  {Fox}, {Leonard}, {Sand}, {Moon}, {Arcavi}, \&
  {Green}}]{Drout2011ApJ...741...97D}
{Drout}, M.~R., {Soderberg}, A.~M., {Gal-Yam}, A., {et~al.} 2011, \apj, 741, 97

\bibitem[{{Fesen} {et~al.}(2017){Fesen}, {Weil}, {Hamilton}, \&
  {H{\"o}flich}}]{Fesen2017ApJ...848..130F}
{Fesen}, R.~A., {Weil}, K.~E., {Hamilton}, A. J.~S., \& {H{\"o}flich}, P.~A.
  2017, \apj, 848, 130

\bibitem[{{Filippenko}(1997)}]{Filippenko1997ARA&A..35..309F}
{Filippenko}, A.~V. 1997, \araa, 35, 309

\bibitem[{{Fox} {et~al.}(2011){Fox}, {Chevalier}, {Skrutskie}, {Soderberg},
  {Filippenko}, {Ganeshalingam}, {Silverman}, {Smith}, \&
  {Steele}}]{Fox2011ApJ...741....7F}
{Fox}, O.~D., {Chevalier}, R.~A., {Skrutskie}, M.~F., {et~al.} 2011, \apj, 741,
  7

\bibitem[{{Fransson} {et~al.}(2002){Fransson}, {Chevalier}, {Filippenko},
  {Leibundgut}, {Barth}, {Fesen}, {Kirshner}, {Leonard}, {Li}, {Lundqvist},
  {Sollerman}, \& {Van Dyk}}]{Fransson2002ApJ...572..350F}
{Fransson}, C., {Chevalier}, R.~A., {Filippenko}, A.~V., {et~al.} 2002, \apj,
  572, 350

\bibitem[{{Fraser}(2020)}]{Fraser2020RSOS....700467F}
{Fraser}, M. 2020, Royal Society Open Science, 7, 200467

\bibitem[{{Fraser} {et~al.}(2021){Fraser}, {Stritzinger}, {Brennan},
  {Pastorello}, {Cai}, {Piro}, {Ashall}, {Brown}, {Burns}, {Elias-Rosa},
  {Kotak}, {Filippenko}, {Galbany}, {Hsiao}, {Jha}, {Reguitti}, {Zhang},
  {Moran}, {Morrell}, {Shappee}, {Tomasella}, {Anderson}, {Barna}, {Ochner},
  {Phillips}, {Tucker}, {Wang}, {Baron}, {Benetti}, {Bersten}, {Brink},
  {Camacho-Neves}, {Davis}, {Dettman}, {Folatelli}, {Gutierrez}, {Hoflich},
  {Holoien}, {Kankare}, {Kumar}, {Lu}, {Mazzali}, {Taubenberger}, {Tinyanont},
  {Kuncarayakti}, {Kwok}, {Shahbandeh}, {Suntzeff}, {Yan}, {Yang}, \&
  {Zheng}}]{Fraser2021arXiv210807278F}
{Fraser}, M., {Stritzinger}, M.~D., {Brennan}, S.~J., {et~al.} 2021, arXiv
  e-prints, arXiv:2108.07278

\bibitem[{{Fuller}(2017)}]{Fuller2017MNRAS.470.1642F}
{Fuller}, J. 2017, \mnras, 470, 1642

\bibitem[{{Fuller} \& {Ro}(2018)}]{Fuller2018MNRAS.476.1853F}
{Fuller}, J. \& {Ro}, S. 2018, \mnras, 476, 1853

\bibitem[{Gal-Yam(2017)}]{Gal-Yam2017hsn..book..195G}
Gal-Yam, A. 2017, Observational and Physical Classification of Supernovae
  (Springer International Publishing), 195–237

\bibitem[{{Gal-Yam} {et~al.}(2014){Gal-Yam}, {Arcavi}, {Ofek}, {Ben-Ami},
  {Cenko}, {Kasliwal}, {Cao}, {Yaron}, {Tal}, {Silverman}, {Horesh}, {De Cia},
  {Taddia}, {Sollerman}, {Perley}, {Vreeswijk}, {Kulkarni}, {Nugent},
  {Filippenko}, \& {Wheeler}}]{Gal-Yam2014Natur.509..471G}
{Gal-Yam}, A., {Arcavi}, I., {Ofek}, E.~O., {et~al.} 2014, \nat, 509, 471

\bibitem[{{Gal-Yam} {et~al.}(2022){Gal-Yam}, {Bruch}, {Schulze}, {Yang},
  {Perley}, {Irani}, {Sollerman}, {Kool}, {Soumagnac}, {Yaron}, {Strotjohann},
  {Zimmerman}, {Barbarino}, {Kulkarni}, {Kasliwal}, {De}, {Yao}, {Fremling},
  {Yan}, {Ofek}, {Fransson}, {Filippenko}, {Zheng}, {Brink}, {Copperwheat},
  {Foley}, {Brown}, {Siebert}, {Leloudas}, {Cabrera-Lavers}, {Garcia-Alvarez},
  {Marante-Barreto}, {Frederick}, {Hung}, {Wheeler}, {Vink{\'o}}, {Thomas},
  {Graham}, {Duev}, {Drake}, {Dekany}, {Bellm}, {Rusholme}, {Shupe},
  {Andreoni}, {Sharma}, {Riddle}, {van Roestel}, \&
  {Knezevic}}]{Gal-Yam2022Natur.601..201G}
{Gal-Yam}, A., {Bruch}, R., {Schulze}, S., {et~al.} 2022, \nat, 601, 201

\bibitem[{{Gan} {et~al.}(2021){Gan}, {Wang}, \&
  {Liang}}]{Gan2021ApJ...914..125G}
{Gan}, W.-P., {Wang}, S.-Q., \& {Liang}, E.-W. 2021, \apj, 914, 125

\bibitem[{{Gangopadhyay} {et~al.}(2020){Gangopadhyay}, {Misra}, {Hiramatsu},
  {Wang}, {Hosseinzadeh}, {Wang}, {Valenti}, {Zhang}, {Howell}, {Arcavi},
  {Anupama}, {Burke}, {Dastidar}, {Itagaki}, {Kumar}, {Kumar}, {Li}, {McCully},
  {Mo}, {Pandey}, {Pellegrino}, {Sai}, {Sahu}, {Sanwal}, {Singh}, {Singh},
  {Zhang}, {Zhang}, \& {Zhang}}]{Gangopadhyay2020ApJ...889..170G}
{Gangopadhyay}, A., {Misra}, K., {Hiramatsu}, D., {et~al.} 2020, \apj, 889, 170

\bibitem[{{Gangopadhyay} {et~al.}(2022){Gangopadhyay}, {Misra}, {Hosseinzadeh},
  {Arcavi}, {Pellegrino}, {Wang}, {Andrew Howell}, {Burke}, {Zhang},
  {Kawabata}, {Singh}, {Dastidar}, {Hiramatsu}, {McCully}, {Mo}, {Chen}, \&
  {Xiang}}]{Gangopadhyay2022ApJ...930..127G}
{Gangopadhyay}, A., {Misra}, K., {Hosseinzadeh}, G., {et~al.} 2022, \apj, 930,
  127

\bibitem[{{Gonz{\'a}lez-Gait{\'a}n} {et~al.}(2015){Gonz{\'a}lez-Gait{\'a}n},
  {Tominaga}, {Molina}, {Galbany}, {Bufano}, {Anderson}, {Gutierrez},
  {F{\"o}rster}, {Pignata}, {Bersten}, {Howell}, {Sullivan}, {Carlberg}, {de
  Jaeger}, {Hamuy}, {Baklanov}, \&
  {Blinnikov}}]{Gonzalez-Gaitan2015MNRAS.451.2212G}
{Gonz{\'a}lez-Gait{\'a}n}, S., {Tominaga}, N., {Molina}, J., {et~al.} 2015,
  \mnras, 451, 2212

\bibitem[{{Gorbikov} {et~al.}(2014){Gorbikov}, {Gal-Yam}, {Ofek}, {Vreeswijk},
  {Nugent}, {Chotard}, {Kulkarni}, {Cao}, {De Cia}, {Yaron}, {Tal}, {Arcavi},
  {Kasliwal}, {Cenko}, {Sullivan}, \& {Chen}}]{Gorbikov2014MNRAS.443..671G}
{Gorbikov}, E., {Gal-Yam}, A., {Ofek}, E.~O., {et~al.} 2014, \mnras, 443, 671

\bibitem[{{Gorbovskoy} {et~al.}(2013){Gorbovskoy}, {Lipunov}, {Kornilov},
  {Belinski}, {Kuvshinov}, {Tyurina}, {Sankovich}, {Krylov}, {Shatskiy},
  {Balanutsa}, {Chazov}, {Kuznetsov}, {Zimnukhov}, {Shumkov}, {Shurpakov},
  {Senik}, {Gareeva}, {Pruzhinskaya}, {Tlatov}, {Parkhomenko}, {Dormidontov},
  {Krushinsky}, {Punanova}, {Zalozhnyh}, {Popov}, {Burdanov}, {Yazev},
  {Budnev}, {Ivanov}, {Konstantinov}, {Gress}, {Chuvalaev}, {Yurkov},
  {Sergienko}, {Kudelina}, {Sinyakov}, {Karachentsev}, {Moiseev}, \&
  {Fatkhullin}}]{Gorbovskoy2013ARep...57..233G}
{Gorbovskoy}, E.~S., {Lipunov}, V.~M., {Kornilov}, V.~G., {et~al.} 2013,
  Astronomy Reports, 57, 233

\bibitem[{{Greiner} {et~al.}(2008){Greiner}, {Bornemann}, {Clemens}, {Deuter},
  {Hasinger}, {Honsberg}, {Huber}, {Huber}, {Krauss}, {Kr{\"u}hler},
  {K{\"u}pc{\"u} Yolda{\c{s}}}, {Mayer-Hasselwander}, {Mican}, {Primak},
  {Schrey}, {Steiner}, {Szokoly}, {Th{\"o}ne}, {Yolda{\c{s}}}, {Klose}, {Laux},
  \& {Winkler}}]{Greiner2008PASP..120..405G}
{Greiner}, J., {Bornemann}, W., {Clemens}, C., {et~al.} 2008, \pasp, 120, 405

\bibitem[{{Hart} {et~al.}(2023){Hart}, {Shappee}, {Hey}, {Kochanek}, {Stanek},
  {Lim}, {Dobbs}, {Tucker}, {Jayasinghe}, {Beacom}, {Boright}, {Holoien},
  {Ong}, {Prieto}, {Thompson}, \& {Will}}]{Hart2023arXiv230403791H}
{Hart}, K., {Shappee}, B.~J., {Hey}, D., {et~al.} 2023, arXiv e-prints,
  arXiv:2304.03791

\bibitem[{{Heger} {et~al.}(2003){Heger}, {Fryer}, {Woosley}, {Langer}, \&
  {Hartmann}}]{Heger2003ApJ...591..288H}
{Heger}, A., {Fryer}, C.~L., {Woosley}, S.~E., {Langer}, N., \& {Hartmann},
  D.~H. 2003, \apj, 591, 288

\bibitem[{{Hosseinzadeh} {et~al.}(2017){Hosseinzadeh}, {Arcavi}, {Valenti},
  {McCully}, {Howell}, {Johansson}, {Sollerman}, {Pastorello}, {Benetti},
  {Cao}, {Cenko}, {Clubb}, {Corsi}, {Duggan}, {Elias-Rosa}, {Filippenko},
  {Fox}, {Fremling}, {Horesh}, {Karamehmetoglu}, {Kasliwal}, {Marion}, {Ofek},
  {Sand}, {Taddia}, {Zheng}, {Fraser}, {Gal-Yam}, {Inserra}, {Laher}, {Masci},
  {Rebbapragada}, {Smartt}, {Smith}, {Sullivan}, {Surace}, \&
  {Wo{\'z}niak}}]{Hosseinzadeh2017ApJ...836..158H}
{Hosseinzadeh}, G., {Arcavi}, I., {Valenti}, S., {et~al.} 2017, \apj, 836, 158

\bibitem[{{Hosseinzadeh} {et~al.}(2019){Hosseinzadeh}, {McCully}, {Zabludoff},
  {Arcavi}, {French}, {Howell}, {Berger}, \&
  {Hiramatsu}}]{Hosseinzadeh2019ApJ...871L...9H}
{Hosseinzadeh}, G., {McCully}, C., {Zabludoff}, A.~I., {et~al.} 2019, \apjl,
  871, L9

\bibitem[{{Hunter} {et~al.}(2009){Hunter}, {Valenti}, {Kotak}, {Meikle},
  {Taubenberger}, {Pastorello}, {Benetti}, {Stanishev}, {Smartt}, {Trundle},
  {Arkharov}, {Bufano}, {Cappellaro}, {Di Carlo}, {Dolci}, {Elias-Rosa},
  {Frandsen}, {Fynbo}, {Hopp}, {Larionov}, {Laursen}, {Mazzali}, {Navasardyan},
  {Ries}, {Riffeser}, {Rizzi}, {Tsvetkov}, {Turatto}, \&
  {Wilke}}]{Hunter2009A&A...508..371H}
{Hunter}, D.~J., {Valenti}, S., {Kotak}, R., {et~al.} 2009, \aap, 508, 371

\bibitem[{{Jacobson-Gal{\'a}n}
  {et~al.}(2024{\natexlab{a}}){Jacobson-Gal{\'a}n}, {Davis}, {Kilpatrick},
  {Dessart}, {Margutti}, {Chornock}, {Foley}, {Arunachalam}, {Auchettl}, {Bom},
  {Cartier}, {Coulter}, {Dimitriadis}, {Dickinson}, {Drout}, {Gagliano},
  {Gall}, {Garretson}, {Izzo}, {Jones}, {LeBaron}, {Miao}, {Milisavljevic},
  {Pan}, {Rest}, {Rojas-Bravo}, {Santos}, {Sears}, {Subrayan}, {Taggart}, \&
  {Tinyanont}}]{Jacobson2024arXiv240419006J}
{Jacobson-Gal{\'a}n}, W.~V., {Davis}, K.~W., {Kilpatrick}, C.~D., {et~al.}
  2024{\natexlab{a}}, arXiv e-prints, arXiv:2404.19006

\bibitem[{{Jacobson-Gal{\'a}n}
  {et~al.}(2024{\natexlab{b}}){Jacobson-Gal{\'a}n}, {Dessart}, {Davis},
  {Kilpatrick}, {Margutti}, {Foley}, {Chornock}, {Terreran}, {Hiramatsu},
  {Newsome}, {Padilla Gonzalez}, {Pellegrino}, {Howell}, {Filippenko},
  {Anderson}, {Angus}, {Auchettl}, {Bostroem}, {Brink}, {Cartier}, {Coulter},
  {de Boer}, {Drout}, {Earl}, {Ertini}, {Farah}, {Farias}, {Gall}, {Gao},
  {Gerlach}, {Guo}, {Haynie}, {Hosseinzadeh}, {Ibik}, {Jha}, {Jones},
  {Langeroodi}, {LeBaron}, {Magnier}, {Piro}, {Raimundo}, {Rest}, {Rest},
  {Rich}, {Rojas-Bravo}, {Sears}, {Taggart}, {Villar}, {Wainscoat}, {Wang},
  {Wasserman}, {Yan}, {Yang}, {Zhang}, \&
  {Zheng}}]{Jacobson2024arXiv240302382J}
{Jacobson-Gal{\'a}n}, W.~V., {Dessart}, L., {Davis}, K.~W., {et~al.}
  2024{\natexlab{b}}, arXiv e-prints, arXiv:2403.02382

\bibitem[{{Jayasinghe} {et~al.}(2019){Jayasinghe}, {Stanek}, {Kochanek},
  {Shappee}, {Holoien}, {Thompson}, {Prieto}, {Dong}, {Pawlak}, {Pejcha},
  {Shields}, {Pojmanski}, {Otero}, {Hurst}, {Britt}, \&
  {Will}}]{Jayasinghe2019MNRAS.485..961J}
{Jayasinghe}, T., {Stanek}, K.~Z., {Kochanek}, C.~S., {et~al.} 2019, \mnras,
  485, 961

\bibitem[{{Karamehmetoglu} {et~al.}(2021){Karamehmetoglu}, {Fransson},
  {Sollerman}, {Tartaglia}, {Taddia}, {De}, {Fremling}, {Bagdasaryan},
  {Barbarino}, {Bellm}, {Dekany}, {Dugas}, {Giomi}, {Goobar}, {Graham}, {Ho},
  {Laher}, {Masci}, {Neill}, {Perley}, {Riddle}, {Rusholme}, \&
  {Soumagnac}}]{Karamehmetoglu2021AA...649A.163K}
{Karamehmetoglu}, E., {Fransson}, C., {Sollerman}, J., {et~al.} 2021, \aap,
  649, A163

\bibitem[{{Karamehmetoglu} {et~al.}(2017){Karamehmetoglu}, {Taddia},
  {Sollerman}, {Wyrzykowski}, {Schmidl}, {Fraser}, {Fremling}, {Greiner},
  {Inserra}, {Kostrzewa-Rutkowska}, {Maguire}, {Smartt}, {Sullivan}, \&
  {Young}}]{Karamehmetoglu2017AA...602A..93K}
{Karamehmetoglu}, E., {Taddia}, F., {Sollerman}, J., {et~al.} 2017, \aap, 602,
  A93

\bibitem[{{Kasliwal} {et~al.}(2010){Kasliwal}, {Kulkarni}, {Gal-Yam}, {Yaron},
  {Quimby}, {Ofek}, {Nugent}, {Poznanski}, {Jacobsen}, {Sternberg}, {Arcavi},
  {Howell}, {Sullivan}, {Rich}, {Burke}, {Brimacombe}, {Milisavljevic},
  {Fesen}, {Bildsten}, {Shen}, {Cenko}, {Bloom}, {Hsiao}, {Law}, {Gehrels},
  {Immler}, {Dekany}, {Rahmer}, {Hale}, {Smith}, {Zolkower}, {Velur},
  {Walters}, {Henning}, {Bui}, \& {McKenna}}]{Kasliwal2010ApJ...723L..98K}
{Kasliwal}, M.~M., {Kulkarni}, S.~R., {Gal-Yam}, A., {et~al.} 2010, \apjl, 723,
  L98

\bibitem[{{Kochanek} {et~al.}(2017){Kochanek}, {Shappee}, {Stanek}, {Holoien},
  {Thompson}, {Prieto}, {Dong}, {Shields}, {Will}, {Britt}, {Perzanowski}, \&
  {Pojma{\'n}ski}}]{Kochanek2017PASP..129j4502K}
{Kochanek}, C.~S., {Shappee}, B.~J., {Stanek}, K.~Z., {et~al.} 2017, \pasp,
  129, 104502

\bibitem[{{Kool} {et~al.}(2021){Kool}, {Karamehmetoglu}, {Sollerman},
  {Schulze}, {Lunnan}, {Reynolds}, {Barbarino}, {Bellm}, {De}, {Duev},
  {Fremling}, {Golkhou}, {Graham}, {Green}, {Horesh}, {Kaye}, {Kim}, {Laher},
  {Masci}, {Nordin}, {Perley}, {Phinney}, {Porter}, {Reiley}, {Rodriguez}, {van
  Roestel}, {Rusholme}, {Sharma}, {Sfaradi}, {Soumagnac}, {Taggart},
  {Tartaglia}, {Williams}, \& {Yan}}]{Kool2021AA...652A.136K}
{Kool}, E.~C., {Karamehmetoglu}, E., {Sollerman}, J., {et~al.} 2021, \aap, 652,
  A136

\bibitem[{{Kr{\"u}hler} {et~al.}(2008){Kr{\"u}hler}, {K{\"u}pc{\"u}
  Yolda{\c{s}}}, {Greiner}, {Clemens}, {McBreen}, {Primak}, {Savaglio},
  {Yolda{\c{s}}}, {Szokoly}, \& {Klose}}]{Kruhler2008ApJ...685..376K}
{Kr{\"u}hler}, T., {K{\"u}pc{\"u} Yolda{\c{s}}}, A., {Greiner}, J., {et~al.}
  2008, \apj, 685, 376

\bibitem[{{Kulkarni} {et~al.}(2021){Kulkarni}, {Harrison}, {Grefenstette},
  {Earnshaw}, {Andreoni}, {Berg}, {Bloom}, {Cenko}, {Chornock}, {Christiansen},
  {Coughlin}, {Wuollet Criswell}, {Darvish}, {Das}, {De}, {Dessart}, {Dixon},
  {Dorsman}, {El-Badry}, {Evans}, {Ford}, {Fremling}, {Gansicke}, {Gezari},
  {Goetberg}, {Green}, {Graham}, {Heida}, {Ho}, {Jaodand}, {Johns-Krull},
  {Kasliwal}, {Lazzarini}, {Lu}, {Margutti}, {Martin}, {Masters}, {McKernan},
  {Naze}, {Nissanke}, {Parazin}, {Perley}, {Phinney}, {Piro}, {Raaijmakers},
  {Rauw}, {Rodriguez}, {Sana}, {Senchyna}, {Singer}, {Spake}, {Stassun},
  {Stern}, {Teplitz}, {Weisz}, \& {Yao}}]{Kulkarni2021arXiv211115608K}
{Kulkarni}, S.~R., {Harrison}, F.~A., {Grefenstette}, B.~W., {et~al.} 2021,
  arXiv e-prints, arXiv:2111.15608

\bibitem[{{Kuncarayakti} {et~al.}(2013){Kuncarayakti}, {Doi}, {Aldering},
  {Arimoto}, {Maeda}, {Morokuma}, {Pereira}, {Usuda}, \&
  {Hashiba}}]{Kuncarayakti2013AJ....146...30K}
{Kuncarayakti}, H., {Doi}, M., {Aldering}, G., {et~al.} 2013, \aj, 146, 30

\bibitem[{{Landolt}(1992)}]{Landolt1992AJ....104..340L}
{Landolt}, A.~U. 1992, \aj, 104, 340

\bibitem[{{Langer}(2012)}]{Langer2012ARA&A..50..107L}
{Langer}, N. 2012, \araa, 50, 107

\bibitem[{{Lipunov} {et~al.}(2012){Lipunov}, {Kornilov}, {Gorbovskoy},
  {Belinski}, {Kuvshinov}, {Tyurina}, {Krylov}, {Shatsky}, {Balanutsa},
  {Chazov}, {Kuznetsov}, {Zimnuhov}, {Tlatov}, {Parkhomenko}, {Dormidontov},
  {Krushinsky}, {Zalozhnyh}, {Popov}, {Yazev}, {Budnev}, {Ivanov},
  {Konstantinov}, {Gress}, {Chvalaev}, {Yurkov}, {Sergienko}, \&
  {Kudelina}}]{Lipunov2012ASInC...7..275L}
{Lipunov}, V., {Kornilov}, V., {Gorbovskoy}, E., {et~al.} 2012, in Astronomical
  Society of India Conference Series, Vol.~7, Astronomical Society of India
  Conference Series, 275

\bibitem[{{Loveday}(1996)}]{Loveday1996MNRAS.278.1025L}
{Loveday}, J. 1996, \mnras, 278, 1025

\bibitem[{{Lyman} {et~al.}(2016){Lyman}, {Bersier}, {James}, {Mazzali},
  {Eldridge}, {Fraser}, \& {Pian}}]{Lyman2016MNRAS.457..328L}
{Lyman}, J.~D., {Bersier}, D., {James}, P.~A., {et~al.} 2016, \mnras, 457, 328

\bibitem[{{Maeda} \& {Moriya}(2022)}]{Maeda2022ApJ...927...25M}
{Maeda}, K. \& {Moriya}, T.~J. 2022, \apj, 927, 25

\bibitem[{{Maeda} {et~al.}(2007){Maeda}, {Tanaka}, {Nomoto}, {Tominaga},
  {Kawabata}, {Mazzali}, {Umeda}, {Suzuki}, \&
  {Hattori}}]{Maeda2007ApJ...666.1069M}
{Maeda}, K., {Tanaka}, M., {Nomoto}, K., {et~al.} 2007, \apj, 666, 1069

\bibitem[{{Matheson} {et~al.}(2000){Matheson}, {Filippenko}, {Chornock},
  {Leonard}, \& {Li}}]{Matheson2000AJ....119.2303M}
{Matheson}, T., {Filippenko}, A.~V., {Chornock}, R., {Leonard}, D.~C., \& {Li},
  W. 2000, \aj, 119, 2303

\bibitem[{{Mattila} {et~al.}(2008){Mattila}, {Meikle}, {Lundqvist},
  {Pastorello}, {Kotak}, {Eldridge}, {Smartt}, {Adamson}, {Gerardy}, {Rizzi},
  {Stephens}, \& {van Dyk}}]{Mattila2008MNRAS.389..141M}
{Mattila}, S., {Meikle}, W.~P.~S., {Lundqvist}, P., {et~al.} 2008, \mnras, 389,
  141

\bibitem[{{Maund} {et~al.}(2016){Maund}, {Pastorello}, {Mattila}, {Itagaki}, \&
  {Boles}}]{Maund2016ApJ...833..128M}
{Maund}, J.~R., {Pastorello}, A., {Mattila}, S., {Itagaki}, K., \& {Boles}, T.
  2016, \apj, 833, 128

\bibitem[{{Meza} \& {Anderson}(2020)}]{Meza2020A&A...641A.177M}
{Meza}, N. \& {Anderson}, J.~P. 2020, \aap, 641, A177

\bibitem[{{Moriya} {et~al.}(2010){Moriya}, {Tominaga}, {Tanaka}, {Nomoto},
  {Sauer}, {Mazzali}, {Maeda}, \& {Suzuki}}]{Moriya2010ApJ...719.1445M}
{Moriya}, T., {Tominaga}, N., {Tanaka}, M., {et~al.} 2010, \apj, 719, 1445

\bibitem[{{Moriya} {et~al.}(2013){Moriya}, {Maeda}, {Taddia}, {Sollerman},
  {Blinnikov}, \& {Sorokina}}]{Moriya2013MNRAS.435.1520M}
{Moriya}, T.~J., {Maeda}, K., {Taddia}, F., {et~al.} 2013, \mnras, 435, 1520

\bibitem[{{Morokuma} {et~al.}(2014){Morokuma}, {Shibata}, {Matsumoto},
  {Tsvetkov}, {Pavlyuk}, {Fukuda}, {Tominaga}, {Tanaka}, {Stritzinger},
  {Hsiao}, {Taddia}, \& {Pastorello}}]{Morokuma2014CBET.3894....1M}
{Morokuma}, T., {Shibata}, T., {Matsumoto}, E., {et~al.} 2014, Central Bureau
  Electronic Telegrams, 3894, 1

\bibitem[{{Mould} {et~al.}(2000){Mould}, {Huchra}, {Freedman}, {Kennicutt},
  {Ferrarese}, {Ford}, {Gibson}, {Graham}, {Hughes}, {Illingworth}, {Kelson},
  {Macri}, {Madore}, {Sakai}, {Sebo}, {Silbermann}, \&
  {Stetson}}]{Mould2000ApJ...529..786M}
{Mould}, J.~R., {Huchra}, J.~P., {Freedman}, W.~L., {et~al.} 2000, \apj, 529,
  786

\bibitem[{{Moustakas} {et~al.}(2023){Moustakas}, {Lang}, {Dey}, {Juneau},
  {Meisner}, {Myers}, {Schlafly}, {Schlegel}, {Valdes}, {Weaver}, \&
  {Zhou}}]{Moustakas2023ApJS..269....3M}
{Moustakas}, J., {Lang}, D., {Dey}, A., {et~al.} 2023, \apjs, 269, 3

\bibitem[{{Nagao} {et~al.}(2023){Nagao}, {Kuncarayakti}, {Maeda}, {Moore},
  {Pastorello}, {Mattila}, {Uno}, {Smartt}, {Sim}, {Ferrari}, {Tomasella},
  {Anderson}, {Chen}, {Galbany}, {Gao}, {Gromadzki}, {Guti{\'e}rrez},
  {Inserra}, {Kankare}, {Magnier}, {M{\"u}ller-Bravo}, {Reguitti}, \&
  {Young}}]{Nagao2023A&A...673A..27N}
{Nagao}, T., {Kuncarayakti}, H., {Maeda}, K., {et~al.} 2023, \aap, 673, A27

\bibitem[{{Nicholls} \& {Stanek}(2019)}]{Nicholls2019TNSTR..20....1N}
{Nicholls}, B. \& {Stanek}, K.~Z. 2019, Transient Name Server Discovery Report,
  2019-20, 1

\bibitem[{{Ouchi} {et~al.}(2021){Ouchi}, {Maeda}, {Anderson}, \&
  {Sawada}}]{Ouchi2021ApJ...922..141O}
{Ouchi}, R., {Maeda}, K., {Anderson}, J.~P., \& {Sawada}, R. 2021, \apj, 922,
  141

\bibitem[{{Pastorello} {et~al.}(2015{\natexlab{a}}){Pastorello}, {Benetti},
  {Brown}, {Tsvetkov}, {Inserra}, {Taubenberger}, {Tomasella}, {Fraser},
  {Rich}, {Botticella}, {Bufano}, {Cappellaro}, {Ergon}, {Gorbovskoy},
  {Harutyunyan}, {Huang}, {Kotak}, {Lipunov}, {Magill}, {Miluzio}, {Morrell},
  {Ochner}, {Smartt}, {Sollerman}, {Spiro}, {Stritzinger}, {Turatto},
  {Valenti}, {Wang}, {Wright}, {Yurkov}, {Zampieri}, \&
  {Zhang}}]{Pastorello2015MNRAS.449.1921P}
{Pastorello}, A., {Benetti}, S., {Brown}, P.~J., {et~al.} 2015{\natexlab{a}},
  \mnras, 449, 1921

\bibitem[{{Pastorello} {et~al.}(2015{\natexlab{b}}){Pastorello}, {Hadjiyska},
  {Rabinowitz}, {Valenti}, {Turatto}, {Fasano}, {Benitez-Herrera}, {Baltay},
  {Benetti}, {Botticella}, {Cappellaro}, {Elias-Rosa}, {Ellman}, {Feindt},
  {Filippenko}, {Fraser}, {Gal-Yam}, {Graham}, {Howell}, {Inserra}, {Kelly},
  {Kotak}, {Kowalski}, {McKinnon}, {Morales-Garoffolo}, {Nugent}, {Smartt},
  {Smith}, {Stritzinger}, {Sullivan}, {Taubenberger}, {Walker}, {Yaron}, \&
  {Young}}]{Pastorello2015MNRAS.449.1954P}
{Pastorello}, A., {Hadjiyska}, E., {Rabinowitz}, D., {et~al.}
  2015{\natexlab{b}}, \mnras, 449, 1954

\bibitem[{{Pastorello} {et~al.}(2008{\natexlab{a}}){Pastorello}, {Mattila},
  {Zampieri}, {Della Valle}, {Smartt}, {Valenti}, {Agnoletto}, {Benetti},
  {Benn}, {Branch}, {Cappellaro}, {Dennefeld}, {Eldridge}, {Gal-Yam},
  {Harutyunyan}, {Hunter}, {Kjeldsen}, {Lipkin}, {Mazzali}, {Milne},
  {Navasardyan}, {Ofek}, {Pian}, {Shemmer}, {Spiro}, {Stathakis},
  {Taubenberger}, {Turatto}, \& {Yamaoka}}]{Pastorello2008MNRAS.389..113P}
{Pastorello}, A., {Mattila}, S., {Zampieri}, L., {et~al.} 2008{\natexlab{a}},
  \mnras, 389, 113

\bibitem[{{Pastorello} {et~al.}(2015{\natexlab{c}}){Pastorello}, {Prieto},
  {Elias-Rosa}, {Bersier}, {Hosseinzadeh}, {Morales-Garoffolo}, {Noebauer},
  {Taubenberger}, {Tomasella}, {Kochanek}, {Falco}, {Basu}, {Beacom},
  {Benetti}, {Brimacombe}, {Cappellaro}, {Danilet}, {Dong}, {Fernandez},
  {Goss}, {Granata}, {Harutyunyan}, {Holoien}, {Ishida}, {Kiyota}, {Krannich},
  {Nicholls}, {Ochner}, {Pojma{\'n}ski}, {Shappee}, {Simonian}, {Stanek},
  {Starrfield}, {Szczygie{\l}}, {Tartaglia}, {Terreran}, {Thompson}, {Turatto},
  {Wagner}, {Wiethoff}, {Wilber}, \&
  {Wo{\'z}niak}}]{Pastorello2015MNRAS.453.3649P}
{Pastorello}, A., {Prieto}, J.~L., {Elias-Rosa}, N., {et~al.}
  2015{\natexlab{c}}, \mnras, 453, 3649

\bibitem[{{Pastorello} {et~al.}(2008{\natexlab{b}}){Pastorello}, {Quimby},
  {Smartt}, {Mattila}, {Navasardyan}, {Crockett}, {Elias-Rosa}, {Mondol},
  {Wheeler}, \& {Young}}]{Pastorello2008MNRAS.389..131P}
{Pastorello}, A., {Quimby}, R.~M., {Smartt}, S.~J., {et~al.}
  2008{\natexlab{b}}, \mnras, 389, 131

\bibitem[{{Pastorello} {et~al.}(2007){Pastorello}, {Smartt}, {Mattila},
  {Eldridge}, {Young}, {Itagaki}, {Yamaoka}, {Navasardyan}, {Valenti}, {Patat},
  {Agnoletto}, {Augusteijn}, {Benetti}, {Cappellaro}, {Boles}, {Bonnet-Bidaud},
  {Botticella}, {Bufano}, {Cao}, {Deng}, {Dennefeld}, {Elias-Rosa},
  {Harutyunyan}, {Keenan}, {Iijima}, {Lorenzi}, {Mazzali}, {Meng}, {Nakano},
  {Nielsen}, {Smoker}, {Stanishev}, {Turatto}, {Xu}, \&
  {Zampieri}}]{Pastorello2007Natur.447..829P}
{Pastorello}, A., {Smartt}, S.~J., {Mattila}, S., {et~al.} 2007, \nat, 447, 829

\bibitem[{{Pastorello} {et~al.}(2015{\natexlab{d}}){Pastorello}, {Tartaglia},
  {Elias-Rosa}, {Morales-Garoffolo}, {Terreran}, {Taubenberger}, {Noebauer},
  {Benetti}, {Cappellaro}, {Ciabattari}, {Dennefeld}, {Dimai}, {Ishida},
  {Harutyunyan}, {Leonini}, {Ochner}, {Sollerman}, {Taddia}, \&
  {Zaggia}}]{Pastorello2015MNRAS.454.4293P}
{Pastorello}, A., {Tartaglia}, L., {Elias-Rosa}, N., {et~al.}
  2015{\natexlab{d}}, \mnras, 454, 4293

\bibitem[{{Pastorello} {et~al.}(2016){Pastorello}, {Wang}, {Ciabattari},
  {Bersier}, {Mazzali}, {Gao}, {Xu}, {Zhang}, {Tokuoka}, {Benetti},
  {Cappellaro}, {Elias-Rosa}, {Harutyunyan}, {Huang}, {Miluzio}, {Mo},
  {Ochner}, {Tartaglia}, {Terreran}, {Tomasella}, \&
  {Turatto}}]{Pastorello2016MNRAS.456..853P}
{Pastorello}, A., {Wang}, X.~F., {Ciabattari}, F., {et~al.} 2016, \mnras, 456,
  853

\bibitem[{{Pastorello} {et~al.}(2015{\natexlab{e}}){Pastorello}, {Wyrzykowski},
  {Valenti}, {Prieto}, {Koz{\l}owski}, {Udalski}, {Elias-Rosa},
  {Morales-Garoffolo}, {Anderson}, {Benetti}, {Bersten}, {Botticella},
  {Cappellaro}, {Fasano}, {Fraser}, {Gal-Yam}, {Gillone}, {Graham}, {Greiner},
  {Hachinger}, {Howell}, {Inserra}, {Parrent}, {Rau}, {Schulze}, {Smartt},
  {Smith}, {Turatto}, {Yaron}, {Young}, {Kubiak}, {Szyma{\'n}ski},
  {Pietrzy{\'n}ski}, {Soszy{\'n}ski}, {Ulaczyk}, {Poleski}, {Pietrukowicz},
  {Skowron}, \& {Mr{\'o}z}}]{Pastorello2015MNRAS.449.1941P}
{Pastorello}, A., {Wyrzykowski}, {\L}., {Valenti}, S., {et~al.}
  2015{\natexlab{e}}, \mnras, 449, 1941

\bibitem[{{Pellegrino} {et~al.}(2022){Pellegrino}, {Howell}, {Terreran},
  {Arcavi}, {Bostroem}, {Brown}, {Burke}, {Dong}, {Gilkis}, {Hiramatsu},
  {Hosseinzadeh}, {McCully}, {Modjaz}, {Newsome}, {Gonzalez}, {Pritchard},
  {Sand}, {Valenti}, \& {Williamson}}]{Pellegrino2022ApJ...938...73P}
{Pellegrino}, C., {Howell}, D.~A., {Terreran}, G., {et~al.} 2022, \apj, 938, 73

\bibitem[{{Perets} {et~al.}(2011){Perets}, {Badenes}, {Arcavi}, {Simon}, \&
  {Gal-yam}}]{Perets2011ApJ...730...89P}
{Perets}, H.~B., {Badenes}, C., {Arcavi}, I., {Simon}, J.~D., \& {Gal-yam}, A.
  2011, \apj, 730, 89

\bibitem[{{Perets} {et~al.}(2010){Perets}, {Gal-Yam}, {Mazzali}, {Arnett},
  {Kagan}, {Filippenko}, {Li}, {Arcavi}, {Cenko}, {Fox}, {Leonard}, {Moon},
  {Sand}, {Soderberg}, {Anderson}, {James}, {Foley}, {Ganeshalingam}, {Ofek},
  {Bildsten}, {Nelemans}, {Shen}, {Weinberg}, {Metzger}, {Piro}, {Quataert},
  {Kiewe}, \& {Poznanski}}]{Perets2010Natur.465..322P}
{Perets}, H.~B., {Gal-Yam}, A., {Mazzali}, P.~A., {et~al.} 2010, \nat, 465, 322

\bibitem[{{Perley} {et~al.}(2020){Perley}, {Fremling}, {Sollerman}, {Miller},
  {Dahiwale}, {Sharma}, {Bellm}, {Biswas}, {Brink}, {Bruch}, {De}, {Dekany},
  {Drake}, {Duev}, {Filippenko}, {Gal-Yam}, {Goobar}, {Graham}, {Graham}, {Ho},
  {Irani}, {Kasliwal}, {Kim}, {Kulkarni}, {Mahabal}, {Masci}, {Modak}, {Neill},
  {Nordin}, {Riddle}, {Soumagnac}, {Strotjohann}, {Schulze}, {Taggart},
  {Tzanidakis}, {Walters}, \& {Yan}}]{Perley2020ApJ...904...35P}
{Perley}, D.~A., {Fremling}, C., {Sollerman}, J., {et~al.} 2020, \apj, 904, 35

\bibitem[{{Perley} {et~al.}(2022){Perley}, {Sollerman}, {Schulze}, {Yao},
  {Fremling}, {Gal-Yam}, {Ho}, {Yang}, {Kool}, {Irani}, {Yan}, {Andreoni},
  {Baade}, {Bellm}, {Brink}, {Chen}, {Cikota}, {Coughlin}, {Dahiwale},
  {Dekany}, {Duev}, {Filippenko}, {Hoeflich}, {Kasliwal}, {Kulkarni}, {Lunnan},
  {Masci}, {Maund}, {Medford}, {Riddle}, {Rosnet}, {Shupe}, {Strotjohann},
  {Tzanidakis}, \& {Zheng}}]{Perley2022ApJ...927..180P}
{Perley}, D.~A., {Sollerman}, J., {Schulze}, S., {et~al.} 2022, \apj, 927, 180

\bibitem[{{Pignata} {et~al.}(2019){Pignata}, {Rodriguez}, {Gromadzki}, \&
  {Yaron}}]{Pignata2019TNSCR..42....1P}
{Pignata}, G., {Rodriguez}, O., {Gromadzki}, M., \& {Yaron}, O. 2019, Transient
  Name Server Classification Report, 2019-42, 1

\bibitem[{{Pilyugin} {et~al.}(2004){Pilyugin}, {V{\'{\i}}lchez}, \&
  {Contini}}]{Pilyugin2004A&A...425..849P}
{Pilyugin}, L.~S., {V{\'{\i}}lchez}, J.~M., \& {Contini}, T. 2004, \aap, 425,
  849

\bibitem[{{Poznanski} {et~al.}(2010){Poznanski}, {Chornock}, {Nugent}, {Bloom},
  {Filippenko}, {Ganeshalingam}, {Leonard}, {Li}, \&
  {Thomas}}]{Poznanski2010Sci...327...58P}
{Poznanski}, D., {Chornock}, R., {Nugent}, P.~E., {et~al.} 2010, Science, 327,
  58

\bibitem[{{Prentice} {et~al.}(2020){Prentice}, {Maguire}, {Boian}, {Groh},
  {Anderson}, {Barbarino}, {Bostroem}, {Burke}, {Clark}, {Dong}, {Fraser},
  {Galbany}, {Gromadzki}, {Guti{\'e}rrez}, {Howell}, {Hiramatsu}, {Inserra},
  {James}, {Kankare}, {Kuncarayakti}, {Mazzali}, {McCully}, {M{\"u}ller-Bravo},
  {Nichol}, {Pellegrino}, {Smartt}, {Sollerman}, {Tartaglia}, {Valenti}, \&
  {Young}}]{Prentice2020MNRAS.499.1450P}
{Prentice}, S.~J., {Maguire}, K., {Boian}, I., {et~al.} 2020, \mnras, 499, 1450

\bibitem[{{Pursiainen} {et~al.}(2023){Pursiainen}, {Leloudas}, {Schulze},
  {Charalampopoulos}, {Angus}, {Anderson}, {Bauer}, {Chen}, {Galbany},
  {Gromadzki}, {Guti{\'e}rrez}, {Inserra}, {Lyman}, {M{\"u}ller-Bravo},
  {Nicholl}, {Smartt}, {Tartaglia}, {Wiseman}, \&
  {Young}}]{Pursiainen2023ApJ...959L..10P}
{Pursiainen}, M., {Leloudas}, G., {Schulze}, S., {et~al.} 2023, \apjl, 959, L10

\bibitem[{{Quimby} {et~al.}(2011){Quimby}, {Kulkarni}, {Kasliwal}, {Gal-Yam},
  {Arcavi}, {Sullivan}, {Nugent}, {Thomas}, {Howell}, {Nakar}, {Bildsten},
  {Theissen}, {Law}, {Dekany}, {Rahmer}, {Hale}, {Smith}, {Ofek}, {Zolkower},
  {Velur}, {Walters}, {Henning}, {Bui}, {McKenna}, {Poznanski}, {Cenko}, \&
  {Levitan}}]{Quimby2011Natur.474..487Q}
{Quimby}, R.~M., {Kulkarni}, S.~R., {Kasliwal}, M.~M., {et~al.} 2011, \nat,
  474, 487

\bibitem[{{Reguitti} {et~al.}(2022){Reguitti}, {Pastorello}, {Pignata},
  {Fraser}, {Stritzinger}, {Brennan}, {Cai}, {Elias-Rosa}, {Fugazza},
  {Gutierrez}, {Kankare}, {Kotak}, {Lundqvist}, {Mazzali}, {Moran}, {Salmaso},
  {Tomasella}, {Valerin}, \& {Kuncarayakti}}]{Reguitti2022AA...662L..10R}
{Reguitti}, A., {Pastorello}, A., {Pignata}, G., {et~al.} 2022, \aap, 662, L10

\bibitem[{{Richmond} {et~al.}(1996){Richmond}, {van Dyk}, {Ho}, {Peng}, {Paik},
  {Treffers}, {Filippenko}, {Bustamante-Donas}, {Moeller}, {Pawellek},
  {Tartara}, \& {Spence}}]{Richmond1996AJ....111..327R}
{Richmond}, M.~W., {van Dyk}, S.~D., {Ho}, W., {et~al.} 1996, \aj, 111, 327

\bibitem[{{Ricker} {et~al.}(2015){Ricker}, {Winn}, {Vanderspek}, {Latham},
  {Bakos}, {Bean}, {Berta-Thompson}, {Brown}, {Buchhave}, {Butler}, {Butler},
  {Chaplin}, {Charbonneau}, {Christensen-Dalsgaard}, {Clampin}, {Deming},
  {Doty}, {De Lee}, {Dressing}, {Dunham}, {Endl}, {Fressin}, {Ge}, {Henning},
  {Holman}, {Howard}, {Ida}, {Jenkins}, {Jernigan}, {Johnson}, {Kaltenegger},
  {Kawai}, {Kjeldsen}, {Laughlin}, {Levine}, {Lin}, {Lissauer}, {MacQueen},
  {Marcy}, {McCullough}, {Morton}, {Narita}, {Paegert}, {Palle}, {Pepe},
  {Pepper}, {Quirrenbach}, {Rinehart}, {Sasselov}, {Sato}, {Seager},
  {Sozzetti}, {Stassun}, {Sullivan}, {Szentgyorgyi}, {Torres}, {Udry}, \&
  {Villasenor}}]{Ricker2015JATIS...1a4003R}
{Ricker}, G.~R., {Winn}, J.~N., {Vanderspek}, R., {et~al.} 2015, Journal of
  Astronomical Telescopes, Instruments, and Systems, 1, 014003

\bibitem[{{Sanders} {et~al.}(2013){Sanders}, {Soderberg}, {Foley}, {Chornock},
  {Milisavljevic}, {Margutti}, {Drout}, {Moe}, {Berger}, {Brown}, {Lunnan},
  {Smartt}, {Fraser}, {Kotak}, {Magill}, {Smith}, {Wright}, {Huang}, {Urata},
  {Mulchaey}, {Rest}, {Sand}, {Chomiuk}, {Friedman}, {Kirshner}, {Marion},
  {Tonry}, {Burgett}, {Chambers}, {Hodapp}, {Kudritzki}, \&
  {Price}}]{Sanders2013ApJ...769...39S}
{Sanders}, N.~E., {Soderberg}, A.~M., {Foley}, R.~J., {et~al.} 2013, \apj, 769,
  39

\bibitem[{{Schlafly} \& {Finkbeiner}(2011)}]{Schlafly2011ApJ...737..103S}
{Schlafly}, E.~F. \& {Finkbeiner}, D.~P. 2011, \apj, 737, 103

\bibitem[{{Schlegel}(1990)}]{Schlegel1990MNRAS.244..269S}
{Schlegel}, E.~M. 1990, \mnras, 244, 269

\bibitem[{{Shappee} {et~al.}(2014){Shappee}, {Prieto}, {Stanek}, {Kochanek},
  {Holoien}, {Jencson}, {Basu}, {Beacom}, {Szczygiel}, {Pojmanski},
  {Brimacombe}, {Dubberley}, {Elphick}, {Foale}, {Hawkins}, {Mullins},
  {Rosing}, {Ross}, \& {Walker}}]{Shappee2014AAS...22323603S}
{Shappee}, B., {Prieto}, J., {Stanek}, K.~Z., {et~al.} 2014, in American
  Astronomical Society Meeting Abstracts, Vol. 223, American Astronomical
  Society Meeting Abstracts \#223, 236.03

\bibitem[{{Shingles} {et~al.}(2021){Shingles}, {Smith}, {Young}, {Smartt},
  {Tonry}, {Denneau}, {Heinze}, {Weiland}, {Flewelling}, {Stalder},
  {Clocchiatti}, {F{\"o}rster}, {Pignata}, {Rest}, {Anderson}, {Stubbs}, \&
  {Erasmus}}]{Shingles2021TNSAN...7....1S}
{Shingles}, L., {Smith}, K.~W., {Young}, D.~R., {et~al.} 2021, Transient Name
  Server AstroNote, 7, 1

\bibitem[{{Shivvers} {et~al.}(2017){Shivvers}, {Zheng}, {Van Dyk}, {Mauerhan},
  {Filippenko}, {Smith}, {Foley}, {Mazzali}, {Kamble}, {Kilpatrick},
  {Margutti}, {Yuk}, {Graham}, {Kelly}, {Andrews}, {Matheson}, {Wood-Vasey},
  {Ponder}, {Brown}, {Chevalier}, {Milisavljevic}, {Drout}, {Parrent},
  {Soderberg}, {Ashall}, {Piascik}, \&
  {Prentice}}]{Shivvers2017MNRAS.471.4381S}
{Shivvers}, I., {Zheng}, W., {Van Dyk}, S.~D., {et~al.} 2017, \mnras, 471, 4381

\bibitem[{{Shivvers} {et~al.}(2016){Shivvers}, {Zheng}, {Mauerhan}, {Kleiser},
  {Van Dyk}, {Silverman}, {Graham}, {Kelly}, {Filippenko}, \&
  {Kumar}}]{Shivvers2016MNRAS.461.3057S}
{Shivvers}, I., {Zheng}, W.~K., {Mauerhan}, J., {et~al.} 2016, \mnras, 461,
  3057

\bibitem[{{Shvartzvald} {et~al.}(2024){Shvartzvald}, {Waxman}, {Gal-Yam},
  {Ofek}, {Ben-Ami}, {Berge}, {Kowalski}, {B{\"u}hler}, {Worm}, {Rhoads},
  {Arcavi}, {Maoz}, {Polishook}, {Stone}, {Trakhtenbrot}, {Ackermann},
  {Aharonson}, {Birnholtz}, {Chelouche}, {Guetta}, {Hallakoun}, {Horesh},
  {Kushnir}, {Mazeh}, {Nordin}, {Ofir}, {Ohm}, {Parsons}, {Pe'er}, {Perets},
  {Perdelwitz}, {Poznanski}, {Sadeh}, {Sagiv}, {Shahaf}, {Soumagnac}, {Tal-Or},
  {Santen}, {Zackay}, {Guttman}, {Rekhi}, {Townsend}, {Weinstein}, \&
  {Wold}}]{Shvartzvald2024ApJ...964...74S}
{Shvartzvald}, Y., {Waxman}, E., {Gal-Yam}, A., {et~al.} 2024, \apj, 964, 74

\bibitem[{{Silverman} {et~al.}(2010){Silverman}, {Kleiser}, {Morton}, \&
  {Filippenko}}]{Silverman2010CBET.2223....1S}
{Silverman}, J.~M., {Kleiser}, I.~K.~W., {Morton}, A.~J.~L., \& {Filippenko},
  A.~V. 2010, Central Bureau Electronic Telegrams, 2223, 1

\bibitem[{{Skrutskie} {et~al.}(2006){Skrutskie}, {Cutri}, {Stiening},
  {Weinberg}, {Schneider}, {Carpenter}, {Beichman}, {Capps}, {Chester},
  {Elias}, {Huchra}, {Liebert}, {Lonsdale}, {Monet}, {Price}, {Seitzer},
  {Jarrett}, {Kirkpatrick}, {Gizis}, {Howard}, {Evans}, {Fowler}, {Fullmer},
  {Hurt}, {Light}, {Kopan}, {Marsh}, {McCallon}, {Tam}, {Van Dyk}, \&
  {Wheelock}}]{Skrutskie2006AJ....131.1163S}
{Skrutskie}, M.~F., {Cutri}, R.~M., {Stiening}, R., {et~al.} 2006, \aj, 131,
  1163

\bibitem[{{Smartt} {et~al.}(2015){Smartt}, {Valenti}, {Fraser}, {Inserra},
  {Young}, {Sullivan}, {Pastorello}, {Benetti}, {Gal-Yam}, {Knapic},
  {Molinaro}, {Smareglia}, {Smith}, {Taubenberger}, {Yaron}, {Anderson},
  {Ashall}, {Balland}, {Baltay}, {Barbarino}, {Bauer}, {Baumont}, {Bersier},
  {Blagorodnova}, {Bongard}, {Botticella}, {Bufano}, {Bulla}, {Cappellaro},
  {Campbell}, {Cellier-Holzem}, {Chen}, {Childress}, {Clocchiatti},
  {Contreras}, {Dall'Ora}, {Danziger}, {de Jaeger}, {De Cia}, {Della Valle},
  {Dennefeld}, {Elias-Rosa}, {Elman}, {Feindt}, {Fleury}, {Gall},
  {Gonzalez-Gaitan}, {Galbany}, {Morales Garoffolo}, {Greggio}, {Guillou},
  {Hachinger}, {Hadjiyska}, {Hage}, {Hillebrandt}, {Hodgkin}, {Hsiao}, {James},
  {Jerkstrand}, {Kangas}, {Kankare}, {Kotak}, {Kromer}, {Kuncarayakti},
  {Leloudas}, {Lundqvist}, {Lyman}, {Hook}, {Maguire}, {Manulis}, {Margheim},
  {Mattila}, {Maund}, {Mazzali}, {McCrum}, {McKinnon}, {Moreno-Raya},
  {Nicholl}, {Nugent}, {Pain}, {Pignata}, {Phillips}, {Polshaw}, {Pumo},
  {Rabinowitz}, {Reilly}, {Romero-Ca{\~n}izales}, {Scalzo}, {Schmidt},
  {Schulze}, {Sim}, {Sollerman}, {Taddia}, {Tartaglia}, {Terreran},
  {Tomasella}, {Turatto}, {Walker}, {Walton}, {Wyrzykowski}, {Yuan}, \&
  {Zampieri}}]{Smartt2015A&A...579A..40S}
{Smartt}, S.~J., {Valenti}, S., {Fraser}, M., {et~al.} 2015, \aap, 579, A40

\bibitem[{{Smith} {et~al.}(2020){Smith}, {Smartt}, {Young}, {Tonry}, {Denneau},
  {Flewelling}, {Heinze}, {Weiland}, {Stalder}, {Rest}, {Stubbs}, {Anderson},
  {Chen}, {Clark}, {Do}, {F{\"o}rster}, {Fulton}, {Gillanders}, {McBrien},
  {O'Neill}, {Srivastav}, \& {Wright}}]{Smith2020PASP..132h5002S}
{Smith}, K.~W., {Smartt}, S.~J., {Young}, D.~R., {et~al.} 2020, \pasp, 132,
  085002

\bibitem[{Smith(2017)}]{Smith2017hsn..book..403S}
Smith, N. 2017, Interacting Supernovae: Types IIn and Ibn (Cham: Springer
  International Publishing), 403--429

\bibitem[{{Smith} {et~al.}(2008){Smith}, {Foley}, \&
  {Filippenko}}]{Smith2008ApJ...680..568S}
{Smith}, N., {Foley}, R.~J., \& {Filippenko}, A.~V. 2008, \apj, 680, 568

\bibitem[{{Smith} {et~al.}(2012){Smith}, {Mauerhan}, {Silverman},
  {Ganeshalingam}, {Filippenko}, {Cenko}, {Clubb}, \&
  {Kandrashoff}}]{Smith2012MNRAS.426.1905S}
{Smith}, N., {Mauerhan}, J.~C., {Silverman}, J.~M., {et~al.} 2012, \mnras, 426,
  1905

\bibitem[{{Spergel} {et~al.}(2007){Spergel}, {Bean}, {Dor{\'e}}, {Nolta},
  {Bennett}, {Dunkley}, {Hinshaw}, {Jarosik}, {Komatsu}, {Page}, {Peiris},
  {Verde}, {Halpern}, {Hill}, {Kogut}, {Limon}, {Meyer}, {Odegard}, {Tucker},
  {Weiland}, {Wollack}, \& {Wright}}]{Spergel2007ApJS..170..377S}
{Spergel}, D.~N., {Bean}, R., {Dor{\'e}}, O., {et~al.} 2007, \apjs, 170, 377

\bibitem[{{Spiro} {et~al.}(2014){Spiro}, {Pastorello}, {Pumo}, {Zampieri},
  {Turatto}, {Smartt}, {Benetti}, {Cappellaro}, {Valenti}, {Agnoletto},
  {Altavilla}, {Aoki}, {Brocato}, {Corsini}, {Di Cianno}, {Elias-Rosa},
  {Hamuy}, {Enya}, {Fiaschi}, {Folatelli}, {Desidera}, {Harutyunyan}, {Howell},
  {Kawka}, {Kobayashi}, {Leibundgut}, {Minezaki}, {Navasardyan}, {Nomoto},
  {Mattila}, {Pietrinferni}, {Pignata}, {Raimondo}, {Salvo}, {Schmidt},
  {Sollerman}, {Spyromilio}, {Taubenberger}, {Valentini}, {Vennes}, \&
  {Yoshii}}]{Spiro2014MNRAS.439.2873S}
{Spiro}, S., {Pastorello}, A., {Pumo}, M.~L., {et~al.} 2014, \mnras, 439, 2873

\bibitem[{{Stetson}(1987)}]{Stetson1987PASP...99..191S}
{Stetson}, P.~B. 1987, \pasp, 99, 191

\bibitem[{{Stritzinger} {et~al.}(2012){Stritzinger}, {Taddia}, {Fransson},
  {Fox}, {Morrell}, {Phillips}, {Sollerman}, {Anderson}, {Boldt}, {Brown},
  {Campillay}, {Castellon}, {Contreras}, {Folatelli}, {Habergham}, {Hamuy},
  {Hjorth}, {James}, {Krzeminski}, {Mattila}, {Persson}, \&
  {Roth}}]{Stritzinger2012ApJ...756..173S}
{Stritzinger}, M., {Taddia}, F., {Fransson}, C., {et~al.} 2012, \apj, 756, 173

\bibitem[{{Strotjohann} {et~al.}(2021){Strotjohann}, {Ofek}, {Gal-Yam},
  {Bruch}, {Schulze}, {Shaviv}, {Sollerman}, {Filippenko}, {Yaron}, {Fremling},
  {Nordin}, {Kool}, {Perley}, {Ho}, {Yang}, {Yao}, {Soumagnac}, {Graham},
  {Barbarino}, {Tartaglia}, {De}, {Goldstein}, {Cook}, {Brink}, {Taggart},
  {Yan}, {Lunnan}, {Kasliwal}, {Kulkarni}, {Nugent}, {Masci}, {Rosnet},
  {Adams}, {Andreoni}, {Bagdasaryan}, {Bellm}, {Burdge}, {Duev}, {Dugas},
  {Frederick}, {Goldwasser}, {Hankins}, {Irani}, {Karambelkar}, {Kupfer},
  {Liang}, {Neill}, {Porter}, {Riddle}, {Sharma}, {Short}, {Taddia},
  {Tzanidakis}, {van Roestel}, {Walters}, \&
  {Zhuang}}]{Strotjohann2021ApJ...907...99S}
{Strotjohann}, N.~L., {Ofek}, E.~O., {Gal-Yam}, A., {et~al.} 2021, \apj, 907,
  99

\bibitem[{{Sun} {et~al.}(2020){Sun}, {Maund}, {Hirai}, {Crowther}, \&
  {Podsiadlowski}}]{Sun2020MNRAS.491.6000S}
{Sun}, N.-C., {Maund}, J.~R., {Hirai}, R., {Crowther}, P.~A., \&
  {Podsiadlowski}, P. 2020, \mnras, 491, 6000

\bibitem[{{Taddia} {et~al.}(2015){Taddia}, {Sollerman}, {Fremling},
  {Pastorello}, {Leloudas}, {Fransson}, {Nyholm}, {Stritzinger}, {Ergon},
  {Roy}, \& {Migotto}}]{Taddia2015A&A...580A.131T}
{Taddia}, F., {Sollerman}, J., {Fremling}, C., {et~al.} 2015, \aap, 580, A131

\bibitem[{{Tody}(1986)}]{Tody1986SPIE..627..733T}
{Tody}, D. 1986, in \procspie, Vol. 627, Instrumentation in astronomy VI, ed.
  D.~L. {Crawford}, 733

\bibitem[{{Tody}(1993)}]{Tody1993ASPC...52..173T}
{Tody}, D. 1993, in Astronomical Society of the Pacific Conference Series,
  Vol.~52, Astronomical Data Analysis Software and Systems II, ed. R.~J.
  {Hanisch}, R.~J.~V. {Brissenden}, \& J.~{Barnes}, 173

\bibitem[{{Tonry} {et~al.}(2018){Tonry}, {Denneau}, {Heinze}, {Stalder},
  {Smith}, {Smartt}, {Stubbs}, {Weiland}, \& {Rest}}]{Tonry2018PASP..130f4505T}
{Tonry}, J.~L., {Denneau}, L., {Heinze}, A.~N., {et~al.} 2018, \pasp, 130,
  064505

\bibitem[{{Vagnozzi}(2019)}]{Vagnozzi2019Atoms...7...41V}
{Vagnozzi}, S. 2019, Atoms, 7, 41

\bibitem[{{Valenti} {et~al.}(2009){Valenti}, {Pastorello}, {Cappellaro},
  {Benetti}, {Mazzali}, {Manteca}, {Taubenberger}, {Elias-Rosa}, {Ferrando},
  {Harutyunyan}, {Hentunen}, {Nissinen}, {Pian}, {Turatto}, {Zampieri}, \&
  {Smartt}}]{Valenti2009Natur.459..674V}
{Valenti}, S., {Pastorello}, A., {Cappellaro}, E., {et~al.} 2009, \nat, 459,
  674

\bibitem[{{Vallely} {et~al.}(2021){Vallely}, {Kochanek}, {Stanek}, {Fausnaugh},
  \& {Shappee}}]{Vallely2021MNRAS.500.5639V}
{Vallely}, P.~J., {Kochanek}, C.~S., {Stanek}, K.~Z., {Fausnaugh}, M., \&
  {Shappee}, B.~J. 2021, \mnras, 500, 5639

\bibitem[{{Vallely} {et~al.}(2018){Vallely}, {Prieto}, {Stanek}, {Kochanek},
  {Sukhbold}, {Bersier}, {Brown}, {Chen}, {Dong}, {Falco}, {Berlind},
  {Calkins}, {Koff}, {Kiyota}, {Brimacombe}, {Shappee}, {Holoien}, {Thompson},
  \& {Stritzinger}}]{Vallely2018MNRAS.475.2344V}
{Vallely}, P.~J., {Prieto}, J.~L., {Stanek}, K.~Z., {et~al.} 2018, \mnras, 475,
  2344

\bibitem[{{von Steiger} \& {Zurbuchen}(2016)}]{vonSteiger2016ApJ...816...13V}
{von Steiger}, R. \& {Zurbuchen}, T.~H. 2016, \apj, 816, 13

\bibitem[{{Wang} {et~al.}(2024{\natexlab{a}}){Wang}, {Hu}, {Wang}, {Yang},
  {Yang}, {Gomez}, {Chen}, {Hu}, {Chen}, {Mo}, {Wang}, {Baade}, {Hoeflich},
  {Wheeler}, {Pignata}, {Burke}, {Hiramatsu}, {Howell}, {McCully},
  {Pellegrino}, {Galbany}, {Hsiao}, {Sand}, {Zhang}, {Uddin}, {Anderson},
  {Ashall}, {Cheng}, {Gromadzki}, {Inserra}, {Lin}, {Morrell},
  {Morales-Garoffolo}, {M{\"u}ller-Bravo}, {Nicholl}, {Gonzalez}, {Phillips},
  {Pineda-Garc{\'\i}a}, {Sai}, {Smith}, {Shahbandeh}, {Srivastav},
  {Stritzinger}, {Yang}, {Young}, {Yu}, \& {Zhang}}]{Wang2024NatAs...8..504W}
{Wang}, L., {Hu}, M., {Wang}, L., {et~al.} 2024{\natexlab{a}}, Nature
  Astronomy, 8, 504

\bibitem[{{Wang} {et~al.}(2024{\natexlab{b}}){Wang}, {Goel}, {Dessart}, {Fox},
  {Shahbandeh}, {Rest}, {Rest}, {Groh}, {Allan}, {Fransson}, {Smith},
  {Hosseinzadeh}, {Filippenko}, {Andrews}, {Bostroem}, {Brink}, {Brown},
  {Burke}, {Chevalier}, {Clayton}, {Dai}, {Davis}, {Foley}, {Gomez}, {Harris},
  {Hiramatsu}, {Howell}, {Jennings}, {Jha}, {Kasliwal}, {Kelly}, {Kool}, {Liu},
  {Ma}, {McCully}, {Miller}, {Murakami}, {Gonzalez}, {Pellegrino}, {Perera},
  {Pierel}, {Rojas-Bravo}, {Siebert}, {Sollerman}, {Szalai}, {Tinyanont}, {Van
  Dyk}, {Zheng}, {Chambers}, {Coulter}, {de Boer}, {Earl}, {Farias}, {Gall},
  {McGill}, {Ransome}, {Taggart}, \& {Villar}}]{Wang2024MNRAS.530.3906W}
{Wang}, Q., {Goel}, A., {Dessart}, L., {et~al.} 2024{\natexlab{b}}, \mnras,
  530, 3906

\bibitem[{{Wang} \& {Li}(2020)}]{Wang2020ApJ...900...83W}
{Wang}, S.-Q. \& {Li}, L. 2020, \apj, 900, 83

\bibitem[{{Wang} {et~al.}(2021){Wang}, {Lin}, {Zhang}, {Zhang}, {Cai}, {Zhang},
  {Filippenko}, {Graham}, {Maeda}, {Mo}, {Xiang}, {Xi}, {Yan}, {Wang}, {Wang},
  {Kawabata}, \& {Zhai}}]{Wang2021ApJ...917...97W}
{Wang}, X., {Lin}, W., {Zhang}, J., {et~al.} 2021, \apj, 917, 97

\bibitem[{{Woosley}(2017)}]{Woosley2017ApJ...836..244W}
{Woosley}, S.~E. 2017, \apj, 836, 244

\bibitem[{{Woosley} {et~al.}(2007){Woosley}, {Blinnikov}, \&
  {Heger}}]{Woosley2007Natur.450..390W}
{Woosley}, S.~E., {Blinnikov}, S., \& {Heger}, A. 2007, \nat, 450, 390

\bibitem[{{Woosley} \& {Weaver}(1995)}]{Woosley1995ApJS..101..181W}
{Woosley}, S.~E. \& {Weaver}, T.~A. 1995, \apjs, 101, 181

\bibitem[{{Yaron} \& {Gal-Yam}(2012)}]{Yaron2012PASP..124..668Y}
{Yaron}, O. \& {Gal-Yam}, A. 2012, \pasp, 124, 668

\bibitem[{{Zampieri} {et~al.}(1998){Zampieri}, {Shapiro}, \&
  {Colpi}}]{Zampieri1998ApJ...502L.149Z}
{Zampieri}, L., {Shapiro}, S.~L., \& {Colpi}, M. 1998, \apjl, 502, L149

\bibitem[{Zhang {et~al.}(2023)Zhang, Lin, Wang, Zhao, Li, Liu, Yan, Xiang,
  Wang, \& Bai}]{ZHANG20232548}
Zhang, J., Lin, H., Wang, X., {et~al.} 2023, Science Bulletin, 68, 2548

\end{thebibliography}
\begin{appendix}

\onecolumn 
{Our observations will be made public via the Weizmann Interactive Supernova Data Repository \citep[WISeREP;][]{Yaron2012PASP..124..668Y}.  }

\section{Photometric tables}
\label{appendix:lightcurves_data}

\begin{longtable}{ccccccc}
\caption{Optical and NIR observed magnitudes of SN\,2018jmt.} \label{table:SN2018jmt_lightcurves_data}\\
\hline \hline
Date & MJD & Phase & Filter & Magnitude & Error & Instrument/Source \\
\hline
\endfirsthead

\multicolumn{7}{c}%
{{\tablename\ \thetable{} -- continued from previous page}} \\
\hline \hline
Date & MJD & Phase & Filter & Magnitude & Error & Instrument/Source \\
\hline
\endhead

\hline \hline
\multicolumn{7}{r}{{Continued on next page}} \\
\endfoot

\hline \hline
\endlastfoot
20181123  &  58445.173  &  -20.49  &  g  &  >17.4  &         &  ASAS-SN    \\   
20181125  &  58447.056  &  -18.60  &  g  &  >18.0  &         &  ASAS-SN    \\   
20181126  &  58448.252  &  -17.41  &  g  &  >18.4  &         &  ASAS-SN    \\   
20181129  &  58451.304  &  -14.36  &  g  &  >18.4  &         &  ASAS-SN    \\   
20181129  &  58451.868  &  -13.79  &  g  &  >18.9  &         &  ASAS-SN    \\   
20181130  &  58452.300  &  -13.36  &  g  &  >18.7  &         &  ASAS-SN    \\   
20181201  &  58453.297  &  -12.36  &  g  &  >18.9  &         &  ASAS-SN    \\   
20181201  &  58453.869  &  -11.79  &  g  &  >18.6  &         &  ASAS-SN    \\   
20181202  &  58454.306  &  -11.35  &  g  &  >18.6  &         &  ASAS-SN    \\   
20181205  &  58457.285  &  -8.38  &  g  &  18.324  &  0.198  &  ASAS-SN    \\ 
20181207  &  58459.282  &  -6.38  &  g  &  18.186  &  0.195  &  ASAS-SN    \\   
20181208 &   58460.278  &  -5.38 & Clear & 16.5    & -   & MASTER \\
20181209  &  58461.113  &  -4.55  &  g  &  17.199  &  0.095  &  ASAS-SN    \\   
20181210  &  58462.251  &  -3.41  &  g  &  17.108  &  0.101  &  ASAS-SN    \\   
20181212  &  58464.300  &  -1.36  &  g  &  16.708  &  0.070  &  ASAS-SN    \\   
20181216  &  58468.135  &  2.47  &  g  &  17.179  &  0.112  &  ASAS-SN    \\   
20181216  &  58468.300  &  2.64  &  V  &  17.153  &  0.012  &  EFOSC    \\   
20181216  &  58468.971  &  3.31  &  g  &  17.581  &  0.122  &  ASAS-SN    \\   
20181217  &  58469.170  &  3.51  &  V  &  17.236  &  0.019  &  EFOSC    \\   
20181217  &  58469.302  &  3.64  &  g  &  17.417  &  0.116  &  ASAS-SN    \\   
20181218  &  58470.890  &  5.23  &  U  &  16.778  &  0.023  &  fa06    \\   
20181218  &  58470.890  &  5.23  &  B  &  17.514  &  0.010  &  fa06    \\   
20181218  &  58470.890  &  5.23  &  V  &  17.433  &  0.021  &  fa06    \\   
20181218  &  58470.905  &  5.24  &  g  &  17.332  &  0.011  &  fa06    \\   
20181218  &  58470.905  &  5.24  &  r  &  17.536  &  0.020  &  fa06    \\   
20181218  &  58470.905  &  5.24  &  i  &  17.660  &  0.020  &  fa06    \\   
20181219  &  58471.870  &  6.21  &  U  &  16.890  &  0.018  &  fa16    \\   
20181219  &  58471.870  &  6.21  &  B  &  17.649  &  0.012  &  fa16    \\   
20181219  &  58471.870  &  6.21  &  V  &  17.499  &  0.016  &  fa16    \\   
20181219  &  58471.885  &  6.22  &  g  &  17.595  &  0.016  &  fa16    \\   
20181219  &  58471.885  &  6.22  &  r  &  17.628  &  0.018  &  fa16    \\   
20181219  &  58471.885  &  6.22  &  i  &  17.761  &  0.022  &  fa16    \\   
20181220  &  58472.180  &  6.52  &  U  &  16.994  &  0.014  &  fa15    \\   
20181220  &  58472.180  &  6.52  &  B  &  17.670  &  0.005  &  fa15    \\   
20181220  &  58472.180  &  6.52  &  V  &  17.522  &  0.009  &  fa15    \\   
20181220  &  58472.195  &  6.53  &  g  &  17.545  &  0.006  &  fa15    \\   
20181220  &  58472.195  &  6.53  &  r  &  17.657  &  0.014  &  fa15    \\   
20181220  &  58472.195  &  6.53  &  i  &  17.834  &  0.020  &  fa15    \\   
20181222  &  58474.095  &  8.43  &  U  &  17.323  &  0.018  &  fa15    \\   
20181222  &  58474.095  &  8.43  &  B  &  17.902  &  0.009  &  fa15    \\   
20181222  &  58474.095  &  8.43  &  V  &  17.746  &  0.026  &  fa15    \\   
20181222  &  58474.105  &  8.44  &  g  &  17.778  &  0.011  &  fa15    \\   
20181222  &  58474.105  &  8.44  &  r  &  17.896  &  0.022  &  fa15    \\   
20181222  &  58474.105  &  8.44  &  i  &  17.970  &  0.022  &  fa15    \\   
20181223  &  58475.035  &  9.38  &  U  &  17.486  &  0.018  &  fa06    \\   
20181223  &  58475.035  &  9.38  &  B  &  18.086  &  0.010  &  fa06    \\   
20181223  &  58475.035  &  9.38  &  V  &  17.845  &  0.014  &  fa06    \\   
20181223  &  58475.055  &  9.39  &  g  &  17.827  &  0.013  &  fa06    \\   
20181223  &  58475.055  &  9.39  &  r  &  17.964  &  0.024  &  fa06    \\   
20181223  &  58475.055  &  9.39  &  i  &  18.001  &  0.024  &  fa06    \\   
20181223  &  58475.300  &  9.64  &  U  &  17.517  &  0.025  &  fa15    \\   
20181223  &  58475.300  &  9.64  &  B  &  18.099  &  0.007  &  fa15    \\   
20181223  &  58475.300  &  9.64  &  V  &  17.930  &  0.009  &  fa15    \\   
20181223  &  58475.315  &  9.65  &  g  &  17.950  &  0.007  &  fa15    \\   
20181223  &  58475.315  &  9.65  &  r  &  18.080  &  0.015  &  fa15    \\   
20181223  &  58475.315  &  9.65  &  i  &  18.121  &  0.018  &  fa15    \\   
20181225  &  58477.015  &  11.35  &  B  &  18.511  &  0.017  &  fa16    \\   
20181225  &  58477.015  &  11.35  &  V  &  18.138  &  0.017  &  fa16    \\   
20181225  &  58477.025  &  11.36  &  g  &  18.314  &  0.021  &  fa16    \\   
20181225  &  58477.025  &  11.36  &  r  &  18.345  &  0.035  &  fa16    \\   
20181225  &  58477.025  &  11.36  &  i  &  18.291  &  0.023  &  fa16    \\   
20181228  &  58480.093  &  14.43  &  g  &  18.806  &  0.287  &  ASAS-SN    \\   
20181231  &  58483.173  &  17.51  &  J  &  18.109  &  0.026  &  SOFI    \\   
20181231  &  58483.173  &  17.51  &  H  &  18.028  &  0.061  &  SOFI    \\   
20181231  &  58483.173  &  17.51  &  K  &  17.713  &  0.073  &  SOFI    \\   
20190101  &  58484.090  &  18.43  &  V  &  18.741  &  0.027  &  EFOSC    \\   
20190101  &  58484.280  &  18.62  &  B  &  19.152  &  0.014  &  fa03    \\   
20190101  &  58484.280  &  18.62  &  V  &  18.824  &  0.014  &  fa03    \\   
20190101  &  58484.295  &  18.63  &  g  &  18.981  &  0.009  &  fa03    \\   
20190101  &  58484.295  &  18.63  &  r  &  18.909  &  0.014  &  fa03    \\   
20190101  &  58484.295  &  18.63  &  i  &  18.930  &  0.025  &  fa03    \\   
20190103  &  58486.155  &  20.49  &  B  &  19.218  &  0.023  &  fa15    \\   
20190103  &  58486.155  &  20.49  &  V  &  18.912  &  0.029  &  fa15    \\   
20190103  &  58486.170  &  20.51  &  g  &  19.145  &  0.017  &  fa15    \\   
20190103  &  58486.170  &  20.51  &  i  &  19.010  &  0.052  &  fa15    \\   
20190105  &  58488.289  &  22.63  &  g  &  19.115  &  0.289  &  ASAS-SN    \\   
20190107  &  58490.135  &  24.47  &  B  &  19.484  &  0.013  &  fa15    \\   
20190107  &  58490.135  &  24.47  &  V  &  19.093  &  0.013  &  fa15    \\   
20190107  &  58490.155  &  24.49  &  g  &  19.244  &  0.014  &  fa15    \\   
20190107  &  58490.155  &  24.49  &  r  &  19.284  &  0.017  &  fa15    \\   
20190107  &  58490.155  &  24.49  &  i  &  19.299  &  0.018  &  fa15    \\   
20190111  &  58494.215  &  28.55  &  B  &  19.656  &  0.018  &  fa15    \\   
20190111  &  58494.215  &  28.55  &  V  &  19.343  &  0.021  &  fa15    \\   
20190111  &  58494.230  &  28.57  &  g  &  19.447  &  0.015  &  fa15    \\   
20190111  &  58494.230  &  28.57  &  r  &  19.484  &  0.026  &  fa15    \\   
20190111  &  58494.230  &  28.57  &  i  &  19.541  &  0.025  &  fa15    \\   
20190114  &  58497.615  &  31.95  &  B  &  19.938  &  0.022  &  fa11    \\   
20190114  &  58497.615  &  31.95  &  V  &  19.533  &  0.031  &  fa11    \\   
20190114  &  58497.635  &  31.97  &  g  &  19.715  &  0.023  &  fa11    \\   
20190114  &  58497.635  &  31.97  &  r  &  19.728  &  0.029  &  fa11    \\   
20190114  &  58497.635  &  31.97  &  i  &  19.819  &  0.026  &  fa11    \\   
20190118  &  58501.305  &  35.64  &  B  &  19.981  &  0.034  &  fa15    \\   
20190118  &  58501.305  &  35.64  &  V  &  19.639  &  0.031  &  fa15    \\   
20190118  &  58501.320  &  35.66  &  g  &  19.781  &  0.028  &  fa15    \\   
20190118  &  58501.320  &  35.66  &  r  &  19.864  &  0.039  &  fa15    \\   
20190118  &  58501.320  &  35.66  &  i  &  20.129  &  0.072  &  fa15    \\   
20190122  &  58505.910  &  40.25  &  B  &  20.239  &  0.057  &  fa14    \\   
20190122  &  58505.910  &  40.25  &  V  &  19.739  &  0.086  &  fa14    \\   
20190122  &  58505.925  &  40.26  &  g  &  20.000  &  0.046  &  fa14    \\   
20190122  &  58505.925  &  40.26  &  r  &  20.037  &  0.069  &  fa14    \\   
20190122  &  58505.925  &  40.26  &  i  &  20.193  &  0.099  &  fa14    \\   
20190124  &  58507.200  &  41.54  &  V  &  19.953  &  0.057  &  EFOSC    \\   
20190126  &  58509.160  &  43.50  &  J  &  20.081  &  0.145  &  SOFI    \\   
20190126  &  58509.160  &  43.50  &  H  &  19.295  &  0.276  &  SOFI    \\   
20190126  &  58509.160  &  43.50  &  K  &  19.204  &  0.184  &  SOFI    \\   
20190127  &  58510.240  &  44.58  &  U  &  20.050  &  0.062  &  EFOSC    \\   
20190127  &  58510.990  &  45.33  &  B  &  20.303  &  0.051  &  fa16    \\   
20190128  &  58511.085  &  45.42  &  B  &  20.327  &  0.046  &  fa15    \\   
20190128  &  58511.085  &  45.42  &  V  &  20.022  &  0.034  &  fa15    \\   
20190128  &  58511.105  &  45.44  &  g  &  20.093  &  0.027  &  fa15    \\   
20190128  &  58511.105  &  45.44  &  r  &  20.236  &  0.033  &  fa15    \\   
20190128  &  58511.105  &  45.44  &  i  &  20.747  &  0.056  &  fa15    \\   
20190202  &  58516.265  &  50.60  &  B  &  20.528  &  0.032  &  fa15    \\   
20190202  &  58516.265  &  50.60  &  V  &  20.263  &  0.044  &  fa15    \\   
20190202  &  58516.285  &  50.62  &  g  &  20.315  &  0.033  &  fa15    \\   
20190202  &  58516.285  &  50.62  &  r  &  20.377  &  0.059  &  fa15    \\   
20190202  &  58516.285  &  50.62  &  i  &  20.680  &  0.069  &  fa15    \\   
20190207  &  58521.185  &  55.52  &  B  &  20.704  &  0.073  &  fa03    \\   
20190207  &  58521.185  &  55.52  &  V  &  20.629  &  0.112  &  fa03    \\   
20190207  &  58521.200  &  55.54  &  g  &  20.291  &  0.041  &  fa03    \\   
20190207  &  58521.200  &  55.54  &  r  &  20.363  &  0.079  &  fa03    \\   
20190207  &  58521.200  &  55.54  &  i  &  >19.6  &         &  fa03    \\   
20190208  &  58522.190  &  56.53  &  U  &  20.447  &  0.072  &  EFOSC    \\   
20190208  &  58522.190  &  56.53  &  V  &  20.494  &  0.093  &  EFOSC    \\   
20190208  &  58522.200  &  56.54  &  z  &  13.737  &  1.249  &  EFOSC    \\   
20190214  &  58528.090  &  62.43  &  g  &  20.835  &  0.042  &  GROND    \\   
20190214  &  58528.090  &  62.43  &  r  &  20.801  &  0.027  &  GROND    \\   
20190214  &  58528.090  &  62.43  &  i  &  20.857  &  0.045  &  GROND    \\   
20190214  &  58528.090  &  62.43  &  z  &  17.171  &  2.236  &  GROND    \\   
20190214  &  58528.090  &  62.43  &  J  &  19.857  &  0.105  &  GRONDIR    \\   
20190214  &  58528.090  &  62.43  &  H  &  19.420  &  0.131  &  GRONDIR    \\   
20190214  &  58528.090  &  62.43  &  K  &  >18.4   &         &  GRONDIR    \\   
20190226  &  58540.207  &  74.55  &  J  &  20.619  &  0.140  &  SOFI    \\   
20190226  &  58540.207  &  74.55  &  H  &  19.812  &  0.302  &  SOFI    \\   
20190226  &  58540.207  &  74.55  &  K  &  19.777  &  0.267  &  SOFI    \\   
20190320  &  58562.147  &  96.49  &  J  &  20.924  &  0.157  &  SOFI    \\   
20190320  &  58562.147  &  96.49  &  H  &  20.300  &  0.201  &  SOFI    \\   
20190320  &  58562.147  &  96.49  &  K  &  19.569  &  0.251  &  SOFI    \\   

\end{longtable}
\begin{longtable}{ccccccc}
\caption{Optical, ATLAS and NIR observed magnitudes of SN\,2019cj.}  \label{table:SN2019cj_lightcurves_data} \\

\hline \hline
Date & MJD & Phase & Filter & Magnitude & Error & Instrument/Source \\
\hline
\endfirsthead

\multicolumn{7}{c}%
{{\tablename\ \thetable{} -- continued from previous page}} \\
\hline \hline
Date & MJD & Phase & Filter & Magnitude & Error & Instrument/Source \\
\hline
\endhead

\hline \hline
\multicolumn{7}{r}{{Continued on next page}} \\
\endfoot

\hline \hline
\endlastfoot
20181129  &  58451.517  &  -40.92  &  o  &  >19.8  &         &  ATLAS    \\   
20181129  &  58451.998  &  -40.44  &  g  &  >19.2  &         &  ASAS-SN    \\   
20181203  &  58455.123  &  -37.32  &  g  &  >19.5  &         &  ASAS-SN    \\   
20181205  &  58457.104  &  -35.34  &  g  &  >19.1  &         &  ASAS-SN    \\   
20181206  &  58458.120  &  -34.32  &  g  &  >19.1  &         &  ASAS-SN    \\   
20181207  &  58459.174  &  -33.27  &  g  &  >19.1  &         &  ASAS-SN    \\   
20181208  &  58460.065  &  -32.38  &  g  &  >19.3  &         &  ASAS-SN    \\   
20181212  &  58464.271  &  -28.17  &  g  &  >19.3  &         &  ASAS-SN    \\   
20181213  &  58465.428  &  -27.01  &  o  &  >18.6  &         &  ATLAS    \\   
20181215  &  58467.429  &  -25.01  &  o  &  >20.1  &         &  ATLAS    \\   
20181216  &  58468.083  &  -24.36  &  g  &  >18.9  &         &  ASAS-SN    \\   
20181217  &  58469.442  &  -23.00  &  o  &  >19.7  &         &  ATLAS    \\   
20181219  &  58471.093  &  -21.35  &  g  &  >18.6  &         &  ASAS-SN    \\   
20181219  &  58471.443  &  -21.00  &  o  &  >20.0  &         &  ATLAS    \\   
20181221  &  58473.434  &  -19.01  &  o  &  >18.7  &         &  ATLAS    \\   
20181225  &  58477.087  &  -15.35  &  g  &  >18.6  &         &  ASAS-SN    \\   
20181226  &  58478.065  &  -14.38  &  g  &  >18.9  &         &  ASAS-SN    \\   
20181227  &  58479.240  &  -13.20  &  g  &  >18.8  &         &  ASAS-SN    \\   
20181227  &  58479.393  &  -13.05  &  o  &  >20.2  &         &  ATLAS    \\   
20181229  &  58481.114  &  -11.33  &  g  &  >19.4  &         &  ASAS-SN    \\   
20181229  &  58481.295  &  -11.15  &  g  &  >18.7  &         &  ASAS-SN    \\   
20181231  &  58483.424  &  -9.02  &  o  &  19.752  &  0.270  &  ATLAS    \\   
20190101  &  58484.074  &  -8.37  &  g  &  >19.1  &         &  ASAS-SN    \\   
20190102  &  58485.237  &  -7.20  &  g  &  18.624  &  0.234  &  ASAS-SN    \\   
20190102  &  58485.413  &  -7.03  &  c  &  18.733  &  0.097  &  ATLAS    \\   
20190103  &  58486.262  &  -6.18  &  g  &  18.343  &  0.179  &  ASAS-SN    \\   
20190104  &  58487.215  &  -5.23  &  g  &  18.142  &  0.170  &  ASAS-SN    \\   
20190104  &  58487.404  &  -5.04  &  o  &  18.422  &  0.054  &  ATLAS    \\   
20190105  &  58488.219  &  -4.22  &  g  &  17.624  &  0.126  &  ASAS-SN    \\   
20190106  &  58489.234  &  -3.21  &  g  &  17.610  &  0.134  &  ASAS-SN    \\   
20190107  &  58490.218  &  -2.22  &  g  &  17.718  &  0.131  &  ASAS-SN    \\   
20190107  &  58490.250  &  -2.19  &  V  &  17.591  &  0.013  &  EFOSC    \\   
20190108  &  58491.280  &  -1.16  &  V  &  17.538  &  0.022  &  EFOSC    \\   
20190108  &  58491.293  &  -1.15  &  g  &  17.748  &  0.131  &  ASAS-SN    \\   
20190108  &  58491.380  &  -1.06  &  o  &  17.989  &  0.029  &  ATLAS    \\   
20190108  &  58491.919  &  -0.52  &  g  &  17.707  &  0.139  &  ASAS-SN    \\   
20190109  &  58492.170  &  -0.27  &  V  &  17.503  &  0.020  &  EFOSC    \\   
20190109  &  58492.279  &  -0.16  &  g  &  17.521  &  0.159  &  ASAS-SN    \\   
20190110  &  58493.207  &  0.77  &  g  &  17.584  &  0.105  &  ASAS-SN    \\   
20190110  &  58493.373  &  0.93  &  c  &  17.841  &  0.044  &  ATLAS    \\   
20190111  &  58494.005  &  1.56  &  U  &  16.906  &  0.021  &  fa16    \\   
20190111  &  58494.005  &  1.56  &  B  &  17.614  &  0.024  &  fa16    \\   
20190111  &  58494.005  &  1.56  &  V  &  17.573  &  0.040  &  fa16    \\   
20190111  &  58494.015  &  1.57  &  g  &  17.475  &  0.025  &  fa16    \\   
20190111  &  58494.015  &  1.57  &  r  &  17.606  &  0.051  &  fa16    \\   
20190111  &  58494.015  &  1.57  &  i  &  17.776  &  0.054  &  fa16    \\   
20190111  &  58494.015  &  1.57  &  z  &  17.932  &  0.125  &  fa16    \\   
20190111  &  58494.277  &  1.84  &  g  &  17.945  &  0.135  &  ASAS-SN    \\   
20190112  &  58495.219  &  2.78  &  g  &  17.477  &  0.103  &  ASAS-SN    \\   
20190112  &  58495.388  &  2.95  &  o  &  17.951  &  0.036  &  ATLAS    \\   
20190113  &  58496.920  &  4.48  &  U  &  17.172  &  0.026  &  fa14    \\   
20190113  &  58496.920  &  4.48  &  B  &  17.893  &  0.034  &  fa14    \\   
20190113  &  58496.920  &  4.48  &  V  &  17.701  &  0.024  &  fa14    \\   
20190113  &  58496.925  &  4.49  &  g  &  17.629  &  0.026  &  fa14    \\   
20190113  &  58496.925  &  4.49  &  r  &  17.656  &  0.020  &  fa14    \\   
20190113  &  58496.925  &  4.49  &  i  &  17.823  &  0.024  &  fa14    \\   
20190113  &  58496.925  &  4.49  &  z  &  17.931  &  0.057  &  fa14    \\   
20190114  &  58497.107  &  4.67  &  J  &  17.612  &  0.040  &  SOFI    \\   
20190114  &  58497.107  &  4.67  &  H  &  17.402  &  0.052  &  SOFI    \\   
20190114  &  58497.107  &  4.67  &  K  &  17.395  &  0.077  &  SOFI    \\   
20190114  &  58497.226  &  4.79  &  g  &  18.029  &  0.183  &  ASAS-SN    \\   
20190114  &  58497.364  &  4.92  &  o  &  17.972  &  0.046  &  ATLAS    \\   
20190114  &  58497.605  &  5.17  &  U  &  17.303  &  0.024  &  fa12    \\   
20190114  &  58497.605  &  5.17  &  B  &  17.845  &  0.017  &  fa12    \\   
20190114  &  58497.605  &  5.17  &  V  &  17.771  &  0.013  &  fa12    \\   
20190114  &  58497.615  &  5.17  &  g  &  17.707  &  0.011  &  fa12    \\   
20190114  &  58497.615  &  5.17  &  r  &  17.725  &  0.016  &  fa12    \\   
20190114  &  58497.615  &  5.17  &  i  &  17.847  &  0.022  &  fa12    \\   
20190114  &  58497.615  &  5.17  &  z  &  18.025  &  0.076  &  fa12    \\   
20190115  &  58498.060  &  5.62  &  V  &  17.718  &  0.022  &  EFOSC    \\   
20190115  &  58498.075  &  5.63  &  U  &  17.393  &  0.041  &  fa03    \\   
20190115  &  58498.075  &  5.63  &  B  &  17.904  &  0.024  &  fa03    \\   
20190115  &  58498.075  &  5.63  &  V  &  17.747  &  0.018  &  fa03    \\   
20190115  &  58498.085  &  5.64  &  g  &  17.698  &  0.016  &  fa03    \\   
20190115  &  58498.085  &  5.64  &  r  &  17.719  &  0.017  &  fa03    \\   
20190115  &  58498.085  &  5.64  &  i  &  17.843  &  0.019  &  fa03    \\   
20190115  &  58498.085  &  5.64  &  z  &  18.225  &  0.068  &  fa03    \\   
20190115  &  58498.143  &  5.70  &  g  &  17.977  &  0.218  &  ASAS-SN    \\   
20190116  &  58499.090  &  6.65  &  U  &  17.430  &  0.025  &  fa15    \\   
20190116  &  58499.090  &  6.65  &  B  &  17.982  &  0.022  &  fa15    \\   
20190116  &  58499.090  &  6.65  &  V  &  17.850  &  0.025  &  fa15    \\   
20190116  &  58499.098  &  6.66  &  g  &  18.186  &  0.259  &  ASAS-SN    \\   
20190116  &  58499.100  &  6.66  &  g  &  17.738  &  0.017  &  fa15    \\   
20190116  &  58499.100  &  6.66  &  r  &  17.811  &  0.020  &  fa15    \\   
20190116  &  58499.100  &  6.66  &  i  &  17.882  &  0.031  &  fa15    \\   
20190116  &  58499.100  &  6.66  &  z  &  18.119  &  0.102  &  fa15    \\   
20190116  &  58499.284  &  6.84  &  g  &  18.238  &  0.179  &  ASAS-SN    \\   
20190117  &  58500.170  &  7.73  &  U  &  17.609  &  0.073  &  fa03    \\   
20190117  &  58500.170  &  7.73  &  B  &  18.152  &  0.029  &  fa03    \\   
20190117  &  58500.170  &  7.73  &  V  &  17.931  &  0.032  &  fa03    \\   
20190117  &  58500.180  &  7.74  &  g  &  17.923  &  0.022  &  fa03    \\   
20190117  &  58500.180  &  7.74  &  r  &  17.832  &  0.018  &  fa03    \\   
20190117  &  58500.180  &  7.74  &  i  &  17.967  &  0.020  &  fa03    \\   
20190117  &  58500.180  &  7.74  &  z  &  18.121  &  0.059  &  fa03    \\   
20190117  &  58500.220  &  7.78  &  J  &  17.662  &  0.047  &  SOFI    \\   
20190117  &  58500.220  &  7.78  &  H  &  17.357  &  0.041  &  SOFI    \\   
20190117  &  58500.220  &  7.78  &  K  &  17.129  &  0.062  &  SOFI    \\   
20190118  &  58501.195  &  8.75  &  U  &  17.692  &  0.036  &  fa15    \\   
20190118  &  58501.195  &  8.75  &  B  &  18.180  &  0.026  &  fa15    \\   
20190118  &  58501.195  &  8.75  &  V  &  17.990  &  0.025  &  fa15    \\   
20190118  &  58501.205  &  8.76  &  g  &  17.906  &  0.019  &  fa15    \\   
20190118  &  58501.205  &  8.76  &  r  &  17.909  &  0.020  &  fa15    \\   
20190118  &  58501.205  &  8.76  &  i  &  18.013  &  0.040  &  fa15    \\   
20190118  &  58501.205  &  8.76  &  z  &  18.114  &  0.083  &  fa15    \\   
20190119  &  58502.865  &  10.42  &  U  &  17.744  &  0.064  &  fa14    \\   
20190119  &  58502.865  &  10.42  &  B  &  18.304  &  0.054  &  fa14    \\   
20190119  &  58502.865  &  10.42  &  V  &  18.115  &  0.049  &  fa14    \\   
20190119  &  58502.875  &  10.43  &  g  &  18.073  &  0.037  &  fa14    \\   
20190119  &  58502.875  &  10.43  &  r  &  18.011  &  0.038  &  fa14    \\   
20190119  &  58502.875  &  10.43  &  i  &  18.074  &  0.037  &  fa14    \\   
20190119  &  58502.875  &  10.43  &  z  &  18.186  &  0.088  &  fa14    \\   
20190120  &  58503.830  &  11.39  &  U  &  17.916  &  0.058  &  fa16    \\   
20190120  &  58503.830  &  11.39  &  B  &  18.391  &  0.043  &  fa16    \\   
20190120  &  58503.830  &  11.39  &  V  &  18.161  &  0.038  &  fa16    \\   
20190120  &  58503.835  &  11.39  &  g  &  18.142  &  0.035  &  fa16    \\   
20190120  &  58503.835  &  11.39  &  r  &  18.015  &  0.031  &  fa16    \\   
20190120  &  58503.835  &  11.39  &  i  &  18.095  &  0.037  &  fa16    \\   
20190120  &  58503.835  &  11.39  &  z  &  18.154  &  0.101  &  fa16    \\   
20190121  &  58504.560  &  12.12  &  U  &  18.157  &  0.187  &  fa11    \\   
20190121  &  58504.560  &  12.12  &  B  &  18.605  &  0.119  &  fa11    \\   
20190121  &  58504.560  &  12.12  &  V  &  18.342  &  0.089  &  fa11    \\   
20190121  &  58504.570  &  12.13  &  g  &  18.257  &  0.080  &  fa11    \\   
20190121  &  58504.570  &  12.13  &  r  &  18.118  &  0.063  &  fa11    \\   
20190121  &  58504.570  &  12.13  &  i  &  18.073  &  0.069  &  fa11    \\   
20190121  &  58504.570  &  12.13  &  z  &  18.148  &  0.160  &  fa11    \\   
20190124  &  58507.059  &  14.62  &  g  &  18.837  &  0.302  &  ASAS-SN    \\   
20190124  &  58507.150  &  14.71  &  V  &  18.435  &  0.027  &  EFOSC    \\   
20190124  &  58507.316  &  14.88  &  o  &  19.413  &  0.389  &  ATLAS    \\   
20190125  &  58508.525  &  16.08  &  U  &  18.849  &  0.044  &  fa12    \\   
20190125  &  58508.525  &  16.08  &  B  &  19.047  &  0.022  &  fa12    \\   
20190125  &  58508.525  &  16.08  &  V  &  18.600  &  0.019  &  fa12    \\   
20190125  &  58508.545  &  16.10  &  g  &  18.682  &  0.016  &  fa12    \\   
20190125  &  58508.545  &  16.10  &  r  &  18.479  &  0.020  &  fa12    \\   
20190125  &  58508.545  &  16.10  &  i  &  18.432  &  0.025  &  fa12    \\   
20190126  &  58509.071  &  16.63  &  g  &  18.582  &  0.239  &  ASAS-SN    \\   
20190126  &  58509.103  &  16.66  &  J  &  17.889  &  0.043  &  SOFI    \\   
20190126  &  58509.103  &  16.66  &  H  &  17.502  &  0.046  &  SOFI    \\   
20190126  &  58509.103  &  16.66  &  K  &  17.020  &  0.063  &  SOFI    \\   
20190129  &  58512.510  &  20.07  &  B  &  >18.2  &         &  fa12    \\   
20190129  &  58512.525  &  20.08  &  g  &  >17.7  &         &  fa12    \\   
20190129  &  58512.525  &  20.08  &  r  &  >17.7  &         &  fa12    \\   
20190129  &  58512.525  &  20.08  &  i  &  >17.3  &         &  fa12    \\   
20190130  &  58513.840  &  21.40  &  B  &  19.952  &  0.041  &  fa14    \\   
20190130  &  58513.840  &  21.40  &  V  &  19.225  &  0.039  &  fa14    \\   
20190130  &  58513.855  &  21.42  &  g  &  19.465  &  0.026  &  fa14    \\   
20190130  &  58513.855  &  21.42  &  r  &  19.094  &  0.037  &  fa14    \\   
20190130  &  58513.855  &  21.42  &  i  &  19.010  &  0.049  &  fa14    \\   
20190203  &  58517.920  &  25.48  &  B  &  20.430  &  0.067  &  fa14    \\   
20190204  &  58518.885  &  26.44  &  B  &  20.524  &  0.046  &  fa16    \\   
20190204  &  58518.885  &  26.44  &  V  &  19.836  &  0.048  &  fa16    \\   
20190204  &  58518.895  &  26.45  &  g  &  19.969  &  0.032  &  fa16    \\   
20190204  &  58518.895  &  26.45  &  r  &  19.513  &  0.028  &  fa16    \\   
20190204  &  58518.895  &  26.45  &  i  &  19.462  &  0.038  &  fa16    \\   
20190205  &  58519.330  &  26.89  &  o  &  19.455  &  0.214  &  ATLAS    \\   
20190208  &  58522.160  &  29.72  &  U  &  20.843  &  0.075  &  EFOSC    \\   
20190208  &  58522.160  &  29.72  &  V  &  20.134  &  0.061  &  EFOSC    \\   
20190208  &  58522.160  &  29.72  &  z  &  19.473  &  0.042  &  EFOSC    \\   
20190209  &  58523.135  &  30.69  &  B  &  21.045  &  0.063  &  fa03    \\   
20190209  &  58523.135  &  30.69  &  V  &  20.213  &  0.035  &  fa03    \\   
20190209  &  58523.160  &  30.72  &  g  &  20.428  &  0.032  &  fa03    \\   
20190209  &  58523.160  &  30.72  &  r  &  19.995  &  0.028  &  fa03    \\   
20190209  &  58523.160  &  30.72  &  i  &  19.847  &  0.048  &  fa03    \\   
20190209  &  58523.323  &  30.88  &  o  &  19.503  &  0.209  &  ATLAS    \\   
20190216  &  58530.885  &  38.44  &  B  &  21.304  &  1.000  &  fa14    \\   
20190216  &  58530.885  &  38.44  &  V  &  21.028  &  0.210  &  fa14    \\   
20190216  &  58530.900  &  38.46  &  g  &  20.610  &  0.118  &  fa14    \\   
20190216  &  58530.905  &  38.46  &  r  &  20.307  &  0.100  &  fa14    \\   
20190216  &  58530.905  &  38.46  &  i  &  20.193  &  0.094  &  fa14    \\   
20190221  &  58535.115  &  42.67  &  B  &  >20.5  &         &  fa03    \\   
20190221  &  58535.115  &  42.67  &  V  &  21.095  &  0.230  &  fa03    \\   
20190221  &  58535.140  &  42.70  &  g  &  >20.7  &         &  fa03    \\   
20190221  &  58535.140  &  42.70  &  r  &  20.942  &  0.187  &  fa03    \\   
20190221  &  58535.140  &  42.70  &  i  &  20.832  &  0.205  &  fa03    \\   
20190226  &  58540.130  &  47.69  &  J  &  19.460  &  0.070  &  SOFI    \\   
20190226  &  58540.130  &  47.69  &  H  &  18.399  &  0.059  &  SOFI    \\   
20190226  &  58540.130  &  47.69  &  K  &  17.586  &  0.080  &  SOFI    \\   
20190301  &  58543.276  &  50.84  &  o  &  20.104  &  0.403  &  ATLAS    \\   
20190308  &  58550.120  &  57.68  &  U  &  >22.3  &         &  EFOSC    \\   
20190308  &  58550.120  &  57.68  &  B  &  22.972  &  0.128  &  EFOSC    \\   
20190308  &  58550.120  &  57.68  &  V  &  22.124  &  0.068  &  EFOSC    \\   
20190308  &  58550.140  &  57.70  &  g  &  22.276  &  0.114  &  EFOSC    \\   
20190308  &  58550.140  &  57.70  &  r  &  21.776  &  0.063  &  EFOSC    \\   
20190308  &  58550.140  &  57.70  &  i  &  21.454  &  0.062  &  EFOSC    \\   
20190308  &  58550.140  &  57.70  &  z  &  21.205  &  0.100  &  EFOSC    \\   
20190309  &  58551.130  &  58.69  &  V  &  >20.5  &         &  EFOSC    \\   
20190320  &  58562.030  &  69.59  &  J  &  20.886  &  0.111  &  SOFI    \\   
20190320  &  58562.030  &  69.59  &  H  &  19.544  &  0.098  &  SOFI    \\   
20190320  &  58562.030  &  69.59  &  K  &  18.365  &  0.079  &  SOFI    \\   

\end{longtable}

\newpage 

\section{Spectroscopic tables}
\label{SpecInfo}

\begin{longtable}{cccccccc}
\caption{Log of spectroscopic observations of SN\,2018jmt.}\label{2018jmtSpecInfo}\\
\hline \hline
Date & MJD & Phase$^a$ & Telescope+Instrument & Grism/Grating+Slit & Spectral range & Resolution & Exp. time \\ 
  &   & (days) &   &        & (\AA)    & (\AA)           & (s)           \\ 
\hline 
20181216 & 58468.3  & +2.6  & NTT+EFOSC2   & gr13+1.0"       & 3640-9230  & 21 & 300        \\
20181217 & 58469.2  & +3.5  & NTT+EFOSC2   & gr11+1.0"       & 3340-7460  & 16 & 1500       \\
20181220 & 58472.7  & +7.0  & COJ 2m+en05  & red/blu+2.0"    & 3150-10870 & 18 & 2700       \\
20181221 & 58473.2  & +7.5  & SOAR+Goodman & 400 l/mm+1.0" & 3390-8720  & 6  & 900        \\
20190101 & 58484.1  & +18.4 & NTT+EFOSC2   & gr11/gr16+1.0"  & 3340-9990  & 16 & 2700/2700       \\
20190124 & 58507.2  & +41.5 & NTT+EFOSC2   & gr13+1.0"       & 3640-9240  & 21 & 2700       \\
20190207 & 58521.9  & +56.2 & SALT+RSS     & PG0300+1.5"     & 3590-8430  & 19 & 1800       \\
\hline
\multicolumn{7}{l}{{$^a$Phases are relative to $g$-band maximum light (MJD = 58465.66 $\pm$ 1.20; 2018-12-13) in observer frame.}} \\
\end{longtable}
\begin{longtable}{@{\extracolsep{0.1em}}cccccccc}
\caption{Log of spectroscopic observations of SN\,2019cj.}
\label{2019cjSpecInfo} \\
\hline \hline
Date & MJD & Phase$^a$ & Telescope+Instrument & Grism/Grating+Slit & Spectral range & Resolution & Exp. time \\ 
  &   & (days) &   &        & (\AA)    & (\AA)           & (s)           \\ 
\hline
20190107 & 58490.3  & $-2.1$  & NTT+EFOSC2  & gr13+1.0"     & 3640-9230  & 21 & 600    \\
20190108 & 58491.3  & $-1.1$  & NTT+EFOSC2  & gr13+1.0"     & 3630-9230  & 21 & 600    \\
20190109 & 58492.2  & $-0.2$  & NTT+EFOSC2  & gr11+1.0"     & 3340-7460  & 16 & 2400   \\
20190112 & 58495.6  & +3.2  & COJ 2m+en05 & red/blu+2.0"  & 3150-10870 & 18 & 2700   \\
20190115 & 58498.1  & +5.7  & NTT+EFOSC2  & gr11+1.0"     & 3340-7460  & 16 & 2700   \\
20190124 & 58507.1  & +14.7 & NTT+EFOSC2  & gr11+1.0"     & 3340-7460  & 16 & 2x2700 \\
20190127 & 58510.9  & +18.5 & SALT+RSS    & PG0300+1.5"    & 3540-8330  & 19 &  1200   \\
\hline
\multicolumn{7}{l}{{$^a$Phases are relative to $V$-band maximum light (MJD = 58492.44 $\pm$ 0.23; 2019-01-09) in observer frame.}} \\
\end{longtable}


\twocolumn

\section{Acknowledgements}
{We gratefully thank the anonymous referee for his/her insightful comments and suggestions that improved the paper.} We thank J. Burke, C. Pellegrino for their LCO data, thank S. J. Smartt for his ePESSTO support, and thank K. Maguire, V. Brinnel, C. Barbarino, A. Razza for conducting part of the ePESSTO observations. Y.-Z. Cai thanks for the helpful discussion with Zhengwei Liu.
Y.-Z. Cai is supported by the National Natural Science Foundation of China (NSFC, Grant No. 12303054) and the Yunnan Fundamental Research Projects (Grant No. 202401AU070063). BW, JJZ and YZC are supported by the International Centre of Supernovae, Yunnan Key Laboratory (No. 202302AN360001).
AP, AR, EC, NER, SB, and GV acknowledge support from the PRIN-INAF 2022 project ``Shedding light on the nature of gap transients: from the observations to the model''.
AR also acknowledges financial support from the GRAWITA Large Program Grant (PI P. D'Avanzo).
KM acknowledges support from the JSPS KAKENHI grant JP20H00174 and JP24H01810. 
{RC acknowledges support from Gemini ANID ASTRO21-0036.}
{T.-W.C., AA acknowledge the Yushan Fellow Program by the Ministry of Education, Taiwan for the financial support (MOE-111-YSFMS-0008-001-P1).}
AGY's research is supported by the EU via ERC grant No. 725161, the ISF GW excellence center, an IMOS space infrastructure grant and a GIF grant, as well as the \text{Andr\'{e}} Deloro Institute for Advanced Research in Space and Optics, The Helen Kimmel Center for Planetary Science, the Schwartz/Reisman Collaborative Science Program and the Norman E Alexander Family M Foundation ULTRASAT Data Center Fund, Minerva and Yeda-Sela;  AGY is the incumbent of the The Arlyn Imberman Professorial Chair.
MN is supported by the European Research Council (ERC) under the European Union's Horizon 2020 research and innovation programme (grant agreement No.~948381) and by UK Space Agency Grant No.~ST/Y000692/1.
F.O.E.\ acknowledges support from the FONDECYT grant nr.\ 1201223.
M.P. acknowledges support from a UK Research and Innovation Fellowship (MR/T020784/1).
Maokai Hu is Supported by the Postdoctoral Fellowship Program of CPSF under Grant Number GZB20240376 and the Shuimu Tsinghua Scholar Program.
Q.W. is supported in part by NASA grants 80NSSC22K0494, 80NSSC21K0242 and 80NSSC19K0112. Q.W. is also partially supported by STScI DDRF fund.
L.G. acknowledges financial support from AGAUR, CSIC, MCIN and AEI 10.13039/501100011033 under projects PID2020-115253GA-I00, PIE 20215AT016, CEX2020-001058-M, and 2021-SGR-01270.
MR acknowledges support from National Agency for Research and Development (ANID) grants ANID-PFCHA/Doctorado Nacional/2020-21202606.
D.-D.Shi acknowledges the support from the National Science Foundation of China (12303015) and the National Science Foundation of Jiangsu Province (BK20231106).
B. Warwick acknowledges the support from UKRI's STFC studentship grant funding, project reference ST/X508871/1.
J.Z. is supported by the National Key R$\&$D Program of China with No. 2021YFA1600404, the National Natural Science Foundation of China (12173082), the science research grants from the China Manned Space Project with No. CMS-CSST-2021-A12, the Yunnan Province Foundation (202201AT070069), the Top-notch Young Talents Program of Yunnan Province, the Light of West China Program provided by the Chinese Academy of Sciences, the International Centre of Supernovae, Yunnan Key Laboratory (No. 202302AN360001).
B. Wang is supported by the National Natural Science Foundation of China (No 12225304) and the Western Light Project of CAS (No. XBZG-ZDSYS-202117).
XFW is supported by the National Natural Science Foundation of China (NSFC grants 12288102, 12033003, 11633002, and 12303047) and the Tencent Xplorer Prize.
XJZ is supported by the National Natural Science Foundation of China (Grant No.~12203004) and by the Fundamental Research Funds for the Central Universities.
This paper includes data collected by the $TESS$ mission. Funding for the $TESS$ mission is provided by the NASA's Science Mission Directorate.
Some of the observations reported in this paper were obtained with the Southern African Large Telescope (SALT). The Inter-University Centre for Astronomy and Astrophysics (IUCAA), India is an official partner of SALT collaboration. RR acknowledges IUCAA SALT collaboration for providing the observing time at SALT under SALT large science proposal "Observing the Transient Universe" with David Buckley as the Principal Investigator. Polish participation in SALT is funded by grant No. MEiN nr 2021/WK/01.
Based on observations collected at the European Southern Observatory under ESO programmes 199.D-0143, 0102.A-9099(A) and data obtained from the ESO Science Archive Facility with DOI(s) under https://doi.org/10.18727/archive/86.
Part of the funding for GROND (both hardware as well as personnel) was generously granted from the Leibniz-Prize to Prof. G. Hasinger (DFG grant HA 1850/28-1).
This work makes use of data from the Las Cumbres Observatory Network and the Global Supernova Project. The LCO team is supported by U.S. NSF grants AST-1911225 and AST-1911151, and NASA.
We thank Las Cumbres Observatory and its staff for their continued support of ASAS-SN. ASAS-SN is funded in part by the Gordon and Betty Moore Foundation through grants GBMF5490 and GBMF10501 to the Ohio State University, and also funded in part by the Alfred P. Sloan Foundation grant G-2021-14192. Development of ASAS-SN has been supported by NSF grant AST-0908816, the Mt. Cuba Astronomical Foundation, the Center for Cosmology and AstroParticle Physics at the Ohio State University, the Chinese Academy of Sciences South America Center for Astronomy (CAS-SACA), and the Villum Foundation.
SDSS is managed by the Astrophysical Research Consortium for the Participating Institutions of the SDSS Collaboration including the Brazilian Participation Group, the Carnegie Institution for Science, Carnegie Mellon University, Center for Astrophysics | Harvard \& Smithsonian (CfA), the Chilean Participation Group, the French Participation Group, Instituto de Astrof\'isica de Canarias, The Johns Hopkins University, Kavli Institute for the Physics and Mathematics of the Universe (IPMU) / University of Tokyo, the Korean Participation Group, Lawrence Berkeley National Laboratory, Leibniz Institut f\"{u}r Astrophysik Potsdam (AIP), Max-Planck-Institut f\"{u}r Astronomie (MPIA Heidelberg), Max-Planck-Institut f\"{u}r Astrophysik (MPA Garching), Max-Planck-Institut f\"{u}r Extraterrestrische Physik (MPE), National Astronomical Observatories of China, New Mexico State University, New York University, University of Notre Dame, Observat\'orio Nacional / MCTI, The Ohio State University, Pennsylvania State University, Shanghai Astronomical Observatory, United Kingdom Participation Group, Universidad Nacional Aut\'onoma de M\'exico, University of Arizona, University of Colorado Boulder, University of Oxford, University of Portsmouth, University of Utah, University of Virginia, University of Washington, University of Wisconsin, Vanderbilt University, and Yale University.
This publication makes use of data products from the Two Micron All Sky Survey, which is a joint project of the University of Massachusetts and the Infrared Processing and Analysis Center/California Institute of Technology, funded by NASA and the National Science Foundation.
This work has made use of data from the Asteroid Terrestrial-impact Last Alert System (ATLAS) project. The Asteroid Terrestrial-impact Last Alert System (ATLAS) project is primarily funded to search for near earth asteroids through NASA grants NN12AR55G, 80NSSC18K0284, and 80NSSC18K1575; byproducts of the NEO search include images and catalogs from the survey area. This work was partially funded by Kepler/K2 grant J1944/80NSSC19K0112 and HST GO-15889, and STFC grants ST/T000198/1 and ST/S006109/1. The ATLAS science products have been made possible through the contributions of the University of Hawaii Institute for Astronomy, the Queen's University Belfast, the Space Telescope Science Institute, the South African Astronomical Observatory, and The Millennium Institute of Astrophysics (MAS), Chile.
This research is based in part on observations obtained at the Southern Astrophysical Research (SOAR) telescope, which is a joint project of the Minist\'{e}rio da Ci\^{e}ncia, Tecnologia, e Inova\c{c}\~{a}o (MCTI) da Rep\'{u}blica Federativa do Brasil, the U.S. National Optical Astronomy Observatory (NOAO), the University of North Carolina at Chapel Hill (UNC), and Michigan State University (MSU).
This research has made use of the NASA/IPAC Extragalactic Database (NED), which is operated by the Jet Propulsion Laboratory, California Institute of Technology, under contract with the National Aeronautics and Space Administration.
{\sc iraf} was distributed by the National Optical Astronomy Observatory, which was managed by the Association of Universities for Research in Astronomy (AURA), Inc., under a cooperative agreement with the U.S. NSF.

\end{appendix}
\end{document}